\documentclass[journal=jpccck, manuscript=article]{achemso}
\usepackage{colortbl}
\usepackage[numbers,sort&compress]{natbib}
\usepackage[utf8]{inputenc}
\usepackage{enumerate}
\usepackage{graphicx}
\usepackage{caption}
\usepackage{subcaption}
\usepackage{amsmath}
\usepackage{verbatim}
\usepackage{hyperref}
\hypersetup{colorlinks, allcolors=black, urlcolor=blue}
\usepackage{booktabs}
\usepackage{tabularx}

\usepackage{hhline}

\SectionNumbersOn

\makeatletter
\DeclareRobustCommand*\textsubscript[1]{\@textsubscript{\selectfont#1}}
\def\@textsubscript#1{{\m@th\ensuremath{_{\mbox{\fontsize\sf@size\z@#1}}}}}
\makeatother

\newcommand{\figref}[1]{Figure~\ref{#1}}

\newcommand{\GW}{G$_0$\!W\!$_0$ }
\newcolumntype{L}[1]{>{\raggedright\let\newline\\\arraybackslash\hspace{0pt}}m{#1}}
\newcolumntype{R}[1]{>{\raggedleft\let\newline\\\arraybackslash\hspace{0pt}}m{#1}}
\newcolumntype{C}[1]{>{\centering\let\newline\\\arraybackslash\hspace{0pt}}m{#1}}

\title{
The Computational 2D Materials Database: High-Throughput Modeling and Discovery of Atomically Thin Crystals}
\author{Sten Haastrup}
\email{sthaa@fysik.dtu.dk}
\author{Mikkel Strange}
\author{Mohnish Pandey}
\author{Thorsten Deilmann}
\author{Per S. Schmidt}
\author{Nicki F. Hinsche}
\author{Morten N. Gjerding}
\author{Daniele Torelli}
\author{Peter M. Larsen}
\author{Anders C. Riis-Jensen}
\author{Jakob Gath}
\author{Karsten W. Jacobsen}
\author{Jens Jørgen Mortensen}
\author{Thomas Olsen}
\author{Kristian S. Thygesen}
\affiliation{CAMD, Department of Physics, Technical University of Denmark}
\alsoaffiliation{Center for Nanostructured Graphene (CNG), Technical University of Denmark}
\date{\today}

\begin{document}

\begin{abstract}
We introduce the Computational 2D Materials Database (C2DB), which organises a variety of structural, thermodynamic, elastic, electronic, magnetic, and optical properties of around 1500 two-dimensional materials distributed over more than 30 different crystal structures. Material properties are systematically calculated by density functional theory and many-body perturbation theory (\GW and the Bethe-Salpeter Equation for $\sim$250 materials) following a semi-automated workflow for maximal consistency and transparency. The C2DB is fully open and can be browsed online at \url{c2db.fysik.dtu.dk} or downloaded in its entirety. In this paper, we describe the workflow behind the database, present an overview of the properties and materials currently available, and explore trends and correlations in the data. Moreover, we identify a large number of new potentially synthesisable 2D materials with interesting properties targeting applications within spintronics, (opto-)electronics, and plasmonics. The C2DB offers a comprehensive and easily accessible overview of the rapidly expanding family of 2D materials and forms an ideal platform for computational modeling and design of new 2D materials and van der Waals heterostructures.
\end{abstract}

\maketitle
\newpage
\tableofcontents
\newpage
\section{Introduction}
\label{sec:intro}
Over the past decade, atomically thin two-dimensional (2D) materials have made their way to the forefront of several research areas including batteries, (electro-)catalysis, electronics, and photonics\cite{ferrari2015science,bhimanapati2015recent}. This development was prompted by the intriguing and easily tunable properties of atomically thin crystals and has been fueled by the constant discovery of new 2D materials and the emergent concepts of lateral\cite{huang2014lateral} and vertical\cite{geim2013van} 2D heterostructures, which opens completely new possibilities for designing materials with tailored and superior properties. 

So far more than fifty compounds have been synthesised or exfoliated as single layers (see Figure \ref{fig:exp}). These include the well known monoelemental crystals (Xenes, e.g. graphene, phosphorene)\cite{molle2017buckled} and their ligand functionalised derivatives (Xanes, e.g. CF, GeH)\cite{bianco2013stability}, transition metal dichalcogenides (TMDCs, e.g. MoS$_2$, TaSe$_2$)\cite{wang2012electronics}, transition metal carbides and -nitrides (MXenes, e.g. Ti$_2$CO$_2$)\cite{naguib2012two}, group III-V semiconductors and insulators (e.g. GaN, BN)\cite{ci2010atomic,al2016two}, transition metal halides (e.g. CrI$_3$)\cite{lin2014towards,huang2017layer}, post-transition metal chalcogenides (e.g. GaS and GaSe)\cite{late2012a, hu2012a} and organic-inorganic hybrid perovskites (e.g. Pb(C$_4$H$_9$NH$_3$)$_2$I$_4$)\cite{dou2015atomically}. However, the already known monolayers are only the tip of a much larger iceberg. Indeed, recent data mining studies indicate that several hundred 2D materials could be exfoliated from known layered bulk crystals\cite{ashton2017topology,cheon2017data,mounet2016novel,choudhary2017a}. In the present work we take a complementary approach to 2D materials discovery based on combinatorial lattice decoration and identify another few hundred previously unknown and potentially synthesisable monolayers.     
 
In the search for new materials with tailored properties or novel functionalities, first-principles calculations are playing an increasingly important role. The continuous increase in computing power and significant advancements of theoretical methods and numerical algorithms have pushed the field to a point where first-principles calculations are comparable to experiments in terms of accuracy and greatly surpass them in terms of speed and cost. For more than a century, experimental databases on e.g. structural, thermal, and electronic properties, have been a cornerstone of materials science, and in the past decade, the experimental data have been augmented by an explosion of computational data obtained from first-principles calculations. Strong efforts are currently being focused on storing and organising the computational data in open repositories\cite{jain2016research,thygesen2016making}. Some of the larger repositories, together containing millions of material entries, are the Materials Project\cite{Jain2013}, the Automatic Flow for Materials Discovery (AFLOWLIB)\cite{curtarolo2012aflow}, the Open Quantum Materials Database (OQMD)\cite{saal2013materials,kirklin2015a}, and the Novel Materials Discovery (NOMAD) Repository\cite{nomad}. 

The advantages of computational materials databases are many. Most obviously, they facilitate open sharing and comparison of research data whilst reducing duplication of efforts. In addition, they underpin the development and benchmarking of new methods by providing easy access to common reference systems\cite{lejaeghere2016reproducibility}. Finally, the databases enable the application of machine learning techniques to identify deep and complex correlations in the materials space and to use them for designing materials with tailored properties and for accelerating the discovery of new materials\cite{hautier2010finding,ghiringhelli2015big,ward2017atomistic}. Among the challenges facing the computational databases is the quality of the stored data, which depends both on the numerical precision (e.g. the employed $k$-point grid and basis set size) and the accuracy of the employed physical models (e.g. the exchange-correlation functional).  Most of the existing computational databases store results of standard density functional theory (DFT) calculations. While such methods, when properly conducted, are quite reliable for ground state properties such as structural and thermodynamic properties, they are generally not quantitatively accurate for excited state properties such as electronic band structures and optical absorption spectra.

Compared to databases of bulk materials, databases of 2D materials are still few and less developed. Early work used DFT to explore the stability and electronic structures of monolayers of group III-V honeycomb lattices\cite{csahin2009monolayer, ciraci_cahangirov_2017} and the class of MX$_2$ transition metal dichalcogenides and oxides\cite{ataca2012stable}. Later, by data-filtering the inorganic crystal structure database (ICSD), 92 experimentally known layered crystals were identified and their electronic band structures calculated at the DFT level\cite{lebegue2013two}. Another DFT study, also focused on stability and band structures, explored around one hundred 2D materials selected from different structure classes\cite{miro2014atlas}. To overcome the known limitations of DFT, a database with many-body G$_0$W$_0$ band structures for 50 semiconducting TMDCs was established\cite{rasmussen2015computational}. Very recently, data mining of the Materials Project and experimental crystal structure databases in the spirit of Ref. \citenum{lebegue2013two}, led to the identification of close to one thousand experimentally known layered crystals from which single layers could potentially be exfoliated\cite{cheon2017data,ashton2017topology,mounet2016novel,choudhary2017a}. These works also computed basic energetic, structural and electronic properties of the monolayers (or at least selected subsets) at the DFT level. 

In this paper, we introduce the open Computational 2D Materials Database (C2DB) which organises a variety of \emph{ab-initio} calculated properties for more than 1500 different 2D materials. The key characteristics of the C2DB are:
\begin{itemize}
\item \textbf{Materials}: the database focuses entirely on 2D materials, i.e. isolated monolayers, obtained by combinatorial lattice decoration of known crystal structure prototypes.
\item \textbf{Consistency}: all properties of all materials are calculated using the same code and parameter settings following the same workflow for maximum transparency, reproducibility, and consistency of the data.
\item \textbf{Properties}: the database contains a large and diverse set of properties covering structural, thermodynamic, magnetic, elastic, electronic, dielectric and optical properties. 
\item \textbf{Accuracy}: Hybrid functionals (HSE06) as well as beyond-DFT many-body perturbation theory (G$_0$W$_0$) are employed to obtain quantitatively accurate band structures, and optical properties are obtained from the random phase approximation (RPA) and Bethe-Salpeter Equation (BSE).   
\item \textbf{Openness}: the database is freely accessible and can be directly downloaded and browsed online using simple and advanced queries.
\end{itemize}

The systematic combinatorial approach used to generate the structures in the database inevitably produces many materials that are unstable and thus unrealistic and impossible to synthesise in reality. Such ``hypothetical'' structures may, however, still be useful in a number of contexts, e.g. for method development and benchmarking, testing and training of machine learning algorithms, identification of trends and structure-property relationships, etc. For this reason we map out the properties of all but the most unstable (and thus chemically unreasonable) compounds. Nevertheless, the reliable assessment of stability and synthesisability of the candidate structures is an essential issue. Using the 55 materials in the C2DB, which have been experimentally synthesised in monolayer form, as a guideline, we set down the criteria that a hypothesised 2D material should fulfill in order for it to be ``likely synthesisable''. On the basis of these criteria, we introduce a simple \emph{stability scale} to quantify a candidate material's dynamic and thermodynamic stability. Out of an initial set of more than 1900 monolayers distributed over 32 different crystal structures, we find 350 in the most stable category. In addition to the 55 experimentally synthesised monolayers, this set also includes around 80 monolayers from experimentally known vdW layered bulk materials, and thus around 200 completely new and potentially synthesisable 2D materials.

In Section~\ref{sec:workflow}, we describe the computational workflow behind the database. The structure and properties of the materials are calculated using well established state-of-the-art methodology. Technical descriptions of the different steps in the workflow are accompanied by illustrative examples and comparisons with literature data. Since \emph{documentation and validation} is the main purpose of the section, we deliberately focus on well known 2D materials like the Xenes and transition metal dichalcogenides where plenty of both computational and experimental reference data is available. It should be clear that the novelty of the present work does not lie in the employed methodology nor in the type of materials properties that we calculate. The significance of our work is rather reflected by the fact that when large and consistently produced data sets are organised and made easily accessible, new scientific opportunities arise. As outlined below, this paper presents several examples of this effect.   

In Section~\ref{sec:overview} we give an overview of the materials and the data contained in the C2DB and provide some specific examples to illustrate its use. Using an extensive set of many-body G$_0$W$_0$ calculations as a reference, we establish the performance of various DFT xc-functionals for predicting band gaps, band edge positions, and band alignment at hetero-interfaces, and we propose an optimal strategy for obtaining accurate band energies at low computational cost. Similarly, the 250 BSE calculations allow us to explore trends in exciton binding energies and perform a statistically significant and unbiased assessment of the accuracy and limitations of the widely used Mott-Wannier model for 2D excitons. From the data on more than 600 semiconductor monolayers, we present strong empirical evidence against an often employed relation between effective masses and band gaps derived from \(\mathbf k\cdot \mathbf p\) perturbation theory. Inspired by the potential of using 2D materials as building blocks for plasmonics and photonics, we propose a model to predict the plasmon dispersion relations in 2D metals from the (intraband) plasma frequency and the onset of interband transitions and use it to identify 2D metals with plasmons in the optical frequency regime. We propose several new magnetic 2D materials (including both metals and semiconductors) with ferromagnetic or anti-ferromagnetic ordering and significant out-of-plane magnetic anisotropy. Finally, we point to new high-mobility 2D semiconductors including some with band gaps in the range of interest for (opto)electronic applications. 

In Section ~\ref{sec:conclusion} we provide our conclusions together with an outlook discussing some opportunities and possible future directions for the C2DB.
 
\section{Workflow}
\label{sec:workflow}
\begin{figure}[ht]
  \centering
  \includegraphics[width=\textwidth]{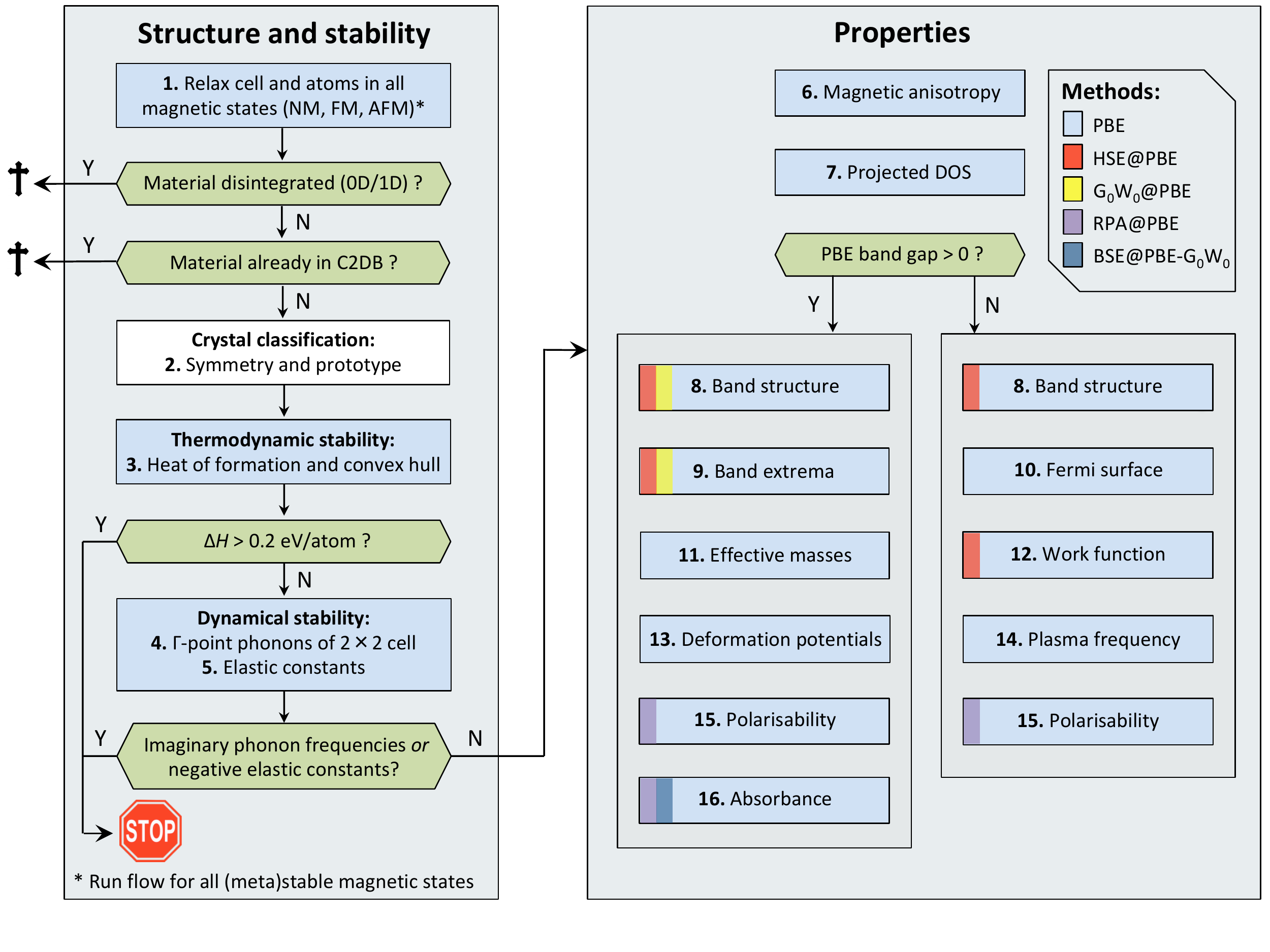}
  \caption{The workflow used to calculate the structure and properties of the materials in C2DB. The cross indicates that the material is not included in the database at all, while the stop sign indicates that no more of the workflow is performed.}
  \label{fig:workflow}
\end{figure}

 The workflow used to generate the data in the C2DB is illustrated in Figure~\ref{fig:workflow}. It consists of two parts: In the first part (left panel) the unit cell and atom positions are optimised for different magnetic configurations: non-magnetic (NM), ferro-magnetic (FM) and antiferro-magnetic (AFM). Materials satisfying certain stability and geometry criteria (indicated by green boxes) are subject to the second part (right panel) where the different properties are computed using DFT and many-body methods. The \GW  band structure and BSE absorbance calculations have been performed only for semiconducting materials with up to four atoms in the unit cell. Per default, properties shown in the online database include spin-orbit coupling (SOC); however, to aid comparison with other calculations, most properties are also calculated and stored without SOC.

All DFT and many-body calculations are performed with the projector augmented wave code GPAW \cite{enkovaara2010electronic} using plane wave basis sets and PAW potentials version 0.9.2. The workflow is managed using the Python based Atomic Simulation Environment (ASE)\cite{larsen2017atomic}. We have developed a library of robust and numerically accurate (convergence verified) ASE-GPAW scripts to perform the various tasks of the workflow, and to create the database afterwards. The library is freely available, under a GPL license. 

Below we describe all steps of the workflow in detail. As the main purpose is to document the workflow, the focus is on technical aspects, including numerical convergence and benchmarking. An overview of the most important parameters used for the different calculations is provided in Table~\ref{tab:parameters}. 

{\renewcommand{\arraystretch}{1.5}
\begin{table}[t]
\caption{Overview of the methods and parameters used for the different steps of the workflow. If a parameter is not specified at a given step, its value equals that of the last step where it was specified.}\label{tab:parameters}
\centering
\begin{tabular}{lp{9cm}}
\toprule
Workflow step(s) & Parameters \\
\midrule
Structure and energetics (1-4)\(^{\dagger}\)         & vacuum = 15\,\AA; $k$-point density = 6.0/\AA$^{-1}$; Fermi smearing = 0.05\,eV; PW cutoff = 800\,eV; xc functional = PBE; maximum force = 0.01\,eV/\AA; maximum stress = 0.002\,eV/\AA$^3$; phonon displacement = 0.01\AA \\

Elastic constants (5)           & $k$-point density = $12.0/\mathrm{\AA}^{-1}$; strain = $\pm$1\%  \\

Magnetic anisotropy (6)           & $k$-point density = $20.0/\mathrm{\AA}^{-1}$; spin-orbit coupling = True\\

PBE electronic properties (7-10 and 12)     & $k$-point density = $12.0/\mathrm{\AA}^{-1}$ ($36.0/\mathrm{\AA}^{-1}$ for step 7) \\

Effective masses (11)     & $k$-point density = $45.0/\mathrm{\AA}^{-1}$; finite difference\\

  Deformation potential (13)     & $k$-point density = 12.0/\AA$^{-1}$; strain = $\pm$1\% \\

Plasma frequency (14)     & $k$-point density = 20.0/\AA$^{-1}$; tetrahedral interpolation \\

HSE band structure (8-12)     & HSE06@PBE; $k$-point density = 12.0/\AA$^{-1}$\\

\GW  band structure  (8, 9)& G$_0$W$_0$@PBE; $k$-point density = $5.0/\mathrm{\AA}^{-1}$; PW cutoff = $\infty$ (extrapolated from 170, 185 and 200~eV); full frequency integration; analytical treatment of $W({q})$ for small $q$; truncated Coulomb interaction \\

RPA polarisability (15)      & RPA@PBE; $k$-point density = $20.0/\mathrm{\AA}^{-1}$; PW cutoff = 50~eV; truncated Coulomb interaction; tetrahedral interpolation \\

BSE absorbance (16)                  & BSE@PBE with \GW  scissors operator; $k$-point density = $20.0/\mathrm{\AA}^{-1}$; PW cutoff = 50~eV; truncated Coulomb interaction; at least 4 occupied and 4 empty bands \\
  \bottomrule
  \multicolumn{2}{L{16cm}}{\(^{\dagger}\)\footnotesize{For the cases with convergence issues, we set a \(k\)-point density of 9.0 and a smearing of 0.02 eV.}}
\end{tabular}
\end{table}
}
\clearpage

\subsection{Structure relaxation}
\label{sec:relaxation}
\begin{figure}[htb]
  \centering
  \includegraphics[width=\textwidth]{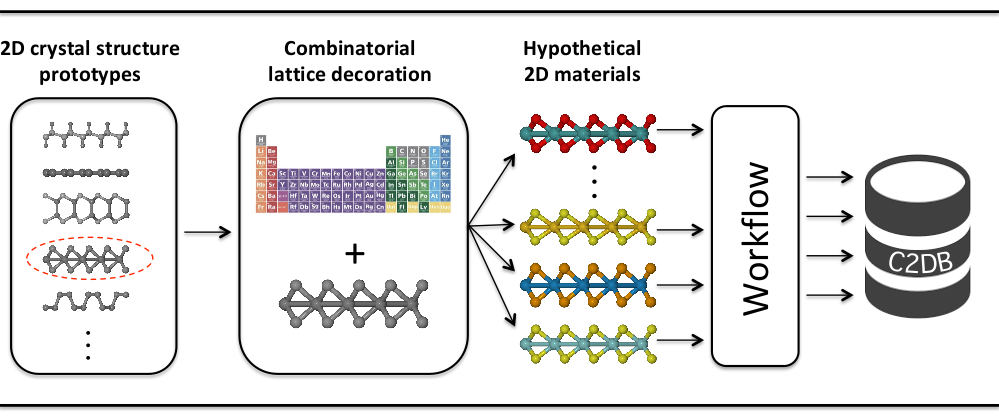}
  \caption{The materials in the C2DB are initially generated by decorating an experimentally known crystal structure prototype with atoms chosen from a (chemically reasonable) subset of the periodic table.}
  \label{fig:database}
\end{figure}

The workflow is initiated with a crystal structure defined by its unit cell (Bravais lattice and atomic basis). The crystal lattice is typically that of an experimentally known prototype (the ``seed structure'') decorated with atoms picked from a subset of the periodic table, see Figure~\ref{fig:database}.
We refer to materials by the chemical formula of their unit cell followed by the crystal structure. The latter is indicated by a representative material of that prototype, as described in Sec~\ref{sec:materials}. For example, monolayer MoS$_2$ in the hexagonal H and T phases are denoted MoS$_2$-MoS$_2$ and MoS$_2$-CdI$_2$, respectively. Now, MoS$_2$ is in fact not stable in the T phase, but undergoes a $2\times 1$ distortion to the so-called T$'$ phase. Because the T$'$ phase is the thermodynamically stable phase of WTe$_2$, we denote MoS$_2$ in the distorted T phase by Mo$_2$S$_4$-WTe$_2$. In the following, we shall refer to the unit cell with which the workflow is initiated, i.e. the unit cell of the seed structure, as the primitive cell or the $1\times 1$ cell, even if this cell is not dynamically stable for the considered material (see Section~\ref{sec:stability}).

The unit cell and internal coordinates of the atoms are relaxed in both a spin-paired (NM), ferromagnetic (FM), and anti-ferromagnetic (AFM) configuration. Calculations for the AFM configuration are performed only for unit cells containing at least two metal atoms. The symmetries of the initial seed structure are kept during relaxation. All relevant computational details are provided in Table \ref{tab:parameters}.

\begin{figure}[htb]
  \centering
  \includegraphics{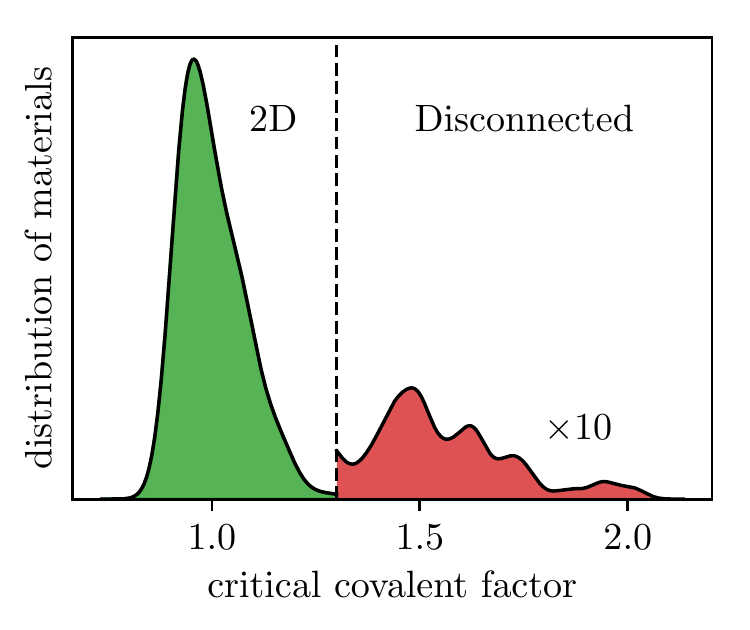}
\caption{The distribution of candidate structures for the C2DB with respect to the critical covalent factor at which they become 2D. Materials in the red region are excluded from the database.}
\label{fig:dimensionality}
\end{figure}

After relaxation, we check that the structure has remained a covalently connected 2D material and not disintegrated into 1D or 0D clusters. This is done by defining clusters of atoms using the covalent radius\cite{B801115J} + 30\% as a measure for covalent bonds between atoms. The dimensionality of a cluster is obtained from the scaling of the number of atoms in a cluster upon repetition of the unit cell following the method described by Ashton \emph{et al.}~\cite{ashton2017topological}. Only materials containing exactly one cluster of dimensionality 2 are given further consideration (an exception is made for the metal-organic perovskites (prototype PbA$_2$I$_4$) for which the metal atom inside the octahedron represents a 0D cluster embedded in a 2D cluster). To illustrate the effect of the covalent radius + 30\% threshold, Figure~\ref{fig:dimensionality} shows the distribution of the candidate structures in the database as a function of the covalent factor needed to fully connect the structure. Most materials have a critical covalent factor below 1.3 and fall in the green shaded region. There is, however, a tail of around 100 disconnected materials (red region); these materials are not included in the database.

We also check that the material is not already contained in the database. This is done by measuring the root mean square distance (RMSD)\cite{larsen2017_structural_analysis} relative to all other materials in the C2DB with the same reduced chemical formula. A threshold of 0.01 \AA~ is used for this test.

In case of multiple metastable magnetic configurations (in practice, if both a FM and AFM ground state are found), these are regarded as different phases of the same material and will be treated separately throughout the rest of the workflow. To indicate the magnetic phase we add the extensions ``FM'' or ``AFM'' to the material name. The total energy of the spin-paired ground state is always stored, even when it is not the lowest. If the energy of the non-magnetic state is higher than the most stable magnetic state by less than 10 meV/atom, the workflow is also performed for the non-magnetic state. This is done in recognition of the finite accuracy of DFT for predicting the correct energetic ordering of different magnetic states.

We have compared the lattice constants of 29 monolayers with those reported in Ref. \citenum{ozccelik2016band}, which were obtained with the VASP code using PBE and very similar numerical settings and find a mean absolute deviation of 0.024 \AA~corresponding to 0.4\%. The small yet finite deviations are ascribed to differences in the employed PAW potentials.

\subsection{Crystal structure classification}
\subsubsection{Symmetry} 
To classify the symmetries of the crystal structure the 3D space group is determined using the crystal symmetry library Spglib\cite{spglib} on the 3D supercell with a tolerance of $10^{-4}$ \AA.

\subsubsection{Prototypes}
\label{sec:prototypes}
\begin{figure}[htb]
  \centering
  \includegraphics{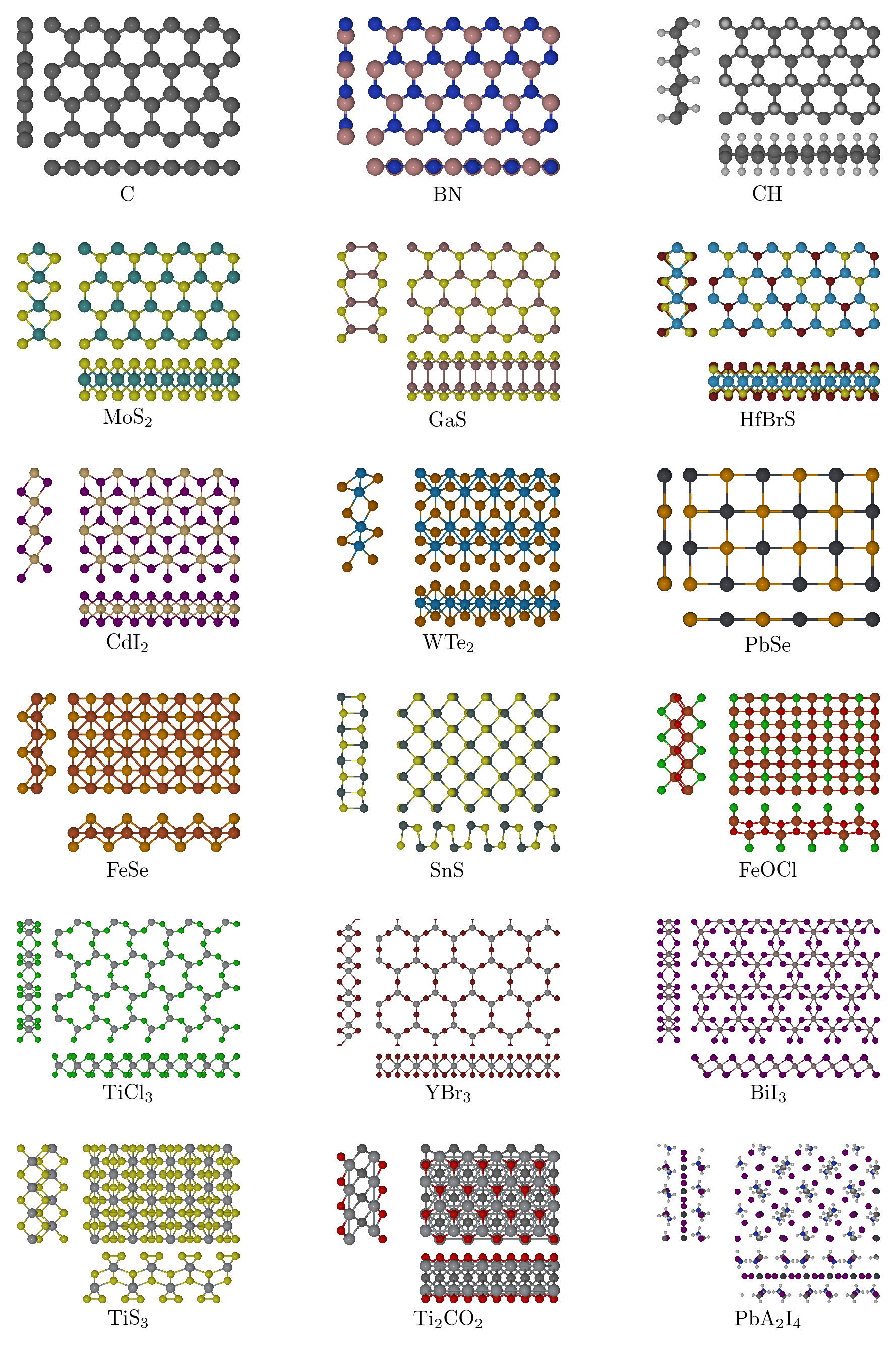}
  \caption{Examples of crystal structure prototypes currently included in the C2DB.  \label{fig:prototype_overview}}
\end{figure}

The materials are classified into crystal structure prototypes based on the symmetry of the crystals. For two materials to belong to the same prototype, we require that they have the same space group, the same stoichiometry, and comparable thicknesses. The last requirement is included to distinguish between materials with the same symmetry and stoichiometry but with different number of atomic layers, see for example monolayer hBN and GaS in Figure~\ref{fig:prototype_overview}. Each prototype is labelled by a specific representative material. For prototypes which have been previously investigated, we comply with the established conventions. However, since the field of 2D materials is still young and because C2DB contains a large number of never-synthesised materials, some of the considered crystal structures fall outside the known prototypes. In these cases we have chosen the representative material to be the one with the lowest energy with respect to the convex hull. Some of the crystal structure prototypes presently contained in the C2DB are shown in Figure~\ref{fig:prototype_overview}.

\clearpage

\subsection{Thermodynamic stability}
The heat of formation, $\Delta H$, is defined as the energy of the material with respect to the standard states of its constituent elements. For example, the heat of formation per atom of a binary compound, A$_x$B$_y$, is 
\begin{align}
\Delta H = (E(\text{A}_x\text{B}_y) - xE(\text{A}) - yE(\text{B}))/(x+y),
\end{align}
where $E(\text{A}_x\text{B}_y)$ is the total energy of the material A$_x$B$_y$,
and $E(\text{A})$ and $E(\text{B})$ are the total energies of the elements A and B in their
standard state. When assessing the stability of a material in the C2DB, it should be kept in mind that the accuracy of the PBE functional for the heat of formation is only around 0.2 eV/atom on average\cite{PhysRevB.91.235201}. Other materials databases, e.g. OQMD, Materials Project, and AFLOW, employ fitted elementary reference energies (FERE)\cite{stevanovic2012correcting} and apply a Hubbard \(U\) term \cite{anisimov1991density} for  the rare earth and transition metal atoms (or a selected subset of them). While such correction schemes in general improve \(\Delta H\) they also introduce some ambiguity, e.g. the dataset from which the FERE are determined or the exact form of the orbitals on which the \(U\) term is applied. Thus in order not to compromise the transparency and reproducibility of the data we use the pure PBE energies.   

For a material to be thermodynamically stable it is necessary but not sufficient that $\Delta H<0$. Indeed, thermodynamic stability requires that $\Delta H$ be negative not only relative to its pure elemental phases but relative to all other competing phases, i.e. its energy must be below the \emph{convex hull}.\cite{C7EE02702H} We stress, however, that in general, but for 2D materials in particular, this definition cannot be directly applied as a criterion for stability and synthesisability. The most important reasons for this are (i) the intrinsic uncertainty on the DFT energies stemming from the approximate xc-functional (ii) substrate interactions or other external effects that can stabilise the monolayer (iii) kinetic barriers that separate the monolayer from other lower energy phases rendering the monolayer (meta)stable for all practical purposes.   

We calculate the energy of the 2D material relative to the convex hull of competing bulk phases, $\Delta H_{\text{hull}}$. The convex hull is currently constructed from the 2836 most stable binary bulk compounds which were obtained from the OQMD\cite{saal2013materials}. The energies of the bulk phases were recalculated with GPAW using the PBE xc-functional and the same numerical settings as applied for the 2D materials (but the structure was not re-optimised). Because the bulk reference structures from OQMD were optimised with the VASP code and with Hubbard $U$-corrections for materials containing 3$d$ elements, and because the PBE misses attractive vdW interaction, the bulk energies could be slightly overestimated relative to the monolayers. As a consequence, monolayers that also exist in a layered bulk phase could have $\Delta H_{\text{hull}}<0$, even if the layered bulk phase is part of the convex hull and thus should be energetically more stable than the monolayer. Comparing our \(\Delta H_{\mathrm{hull}}\) values for 35 compounds with the exfoliation energies calculated in Ref. \citenum{mounet2016novel} employing vdW compliant xc-functionals for both bulk and monolayer, we estimate the errors in the convex hull energies to be below 0.1 eV/atom. 

\begin{figure}
  \center
  \includegraphics{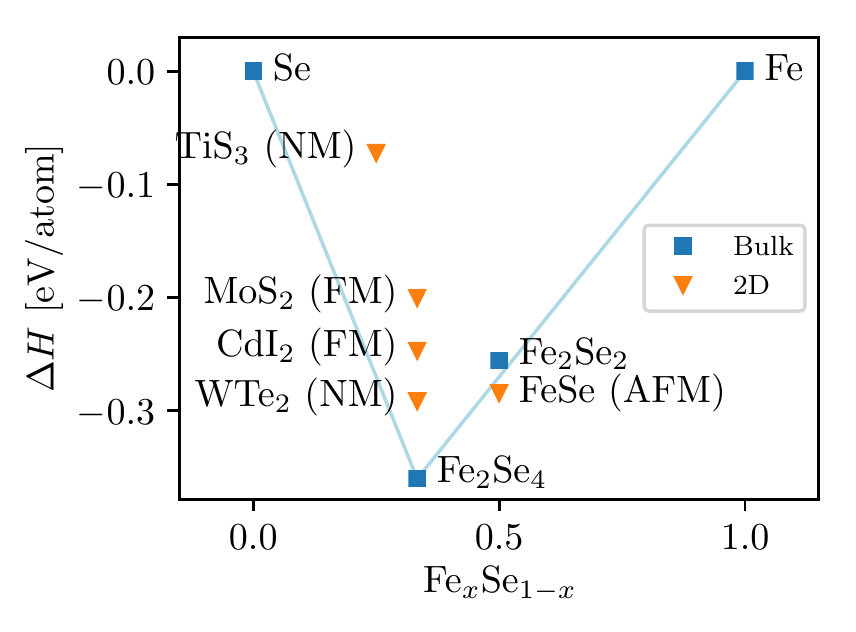}
  \caption{Convex hull for Fe$_x$Se$_{1-x}$. The convex hull as defined by the bulk phases is represented by the blue lines.
    Blue squares denote bulk binary reference phases while orange triangles represent 2D materials. The labels for the 2D materials refer to the crystal prototype and magnetic order.}
  \label{fig:hull}
\end{figure}

As an example, the convex hull for Fe$_x$Se$_{1-x}$ is shown in Figure~\ref{fig:hull}. The convex hull as defined by the bulk binaries is indicated by the blue lines. The labels for the 2D materials refer to the crystal prototype and magnetic order. Clearly, most 2D materials lie above the convex hull and are thus predicted to be thermodynamically unstable in freestanding form under standard conditions. However, as mentioned above, depending on the material, errors on the PBE formation energies can be sizable and thus the hull diagram should only be taken as guideline. Nevertheless, in the present example we find that FeSe (which is itself a prototype) with anti-ferromagnetic ordering lies slightly below the convex hull and is thus predicted to be thermodynamically stable. This prediction is consistent with the recent experimental observation that monolayer FeSe deposited on SrTiO$_3$ exhibits AFM order\cite{zhou2018antiferromagnetic}.

\subsection{Phonons and dynamic stability}
\label{sec:stability}
Due to the applied symmetry constraints and/or the limited size of the unit cell, there is a risk that the structure obtained after relaxation does not represent a local minimum of the potential energy surface, but only a saddle point. We test for dynamical stability by calculating the $\Gamma$-point phonons of a $2 \times 2$ repeated cell (without re-optimising the structure) as well as the elastic constants (see Section \ref{sec:elastic}). These quantities represent second-order derivatives of the total energy with respect to atom displacements and unit cell lengths, respectively, and negative values for either quantity indicate a structural instability.  

The $\Gamma$-point phonons of the $2 \times 2$ supercell are obtained using the finite displacement method\cite{alfe2009a}. We displace each atom in the primitive cell by $\pm 0.01$ \AA~, and calculate the forces induced on all the atoms in the supercell. From the forces we construct the dynamical matrix, which is diagonalised to obtain the $\Gamma$-point phonons of the $2\times 2$ cell (or equivalently the $\Gamma$-point and zone boundary phonons of the primitive cell). The eigenvalues of the dynamical matrix correspond to the square of the mass-renormalised phonon frequencies, $\tilde{\omega}$. Negative eigenvalues are equivalent to imaginary frequencies and signal a saddle point. 

Our procedure explicitly tests for stability against local distortions of periodicities up to $2 \times 2$ and thus provides a necessary, but not sufficient condition for dynamic stability. We stress, however, that even in cases where a material would spontaneously relax into a structure with periodicity larger than $2 \times 2$, the $\Gamma$-point dynamical matrix of the $2\times 2$ cell could exhibit negative eigenvalues. Our test is thus more stringent than it might seem at first glance. In principle, a rigorous test for dynamic stability would require the calculation of the full phonon band structure. Mathematically, the instabilities missed by our approach are those that result in imaginary phonons in the interior of the BZ but \emph{not} at the zone boundary. Physically, such modes could be out of plane buckling or charge density wave-driven reconstructions with periodicities of several unit cells. In general, these types of instabilities are typically rather weak (as measured by the magnitude of the imaginary frequency) as compared to more local distortions such as the T to T$'$ distortion considered below. Moreover, they could well be a special property of the isolated monolayer and become stabilised by the ubiquitous interactions of the 2D material with its environment, e.g. substrates. This is in fact supported by the full phonon calculations by Mounet \emph{et al.} for $\sim 250$ isolated monolayers predicted to be easily exfoliable from experimentally known layered bulk phases\cite{mounet2016novel}. Indeed, most of the instabilities revealed by their calculations are of the type described above and would thus be missed by our test. However, these instabilities cannot be too critical as the monolayers are known to be stable in the vdW bonded layered bulk structure.

\begin{figure}
  \center
  \includegraphics[width=\textwidth]{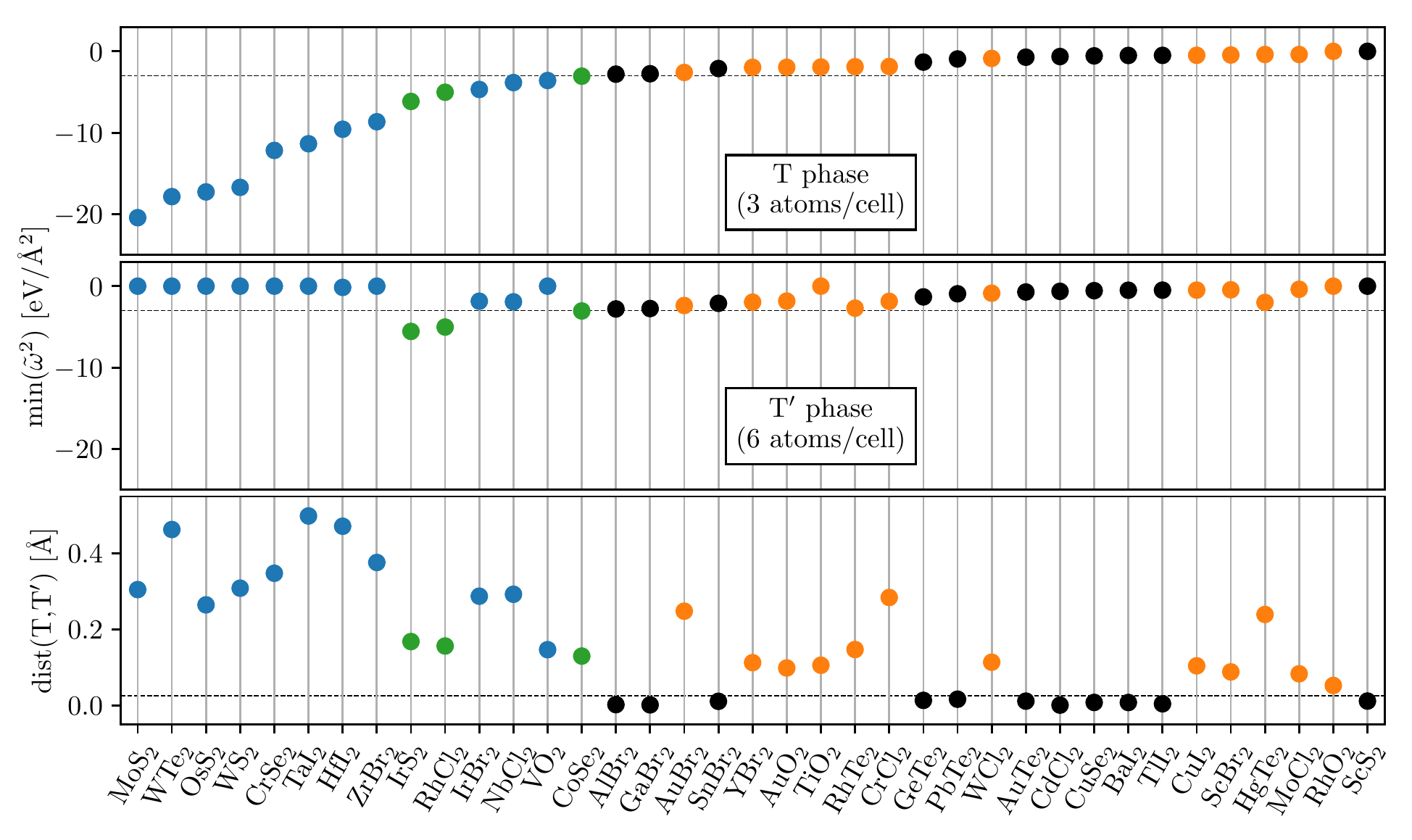}   \hspace*{1.cm}~\includegraphics[width=.92\textwidth]{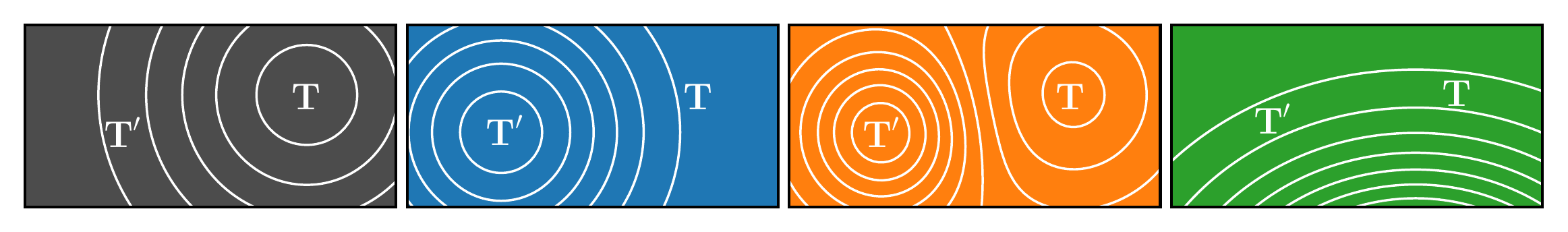}
  \caption{Dynamical stability of a set of transition metal dichalcogenides and -oxides in the T and T$'$ phases (CdI$_2$ and WTe$_2$ prototypes), respectively. The first and second panels show the minimum eigenvalue of the $\Gamma$-point dynamical matrix of the $2\times 2$ unit cell (containing 12 and 24 atoms for the T and T$'$ phase, respectively. The lower panel shows the root mean square distance (RMSD) between the relaxed structures. The color indicates whether the material is dynamically stable in the T phase (black), the T$'$ phase (blue), both phases (orange) or neither of the phases (green).}
  \label{fig:stability}
\end{figure}

As an example, Figure~\ref{fig:stability} compares the dynamical stability of a subset of transition metal dichalcogenides and -oxides in the T and T$'$ phases (CdI$_2$ and WTe$_2$ prototypes). The two upper panels show the smallest eigenvalue of the $\Gamma$-point dynamical matrix of the $2\times 2$ cell. Only materials above the dashed line are considered dynamically stable (for this example we do not consider the sign of the elastic constants which could further reduce the set of dynamically stable materials). Since the unit cell of the T$'$ phase contains that of the T phase it is likely that a material initially set up in the T$'$ phase relaxes back to the T phase. To identify these cases, and thereby avoid the presence of duplicates in the database, the third panel shows the root mean square distance (RMSD) between the structures obtained after relaxations starting in the T- and T$'$ phase, respectively. Structures below the dashed line are considered identical. The color of each symbol refers to the four different potential energy surfaces illustrated at the bottom of the figure.

\subsubsection{Stability criteria}
\label{sec:stabilityscale}

\begin{figure}[htb]
  \centering
  \includegraphics{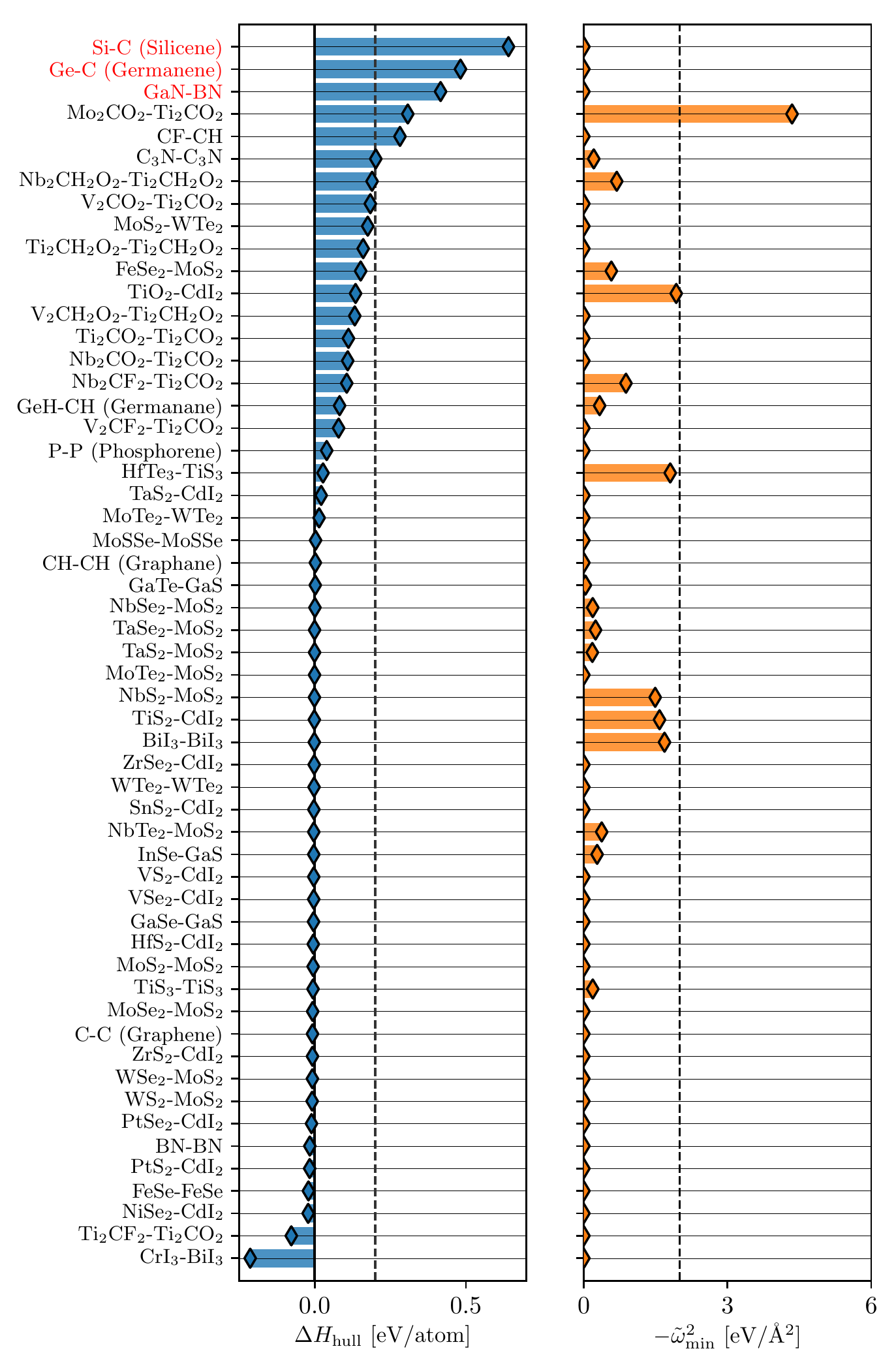}
  \caption{The calculated energy above the convex hull and minimum eigenvalue of the dynamical matrix (evaluated at the $\Gamma$-point for the $2\times 2$ cell) for the 55 materials in the C2DB that have been synthesised or exfoliated in monolayer form, see Refs. \citenum{novoselov2004electric,sone2014epitaxial,davila2014germanene,li2014black,bianco2013stability,yang2017c3n,ci2010atomic,kappera2014phase,mak2010atomically,tonndorf2013photoluminescence,lu2017janus,naylor2016monolayer,wang2017structural,okada2014direct,tonndorf2013photoluminescence,fei2017edge,xi2015strongly,wang2017chemical,zhang2015controlled,manas2016raman,xu2015ultrasensitive,fu2016controlled,ryu2018persistent,zhao2016colloidal,feng2011metallic,wang2015monolayer,zhao2016extraordinarily,wan2015flexible,shao2017epitaxial,tongay2014monolayer,oyedele2017pdse2,zheng2018inse,del2015photoluminescence,zhou2014epitaxy,al2016two,island2014ultrahigh,wang2016tunable,song2011molecular,liu2016metal,cai2017rapid,huang2017layer,xu2015van,kong2010few,mashtalir2015amine,liu2017preparation,lukatskaya2013cation,melchior2018high,liu2017vertically,li2017synthesis,elias2009control,nair2010fluorographene}\label{fig:exp}. The three materials highlighted in red have only been synthesised on metallic substrates. The black dashed lines indicate the thresholds used to categorise the thermodynamic and dynamical stability of materials in the C2DB.}
\end{figure}

To assess the stability of the materials in the C2DB, we turn to the set of experimentally synthesised/exfoliated monolayers. For these materials, the calculated energy above the convex hull and minimum eigenvalue of the dynamical matrix are shown on Figure~\ref{fig:exp}. It is clear that all but five known monolayers have a hull energy below 0.2 eV/atom, and three of these have only been synthesised on a metal substrate. Turning to the dynamical stability, all but one of the experimentally known monolayers have a minimum eigenvalue of the dynamical matrix above \(-2\, \mathrm{eV}/\mathrm{\AA}^2\), and 75\% have a minimum eigenvalue above \(-1\times10^{-5}\, \mathrm{eV}/\mathrm{\AA}^2\) (shown in light grey).

Guided by these considerations, we assign each material in the C2DB a stability level (low, medium or high) for both dynamical and thermodynamic stability, as illustrated in Table \ref{tab:stabilityscale}. For ease of reference, we also define the overall stability level of a given material as the lower of the dynamical and thermodynamic stability levels. If a material has ``low'' overall stability (colored red in the table), we consider it unstable and do not carry out the rest of the workflow. Materials with "high" overall stability are considered likely to be stable and thus potentially synthesisable. Materials in the "medium" stability category, while unlikely to be stable as freestanding monolayers, cannot be discarded and might be metastable and possible to synthesise under the right conditions. For example, free-standing silicene has a heat of formation of 0.66 eV/atom, but can be grown on a silver substrate. Likewise, the T$'$ phase of MoS$_2$ (WTe$_2$ prototype) has an energy of 0.27 eV/atom higher than the thermodynamically stable H phase, but can be stabilised by electron doping.

\clearpage

\begin{table}[htb]
  \begin{center}
    \renewcommand*{\arraystretch}{1.2}
    \setlength{\arrayrulewidth}{1pt}
    \definecolor{sr}{RGB}{214, 39, 40}
    \definecolor{sy}{RGB}{179, 179, 54}
    \definecolor{sg}{RGB}{44, 160, 44}
    \newcommand{\red}[1]{\textcolor{sr}{\textbf{#1}}}
    \newcommand{\yellow}[1]{\textcolor{sy}{\textbf{#1}}}
    \newcommand{\green}[1]{\textcolor{sg}{\textbf{#1}}}
    \newcolumntype{M}[1]{>{\centering\arraybackslash}m{#1}}
    \newcolumntype{N}{@{}m{0pt}@{}}
    \begin{tabular}{rM{2.5cm}M{3.2cm}M{3cm}cN}
      \toprule
      & \multicolumn{3}{c}{Dynamic stability} \\
      \cmidrule(l){2-4}
      & \multicolumn{1}{M{3cm}}{$\left| \tilde \omega_{\mathrm{min}}^2\right| > 2$ or $C_{ii}<0$} & \multicolumn{1}{M{3.2cm}}{\(10^{-5}  < \left|\tilde\omega_{\mathrm{min}}^{2}\right| < 2\) \newline \(C_{ii} > 0\)} & \multicolumn{1}{C{3.2cm}}{$\quad \left|\tilde\omega_{\mathrm{min}}^2\right| < 10^{-5}$ \newline \(\ \ C_{ii} > 0 \)} & \multicolumn{1}{C{2cm}}{\newline Total}\\
      \midrule
      $\Delta H > 0.2$ eV                & \red{\phantom{1}6.0\%} & \red{\phantom{1}4.2\%} & \red{\phantom{1}1.7\%}    & 12.0\% & \\
      $\Delta H <  0.2 \textrm{ eV}$     & \red{14.9\%}           & \yellow{10.9\%}        & \yellow{\phantom{1}6.4\%} & 32.2\% & \\
      $\Delta H_\mathrm{hull} < 0.2$ eV    & \red{11.4\%} & \yellow{24.1\%}        & \green{20.3\%}            & 55.8\% & \\
      \midrule
      Total                              & 32.3\%                 & 39.2\%                 & 28.5\%                               \\
      \bottomrule
    \end{tabular}

\end{center}
  \caption{The materials in the C2DB distributed over the nine stability categories defined by the three levels (high, medium and low) of dynamical stability
(columns) and thermodynamic stability (rows).  The overall stability of the materials is defined as the lower of the two separate stability scales. Materials with low overall stability (red) are considered unstable.\label{tab:stabilityscale}.}
\end{table}

\subsection{Elastic constants}
\label{sec:elastic}
The elastic constants of a material are defined by the generalised Hooke's law,
\begin{equation}
  \label{eq:hooke}
  \sigma_{ij} = C_{ijkl} \varepsilon_{kl}
\end{equation}
where $\sigma_{ij}$, $C_{ijkl}$ and $\varepsilon_{kl}$ are the stress, stiffness and strain tensors, respectively, and where we have used the Einstein summation convention. In two dimensions, the stress and strain tensors have three independent components, namely planar stress/strain in the $x$ and $y$ directions, as well as shear stress/strain. The stiffness tensor is a symmetric linear map between these two tensors, and therefore has up to six independent components. Disregarding shear deformations, the relationship between planar strain and stress is
\begin{equation}
  \begin{bmatrix}
    \sigma_{xx} \\
    \sigma_{yy} 
  \end{bmatrix}
    =
    \begin{bmatrix}
    C_{11} & C_{12}\\
    C_{12} & C_{22} 
    \end{bmatrix}
    \begin{bmatrix}
    \epsilon_{xx} \\
    \epsilon_{yy} 
    \end{bmatrix}
\end{equation}
For all materials in the C2DB, we calculate the planar elastic stiffness coefficients $C_{11}$, $C_{22}$, and $C_{12}$. These are calculated using a central difference approximation to the derivative of the stress tensor: the material is strained along one of the coordinate axes, $x$ or $y$, and the stress tensor is calculated after the ions have relaxed. We use strains of $\pm 1\%$ which we have found to be sufficiently large to eliminate effects of numerical noise and sufficiently small to stay within the linear response regime. 

Table~\ref{tab:elastic} shows the calculated planar stiffness coefficients of a set of 2D materials. As can be seen the values from the C2DB are in very good agreement with previously published PBE results. For the isotropic materials MoS$_2$, WSe$_2$ and WS$_2$, $C_{11}$ and $C_{22}$ should be identical, and we see a variation of up to 0.6\%. This provides a test of how well converged the values are with respect to numerical settings.

\begin{table}[htb]
  \centering
\begin{tabular}[h]{lrrrrrr}
  \toprule
              & \multicolumn{2}{c}{$C_{11}$ (N/m)} & \multicolumn{2}{c}{$C_{22}$ (N/m)} & \multicolumn{2}{c}{$C_{12}$ (N/m)}                                    \\
  \cmidrule(l){2-3} \cmidrule(l){4-5} \cmidrule(l){6-7}
              & C2DB                               & Literature                         & C2DB  & Literature                 & C2DB & Literature                \\
  \midrule
  P (phosphorene) & 101.9 & 105.2 \cite{wang2015a}     & 25.1  & 26.2\cite{wang2015a}       & 16.9 & 18.4\cite{wang2015a}      \\
  MoS$_2$             & 131.4 & 132.6\cite{choudhary2017a} & 131.3 & 132.6\cite{choudhary2017a} & 32.6 & 32.7\cite{choudhary2017a} \\
  WSe$_2$             & 120.6 & 119.5\cite{choudhary2017a} & 121.3 & 119.5\cite{choudhary2017a} & 22.8 & 22.7\cite{choudhary2017a} \\
  WS$_2$              & 146.3 & 145.3\cite{choudhary2017a} & 146.7 & 145.3\cite{choudhary2017a} & 32.2 & 31.5\cite{choudhary2017a} \\
  \midrule
  MAD                 & 1.7   & -\phantom{0.0}                          & 1.4   & -\phantom{0.0}                          & 0.6  &         -\phantom{0.0}                  \\
  \bottomrule
\end{tabular}
\caption{Planar elastic stiffness coefficients (in N/m) calculated at the PBE level. The results of this work are compared to previous calculations from the literature and the mean absolute deviation (MAD) is shown.}\label{tab:elastic}
\end{table}

\subsection{Magnetic anisotropy}
The energy dependence on the direction of magnetisation, or magnetic anisotropy (MA), arises from spin-orbit coupling (SOC). According to the magnetic force theorem \cite{Wang1996} this can be evaluated from the eigenvalue differences such that the correction to the energy becomes
\begin{equation}\label{eq:anisotropy}
\Delta E(\mathbf{\hat n}) = \sum_{\mathbf{k}n}f(\varepsilon_{\mathbf{k}n}^\mathbf{\hat n})\varepsilon_{\mathbf{k}n}^\mathbf{\hat n}-\sum_{\mathbf{k}n}f(\varepsilon_{\mathbf{k}n}^0)\varepsilon_{\mathbf{k}n}^0,
\end{equation}
where  $\varepsilon_{\mathbf{k}n}^\mathbf{\hat n}$ and $f(\varepsilon_{\mathbf{k}n}^\mathbf{\hat n})$ are the eigenenergies and occupation numbers, respectively, obtained by diagonalising the Kohn-Sham Hamiltonian including SOC in a basis of collinear spinors aligned along the direction  $\mathbf{\hat n}$, while $\varepsilon_{\mathbf{k}n}^0$ and $f(\varepsilon_{\mathbf{k}n}^0)$ are the bare Kohn-Sham eigenenergies and occupation numbers without SOC. 

For all magnetic materials we have calculated the energy difference between out-of-plane and in-plane magnetisation $E_{\text{MA}}(i)=\Delta E(\mathbf{\hat z})-\Delta E(i)$, ($i=\mathbf{\hat x}, \mathbf{\hat y}$). Negative values of $E_{\text{MA}}(i)$ thus indicate that there is an out-of-plane easy axis of magnetisation.

Calculations for the ground state have been performed with plane-wave cutoff and energetic convergence threshold set to 800 eV and 0.5 meV/atom respectively. For all calculations we have used a $\Gamma$-centered Monkhorst-Pack $k$-point with a density of 20/$\mathrm{\AA}^{-1}$. The SOC contribution is introduced via a non-self-consistent diagonalisation of the Kohn-Sham Hamiltonian evaluated in the projector-augmented wave formalism\cite{olsen2016designing}.

\subsection{Projected density of states}
The projected density of states (PDOS) is a useful tool for identifying which atomic orbitals comprise a band. It is defined as
\begin{align}\label{eq:pdos_paw}
\rho^S_l(\varepsilon) = \sum_{a \in S} \sum_{\mathbf{k}n}\sum_{m} |\langle \phi_{l, m}^a|\psi_{\mathbf{k}n}\rangle|^2 \delta(\varepsilon - \varepsilon_{\mathbf{k}n}),
\end{align}
where $\psi_{\mathbf{k}n}$ are the Kohn-Sham wave functions with eigenvalues $\varepsilon_{\mathbf{k}n}$ and $\phi_{l,m}^a$ are the spin-paired Kohn-Sham orbitals of atomic species $S$ with angular momentum $l$ ($s,p,d,f$). We sum over all atoms belonging to species $S$ so every atomic species has one entry per angular momentum channel. In the PAW formalism this can be approximated as
\begin{align}\label{eq:pdos_paw2}
\rho^S_l(\varepsilon) = \sum_{a \in S} \sum_{\mathbf{k}n}\sum_{m} |\langle \tilde{p}_{l, m}^a|\tilde{\psi}_{\mathbf{k}n}\rangle|^2 \delta(\varepsilon - \varepsilon_{\mathbf{k}n})	
\end{align}
where $\tilde{\psi}_{\mathbf{k}n}$ are the pseudo wave functions and $\tilde{p}_{l,m}^a$ are the PAW projectors associated with the atomic orbitals $\phi_{l, m}^a$. The PDOS is calculated from Eq. \eqref{eq:pdos_paw2} using linear tetrahedron interpolation\cite{MacDonald1979} (LTI) of energy eigenvalues obtained from a ground state calculation with a $k$-point sampling of $36/\mathrm{\AA}^{-1}$. In contrast to other techniques for calculating the PDOS using smearing, the PDOS yielded by the LTI method returns exactly zero at energies with no states. Examples of PDOS are shown in  Figure~\ref{fig:PBE_vo2} (right) for respectively the ferromagnetic metal VO$_2$ and the semiconductor  WS$_2$ in the H phase (MoS$_2$ prototype).

\subsection{Band structures}
Electronic band structures are calculated along the high symmetry paths shown in Figure~\ref{fig:bandpaths} for the five different types of 2D Bravais lattices. The band energies are computed within DFT using three different xc-functionals, namely PBE, HSE06, and GLLBSC. These single-particle approaches are complemented by many-body \GW  calculations for materials with a finite gap and up to four atoms in the unit cell (currently around 250 materials). For all methods, SOC is included by non-selfconsistent diagonalisation in the full basis of Kohn-Sham eigenstates. Band energies always refer to the vacuum level defined as the asymptotic limit of the Hartree potential, see Figure~\ref{fig:pot}. Below we outline the employed methodology while Section~\ref{sec:trend-band-gap} provides an overview and comparison of the band energies obtained with the different methods.

\subsubsection{PBE band structure}
\begin{figure}
\centering
\includegraphics[width=0.8\textwidth]{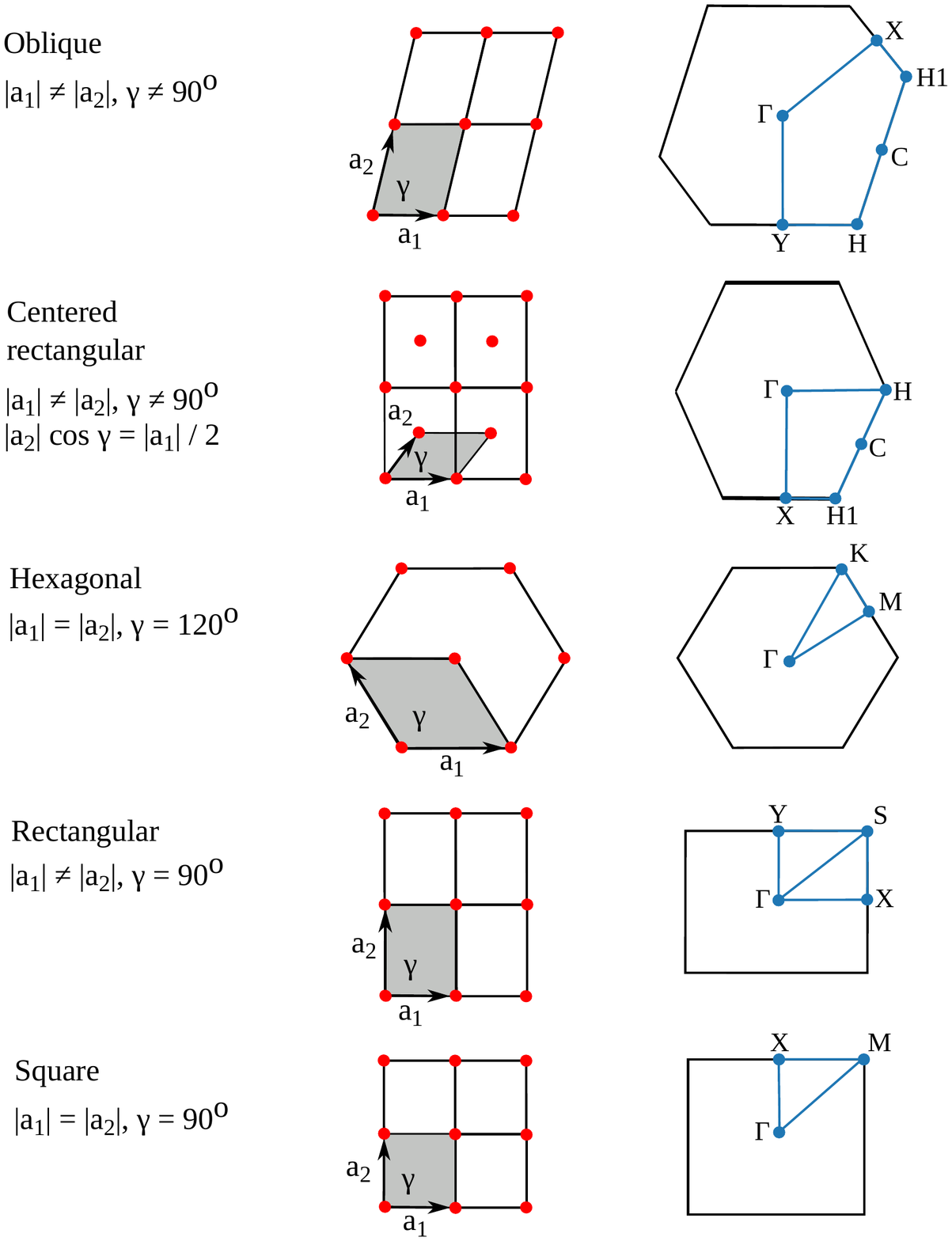}
\caption{Overview of the five 2D Bravais lattices and corresponding Brillouin zones. The unit vectors $\textbf{a}_1$ and $\textbf{a}_2$ are shown together with the angle between them $\gamma$. The primitive unit cell is indicated in gray. High symmetry points of the BZ and the path along which the band structure is evaluated, are indicated in blue.}
\label{fig:bandpaths}
\end{figure}

\begin{figure}
\centering
    \begin{subfigure}[b]{0.52\columnwidth}
        \includegraphics{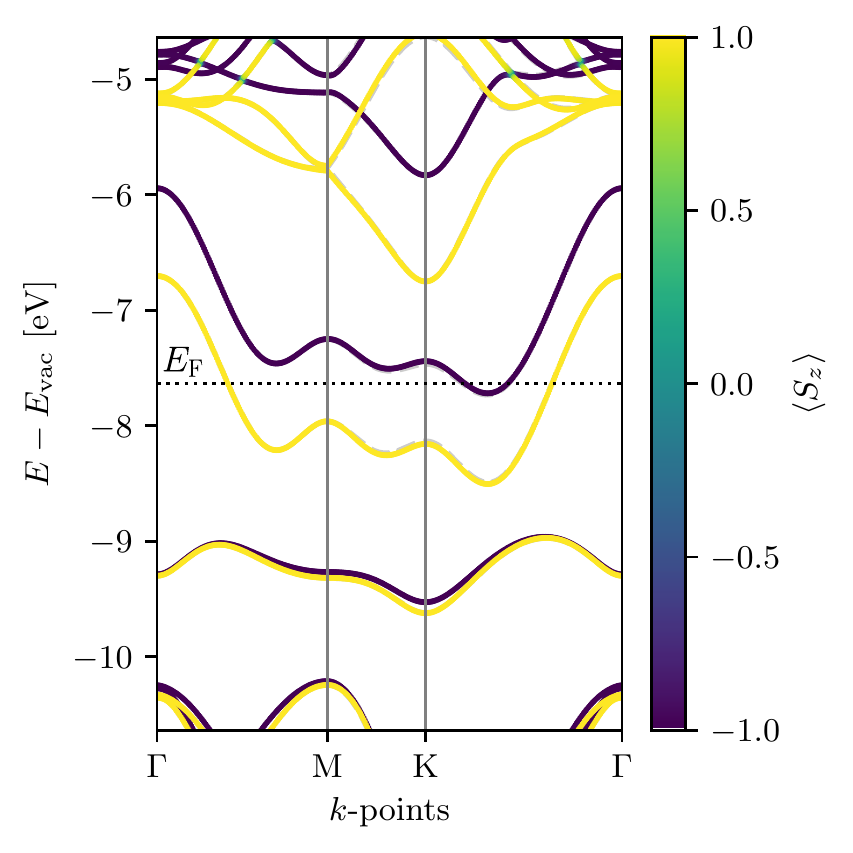}
      \end{subfigure}%
      \begin{subfigure}[b]{0.5\columnwidth}
        \includegraphics{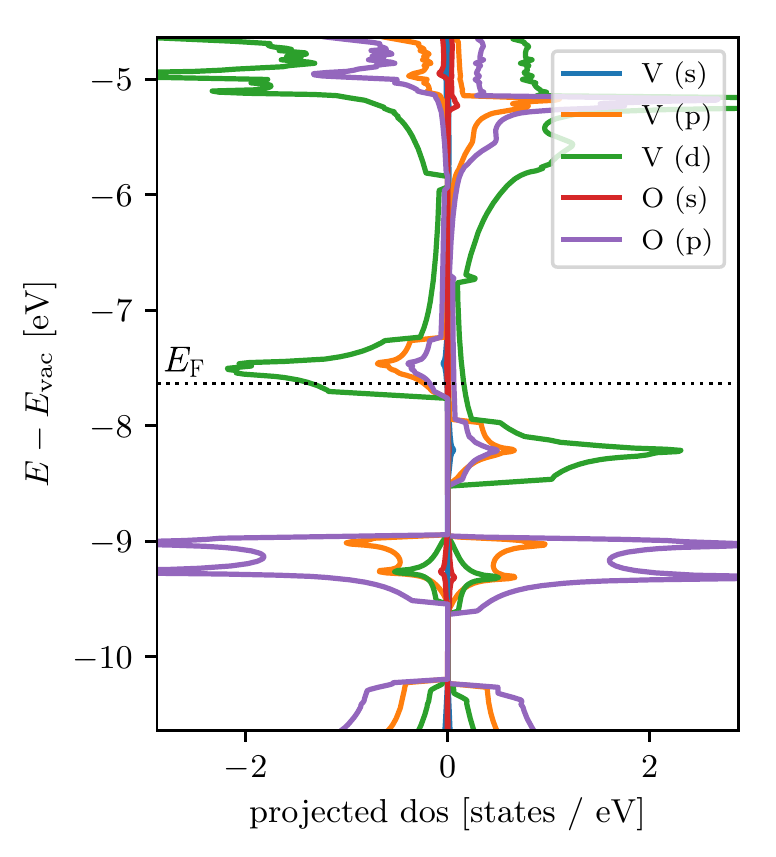}
    \end{subfigure}\\ 
    \begin{subfigure}[b]{0.52\columnwidth}
      \hspace{0.9mm}
        \includegraphics{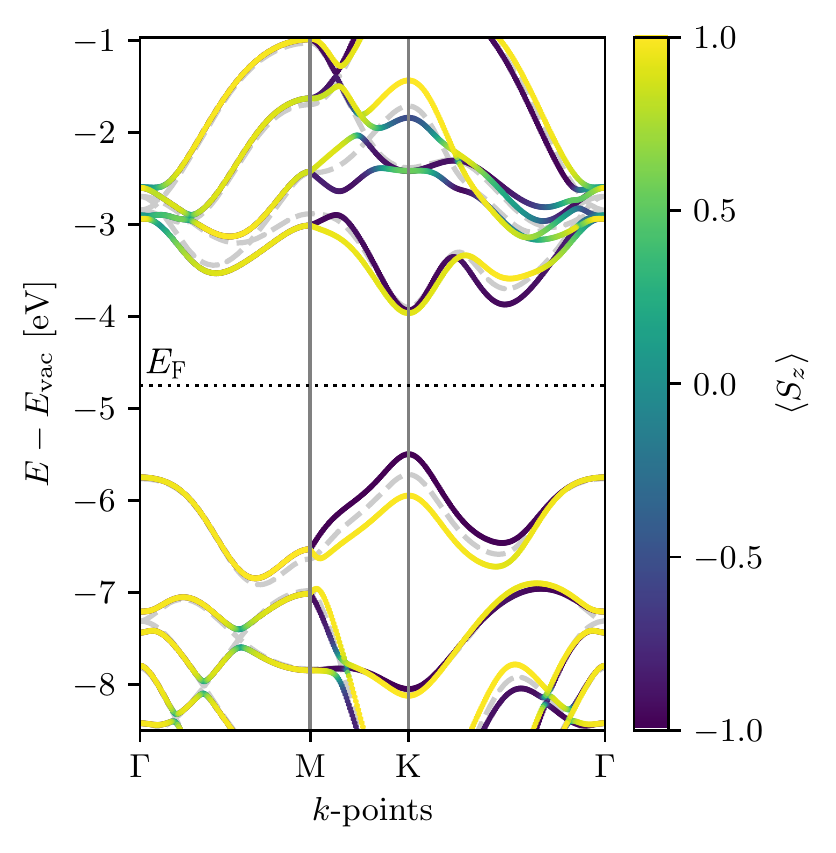}
      \end{subfigure}%
      \begin{subfigure}[b]{0.5\columnwidth}
      \hspace{0.6mm}
        \includegraphics{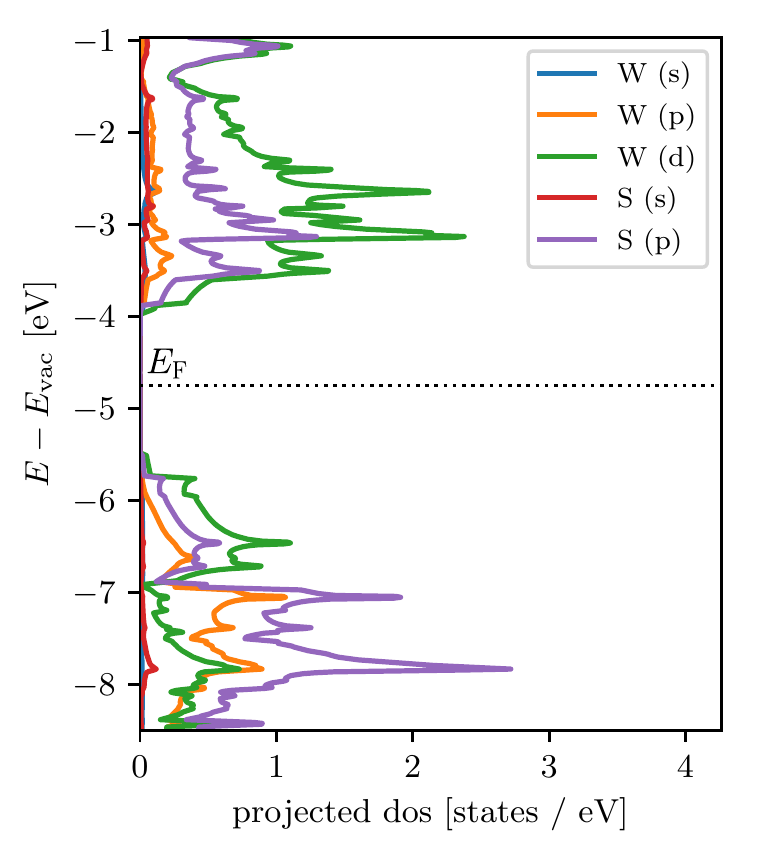}
    \end{subfigure}
    \caption{Band structure (left) and projected density of states (right) for VO$_2$ (top) and WS$_2$ (bottom) in the MoS$_2$ prototype,  calculated with the PBE xc-functional. The $z$-component of the spin is indicated by the color code on the band structure.}
    \label{fig:PBE_vo2}
\end{figure}

The electron density is determined self-consistently on a uniform $k$-point grid of density 12.0/$\mathrm{\AA}^{-1}$. From this density, the PBE band structure is computed non-selfconsistently at 400 $k$-points distributed along the band path (see Figure~\ref{fig:bandpaths}). Examples of PBE band structures are shown in Figure~\ref{fig:PBE_vo2} for the ferromagnetic metal VO$_2$ and the semiconductor WS$_2$ both in the MoS$_2$ prototype structure. The expectation value of the out-of-plane spin component, $\langle \chi_{n\mathbf{k}\sigma}|\hat S_z|\chi_{n\mathbf{k}\sigma}\rangle$, is evaluated for each spinorial wave function, $\chi_{n\mathbf{k}\sigma}=(\psi_{n\mathbf{k}\uparrow},\psi_{n\mathbf{k}\downarrow})$, and is indicated by the color of the band. For materials with inversion symmetry, the SOC cannot induce band splitting, meaning that $\langle \chi_{n\mathbf{k}\sigma}|\hat S_z|\chi_{n\mathbf{k}\sigma}\rangle$ is ill-defined and no color coding is used. The band structure without SOC is indicated by a dashed grey line. We have compared our PBE+SOC band gaps of 29 different monolayers with those obtained with the VASP code in Ref. \citenum{ozccelik2016band} and find a mean absolute deviation of 0.041 eV.

\subsubsection{HSE band structure}
\begin{figure}\centering
    \begin{subfigure}[b]{0.5\columnwidth}
        \includegraphics[width=\columnwidth]{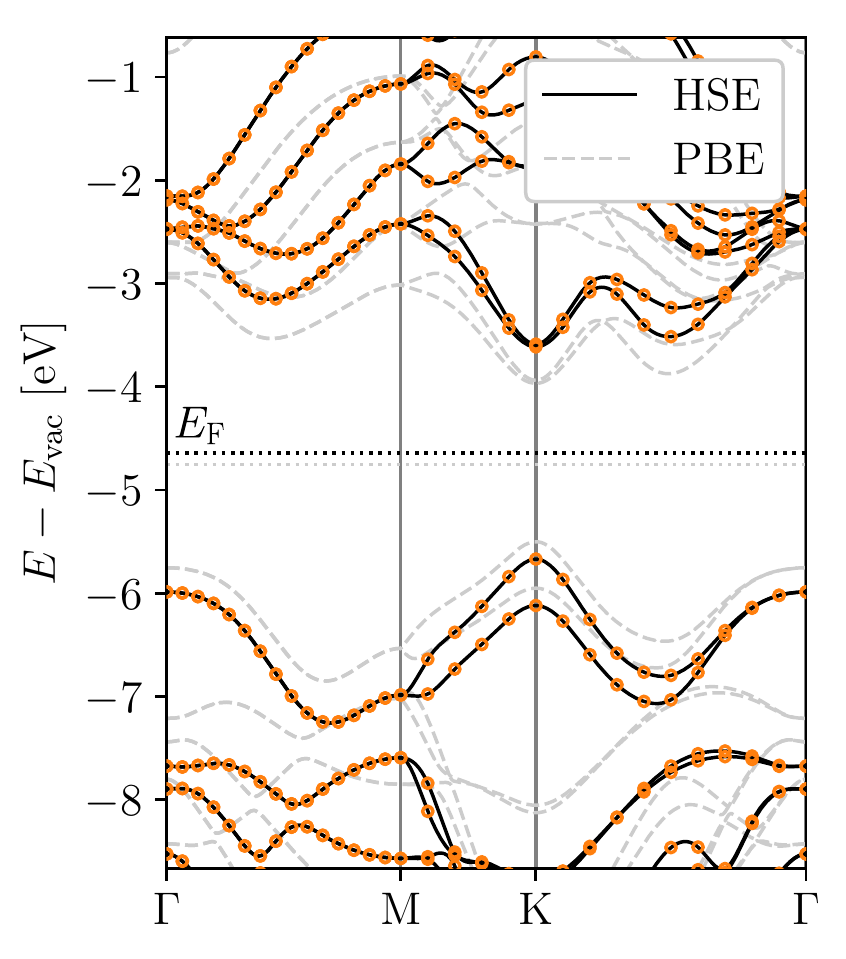}
      \end{subfigure}%
      \begin{subfigure}[b]{0.5\columnwidth}
        \includegraphics[width=\columnwidth]{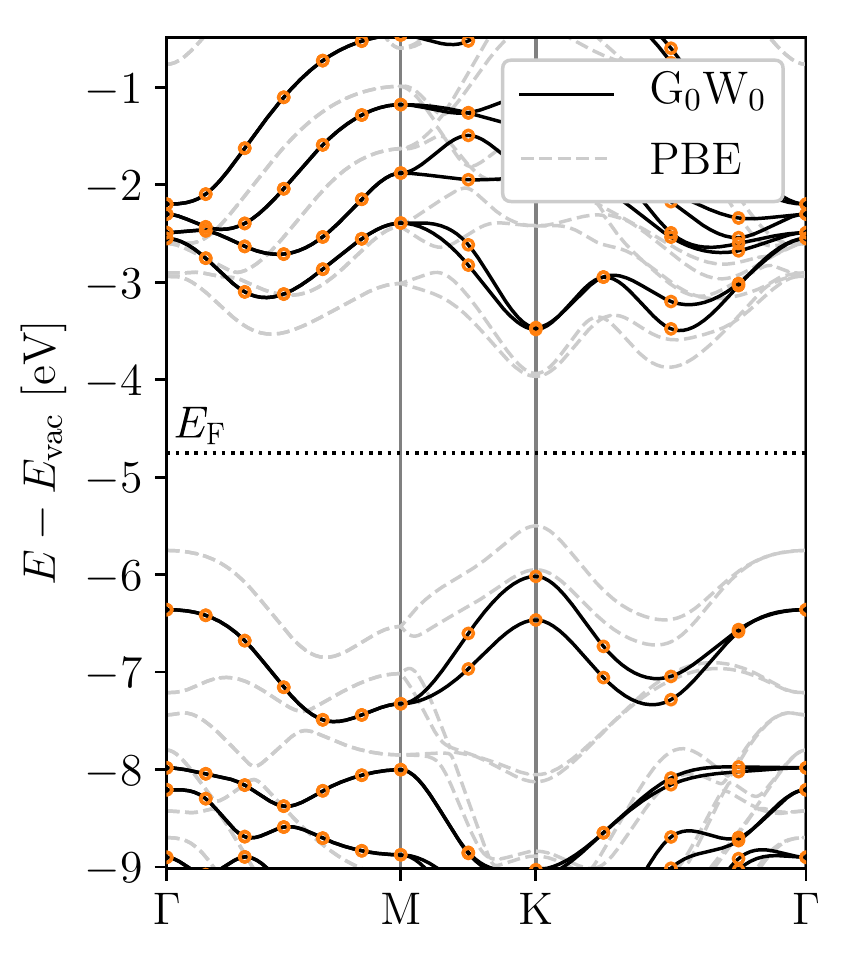}
    \end{subfigure}
\caption{\label{fig:gw-hse-bs} Band structure of WS$_2$ calculated with the HSE06 functional (left) and \GW  (right). For comparison the PBE result is also shown (grey dashed). Spin-orbit coupling (SOC) is included in all calculations. The band energies refer to the vacuum level. The points show the calculated eigenvalues from which the band structure is interpolated. The relatively coarse $k$-point grid used for \GW is justified by the analytical treatment of the screened interaction $W(q)$ around $q=0$, see Figure \ref{fig:gw-conv}.}
\end{figure}

The band structure is calculated non-selfconsistently using the range-separated hybrid functional HSE06\cite{heyd2003_hybrid} on top of a PBE calculation with $k$-point density 12.0/$\mathrm{\AA}^{-1}$ and 800 eV plane wave cutoff. We have checked for selected systems that the HSE band structure is well converged with these settings. The energies along the band path are obtained by spline interpolation from the uniform $k$-point grid. As an example, the HSE band structure of WS$_2$ is shown in the left panel of Figure~\ref{fig:gw-hse-bs} (black line) together with the PBE band structure (grey dashed). The PBE band gap increases from 1.52 eV to 2.05 eV with the HSE06 functional in good agreement with earlier work reporting band gaps of 1.50 eV (PBE) and 1.90 eV (HSE)\cite{kosmider2013electronic} and 1.55 eV (PBE) and 1.98 eV (HSE)\cite{kang2013band}, respectively. A more systematic comparison of our results with the HSE+SOC band gaps obtained with the VASP code in Ref. \citenum{ozccelik2016band} for 29 monolayers yield a mean absolute deviation of 0.14 eV. We suspect this small but non-zero deviation is due to differences in the employed PAW potentials and the non-selfconsistent treatment of the HSE in our calculations.

\subsubsection{GLLBSC fundamental gap}
For materials with a finite PBE band gap, the fundamental gap (i.e. the difference between the ionisation potential and electron affinity) also sometimes referred to as the quasiparticle gap, is calculated self-consistently using the GLLBSC\cite{kuisma_kohn-sham_2010} xc-functional with a Monkhorst-Pack $k$-point grid of density 12.0/$\mathrm{\AA}^{-1}$. The GLLBSC is an orbital-dependent exact exchange-based functional, which evaluates the fundamental gap as the sum of the Kohn-Sham gap and the xc-derivative discontinuity, $E_{\mathrm{gap}}=\varepsilon_{\mathrm{gap}}^{\mathrm{KS}}+\Delta_{\mathbf{xc}}$. The method has been shown to yield excellent quasiparticle band gaps at very low computational cost for both bulk\cite{kuisma_kohn-sham_2010,castelli2012computational} and 2D semiconductors\cite{rasmussen2015computational}.

In the exact Kohn-Sham theory, $\varepsilon_{\mathrm{v}}^{\mathrm{KS}}$ should equal the exact ionisation potential and thus $\Delta_{\mathrm{xc}}$ should be used to correct only the conduction band energies\cite{baerends2017kohn}. Unfortunately, we have found that in practice this procedure leads to up-shifted band energies (compared with the presumably more accurate \GW  results, see Figure \ref{fig:bs-trends}). Consequently, we store only the fundamental gap and $\Delta_{\mathrm{xc}}$ in the database. However, as will be shown in Section~\ref{sec:trend-band-gap} the center of the gap is in fact reasonably well described by PBE suggesting that efficient and fairly accurate predictions of the absolute band edge energies can be obtained by a symmetric GLLBSC correction of the PBE band edges. 
 
\subsubsection{G\textsubscript{0}\!W\!\textsubscript{0} band structure}
\begin{figure}\centering
\includegraphics[width=1.0\columnwidth]{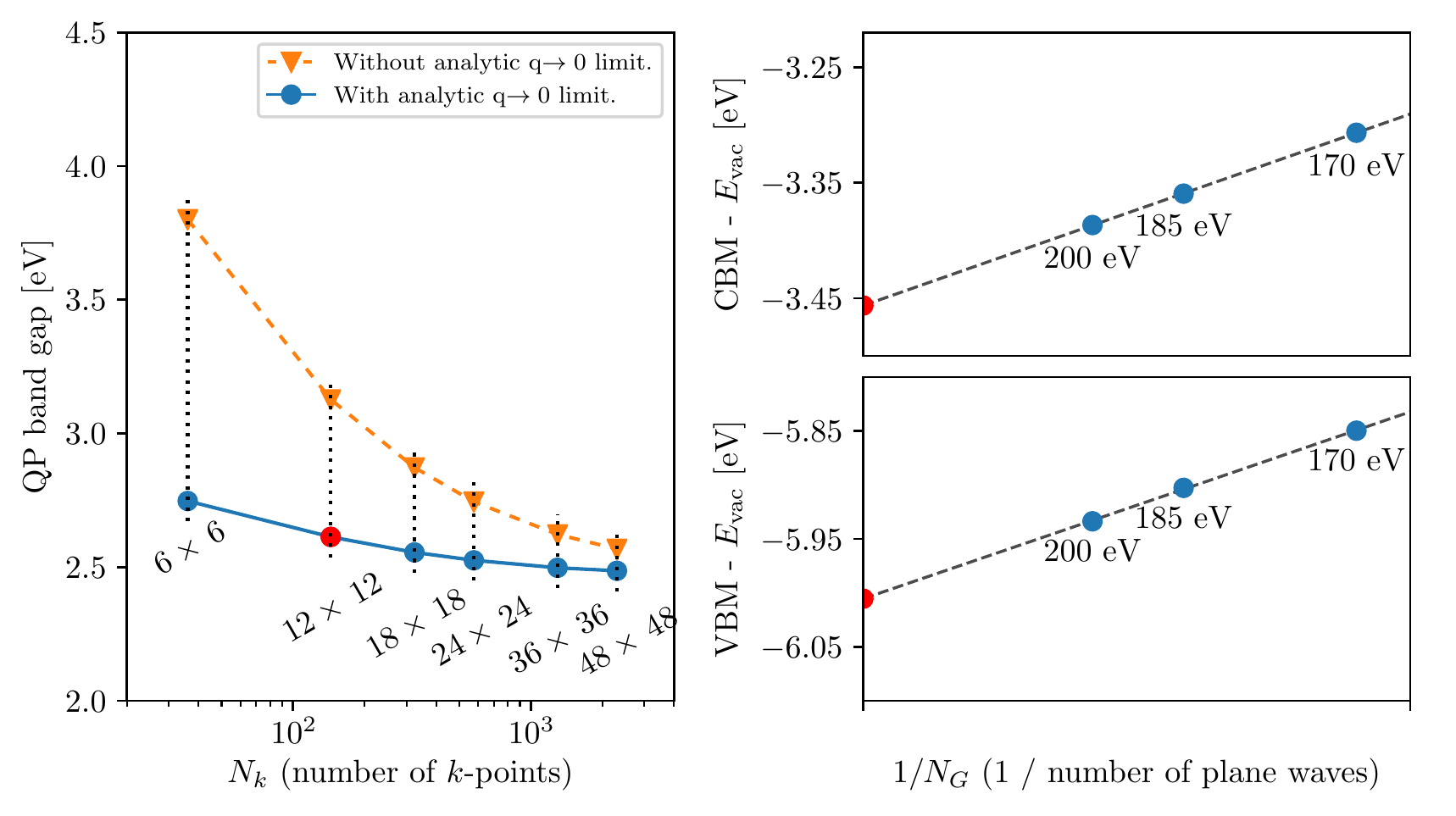}
\caption{\label{fig:gw-conv} Left: Convergence of the QP band gap of MoS$_2$ as a function of $k$-point sampling with and without the analytical treatment of $W(q)$ around $q=0$. It is clear that the analytical treatment of the screened interaction significantly improves the $k$-point convergence. Right: The convergence of the CBM and VBM versus the number of plane waves. The band energies are obtained by extrapolation of three calculations performed with PW cutoff up to 200 eV. In all panels, the red dot indicates the data point calculated by the workflow and available in the C2DB.}
\end{figure}

For materials with finite PBE band gap the quasiparticle (QP) band structure is calculated using the \GW  approximation on top of PBE following our earlier work~\cite{falcogw,filip2dcorr}. Currently, this resource demanding step is performed only for materials with up to four atoms in the unit cell. The number of plane waves and the number of unoccupied bands included in the calculation of the non-interacting density response function and the GW self-energy are always set equal. The individual QP energies are extrapolated to the infinite basis set limit from calculations at plane wave cutoffs of 170, 185 and 200 eV, following the standard $1/N_\mathrm{G}$ dependence\cite{tiago_effect_2004,klimes_predictive_2014}, see Figure~\ref{fig:gw-conv}(right). The screened Coulomb interaction is represented on a non-linear real frequency grid ranging from 0 eV to 230 eV and includes around 250 frequency points. The exchange contribution to the self-energy is calculated using a Wigner-Seitz truncation scheme\cite{wigner_seitz} for a more efficient treatment of the long range part of the exchange potential. For the correlation part of the self-energy, a 2D truncation of the Coulomb interaction is used\cite{2dcoul1,2dcoul2}. We stress that the use of a truncated Coulomb interaction is essential to avoid unphysical screening from periodically repeated layers which otherwise leads to significant band gap reductions. 

Importantly, the use of a truncated Coulomb interaction leads to much slower $k$-point convergence because of the strong $q$-dependence of the 2D dielectric function around $q=0$. We alleviate this problem by using an analytical expression for the screened interaction when performing the BZ integral around $q=0$\cite{filip2dcorr}. This allows us to obtain well converged results with a relatively low $k$-point density of 5.0/$\mathrm{\AA}^{-1}$ (corresponding to $12\times12$ for MoS$_2$). For example, with this setting the \GW  band gap of MoS$_2$ is converged to within 0.05 eV, see Figure~\ref{fig:gw-conv}(left). In comparison, standard BZ sampling with no special treatment of the $q=0$ limit, requires around $40\times40$ $k$-points to reach the same accuracy.  

Figure~\ref{fig:gw-hse-bs} (right) shows the PBE and \GW  band structures of WS$_2$ (including SOC). The \GW  self-energy opens the PBE band gap by 1.00 eV and the HSE gap by 0.47 eV, in good agreement with previous studies\cite{lee2017strain}. We note in passing that our previously published \GW band gaps for 51 monolayer TMDCs\cite{rasmussen2015computational} are in good agreement with the results obtained using the workflow described here. The mean absolute error between the two data sets is around 0.1 eV and can be ascribed to the use of PBE rather than LDA as starting point and the use of the analytical expression for $W$ around $q=0$. 

A detailed comparison of our results with previously published \GW  data is not meaningful because of the rather large differences in the employed implementations/parameter settings. In particular, most reported calculations do not employ a truncated Coulomb interaction and thus suffer from spurious screening effects, which are then corrected for in different ways. Moreover, they differ in the amount of vacuum included the supercell, the employed $k$-point grids and basis sets, the in-plane lattice constants, and the DFT starting points. For example, published values for the QP band gap of monolayer MoS$_2$ vary from from 2.40 to 2.90 eV\cite{ramasubramaniam_large_2012,chewchancham_2012,komsa_2012,molina_2013,Shi_2013,conley_bandgap_2013,huser2013dielectric,qiu2016screening} (see Ref. \citenum{huser2013dielectric} for a detailed overview). The rather large variation in published GW results for 2D materials is a result of the significant numerical complexity of such calculations and underlines the importance of establishing large and consistently produced benchmark data sets like the present.

For bulk materials, self-consistency in the Green's function part of the self-energy, i.e. the GW$_0$ method, has been shown to increase the \GW  band gaps and improve the agreement with experiments\cite{shishkin2007self}. The trend of band gap opening is also observed for 2D materials\cite{filip2dcorr,schmidt2017simple,qiu2016screening,qiu2016screening}, however, no systematic improvement with respect to experiments has been established\cite{schmidt2017simple}. For both bulk and 2D materials, the fully self-consistent GW self-energy systematically overestimates the band gap\cite{shishkin2007self,schmidt2017simple} due to the neglect of vertex corrections\cite{shishkin2007accurate,schmidt2017simple}. In \GW  the neglect of vertex corrections is partially compensated by the smaller band gap of the non-interacting Kohn-Sham Green's function compared to the true interacting Green's function. In this case, the vertex corrections affect mainly the absolute position of the bands relative to vacuum while the effect on the band gap is relatively minor\cite{schmidt2017simple}.

In Table \ref{tab:gaps} we compare calculated band gaps from C2DB with experimental band gaps for three monolayer TMDCs and phosphorene. The experimental data has been corrected for substrate interactions\cite{schmidt2017simple,wang2015highly}, but not for zero-point motion, which is expected to be small ($<0.1 \mathrm{ eV}$). The \GW  results are all within 0.2 eV of the experiments. A further (indirect) test of the \GW  band gaps against experimental values is provided by the comparison of our BSE spectra against experimental photoluminescence data in Table~\ref{tab:bse}, where we have used a \GW  scissors operator. Finally, we stress that the employed PAW potentials are not norm-conserving, and this can lead to errors for bands with highly localised states (mainly 4$f$ and 3$d$ orbitals), as shown in Ref. \citenum{klimevs2014predictive}. Inclusion of vertex corrections and use of norm conserving potentials will be the focus of future work on the C2DB.

\begin{table}[htb]
  \begin{center}
\caption{Comparison between calculated and experimental band gaps for four freestanding monolayers. The experimental values have been corrected for substrate screening. MAD refers to the mean absolute deviation between the predicted values and the measured values.\label{tab:gaps}}
\begin{tabular}[h]{lrrrrr}
  \toprule
           & \multicolumn{5}{c}{Band gap (eV)}          \\
  \cmidrule(l){2-5} \cmidrule(l){2-6}
  Material & PBE & HSE06 & GLLBSC & \GW  & Experiment \\
  \midrule
MoS$_2$                 & 1.58 & 2.09 & 2.21 & 2.53 & 2.50\cite{klots2014probing} \\
MoSe$_2$                & 1.32 & 1.80 & 1.88 & 2.12 & 2.31\cite{ugeda2014a}       \\
WS$_2$                  & 1.53 & 2.05 & 2.16 & 2.53 & 2.72\cite{hill2016band}     \\
P (phosphorene)     & 0.90 & 1.51 & 1.75 & 2.03 & 2.20\cite{wang2015highly}    \\
\midrule
  MAD w.r.t. experiment & 1.10 & 0.57 & 0.43 & 0.15 & -\phantom{2.2 } \\
  \bottomrule
\end{tabular}
\end{center}
\end{table}

\subsection{Band extrema}
\label{sec:extrema}
\begin{figure}
\includegraphics{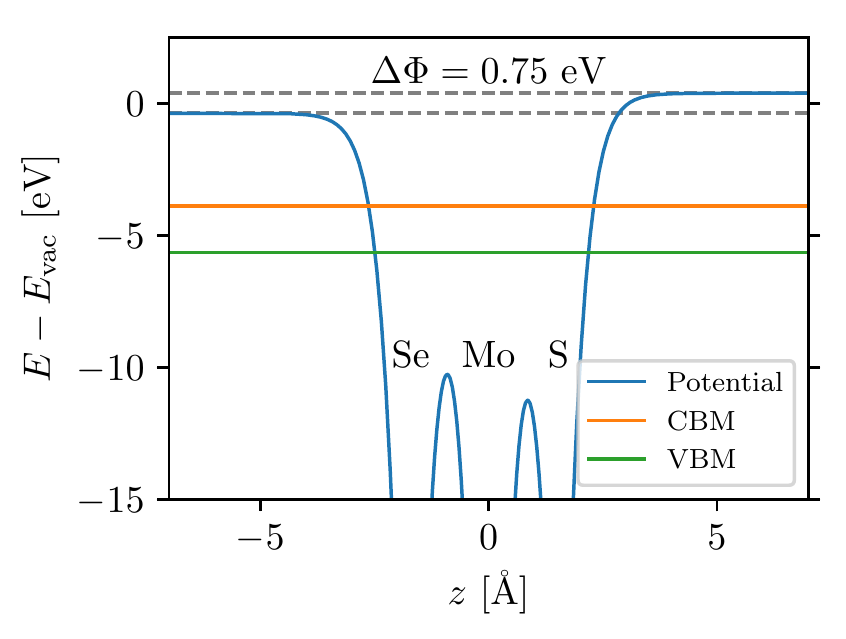}
\caption{Electrostatic potential profile perpendicular to monolayer MoSSe (averaged in plane). The position of the VBM and CBM are indicated together with the splitting of the vacuum levels caused by the out-of-plane dipole moment of the MoSSe layer.}
\label{fig:pot}
\end{figure}

For materials with a finite band gap, the positions of the valence band maximum (VBM) and conduction band minimum (CBM) within the BZ are identified together with their energies relative to the vacuum level. The latter is defined as the asymptotic value of the electrostatic potential, see Figure~\ref{fig:pot}. The PBE electrostatic potential is used to define the vacuum level in the non-selfconsistent HSE and G$_0$W$_0$ calculations. For materials with an out-of-plane dipole moment, a dipole correction is applied during the selfconsistent DFT calculation, and the vacuum level is defined as the average of the asymptotic electrostatic potentials on the two sides of the structure. The PBE vacuum level shift is also stored in the database. 

\subsection{Fermi surface}
The Fermi surface is calculated using the PBE xc-functional including SOC for all metallic compounds in the database. Based on a ground state calculation with a $k$-point density of at least 20/\AA$^{-1}$, the eigenvalues are interpolated with quadratic splines and plotted within the first BZ. Figure~\ref{fig:bz} (left) shows an example of the Fermi surface for VO$_2$-MoS$_2$ with color code indicating the out-of-plane spin projection $\langle S_z \rangle$. The band structure refers to the ferromagnetic ground state of VO$_2$-MoS$_2$, which has a magnetic moment of 0.70 $\mu_\text{B}$ per unit cell, characterised by alternating spin-polarised lobes with $\langle S_z \rangle = \pm 1$.

\def\imagebox#1#2{\vtop to #1{\null\hbox{#2}\vfill}}

\begin{figure*}
    \begin{subfigure}[t]{3.4in}
      \imagebox{2.6in}{\includegraphics{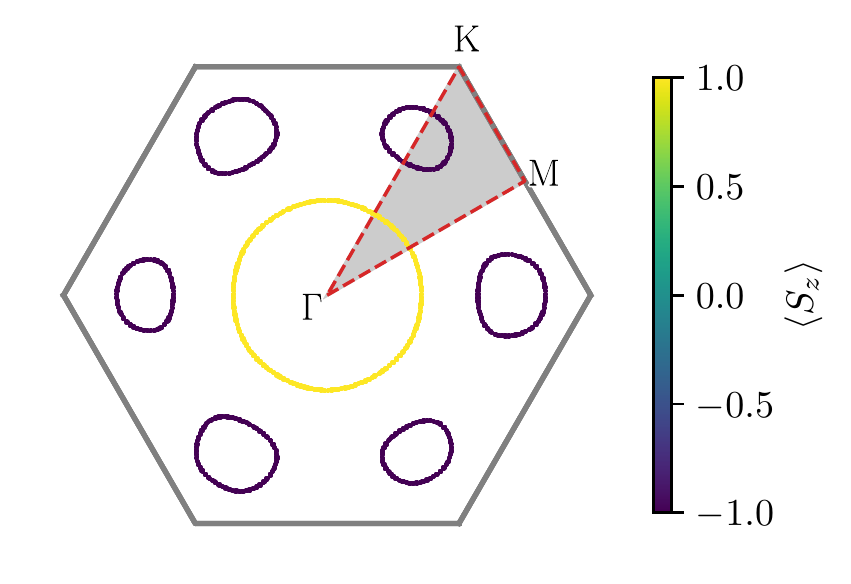}}
    \end{subfigure}    %
    \begin{subfigure}[t]{2.62in}
        \imagebox{2.6in}{\includegraphics{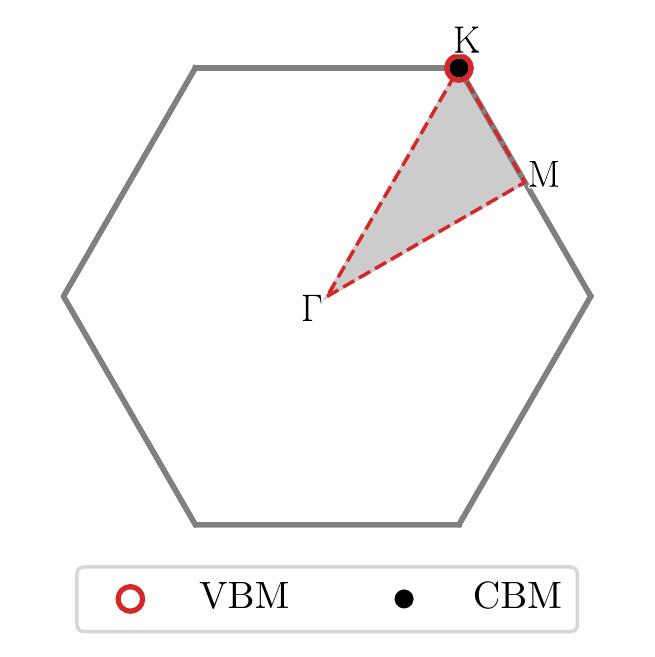}}
    \end{subfigure}
\caption{Left: Brillouin zone and Fermi surface calculated with PBE and spin-orbit coupling for VO$_2$ in the MoS2$_2$ crystal structure. The Fermi surface is colored by the spin projection along the $z$-axis. Right: Brillouin zone, valence band maximum (VBM) and conduction band minimum (CBM) for WS$_2$ in the MoS$_2$ crystal structure. The grey areas in both plots mark the irreducible Brillouin zone.}
\label{fig:bz}
\end{figure*}

\subsection{Effective masses}
\label{sec:mass}
For materials with a finite PBE gap, the effective electron and hole masses are calculated from the PBE eigenvalues; initially these are calculated on an ultrafine $k$-point mesh of density $45.0/\mathrm{\AA}^{-1}$ uniformly distributed inside a circle of radius 0.015 $\mathrm{\AA}^{-1}$ centered at the VBM and CBM, respectively. The radius is chosen to be safely above the noise level of the calculated eigenvalues but still within the harmonic regime; it corresponds to a spread of eigenvalues of about 1 meV within the circle for an effective mass of 1 $m_0$. For each band within an energy window of 100 meV above/below the CBM/VBM, the band curvature is obtained by fitting a third order polynomial. Even though the masses represent the second derivative of the band energies, we have found that the inclusion of 3rd order terms stabilises the fitting procedure and yields masses that are less sensitive to the details of the employed $k$-point grids. For each band the mass tensor is diagonalised to yield heavy and light masses in case of anisotropic band curvatures. The masses (in two directions) and the energetic splitting of all bands within 100 meV of the band extremum are calculated both with and without SOC and stored in the database. Other approaches exist for calculating effective masses, such as \(\mathbf{k} \cdot \mathbf{p}\) perturbation theory (see e.g. \citenum{kormanyos2015a} and references therein); the present scheme was chosen for its simplicity and ease of application to a wide range of different materials.

In addition to the effective masses at the VBM and CBM, the exciton reduced mass is calculated by applying the above procedure to the direct valence-conduction band transition energies, $\varepsilon_{\mathrm{v-c}}(\mathbf{k})=\varepsilon_{\mathrm{c}}(\mathbf{k})-\varepsilon_{\mathrm{v}}(\mathbf{k})$. For direct band gap materials the exciton reduced mass is related to the electron and hole masses by $1/\mu_{\mathrm{ex}}=1/m^*_{\mathrm{e}}+1/m^*_{\mathrm{h}}$, but in the more typical case of indirect band gaps, this relation does not hold.

\begin{figure}[htb]
\centering
    \begin{subfigure}[b]{0.50\columnwidth}
        \includegraphics[width=\columnwidth]{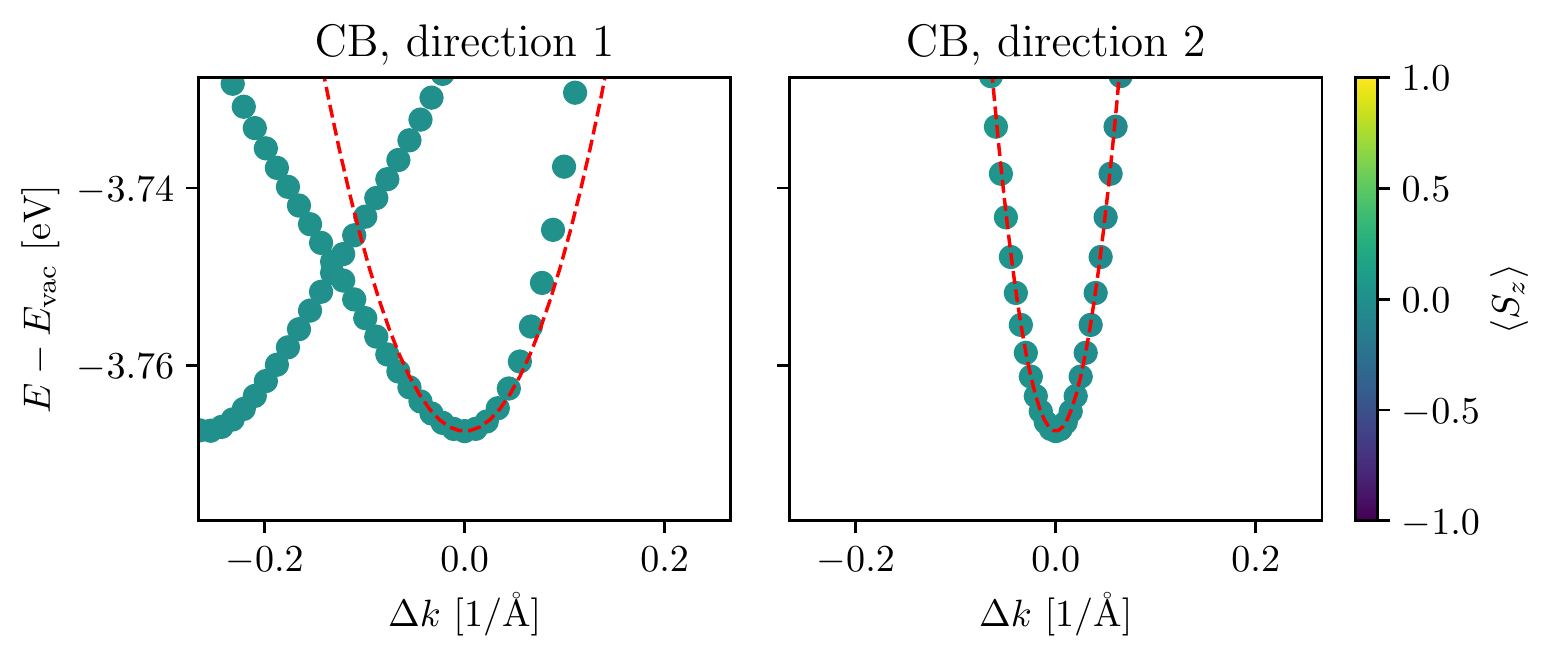}
    \end{subfigure}
    ~ 
    \begin{subfigure}[b]{0.50\columnwidth}
        \includegraphics[width=\columnwidth]{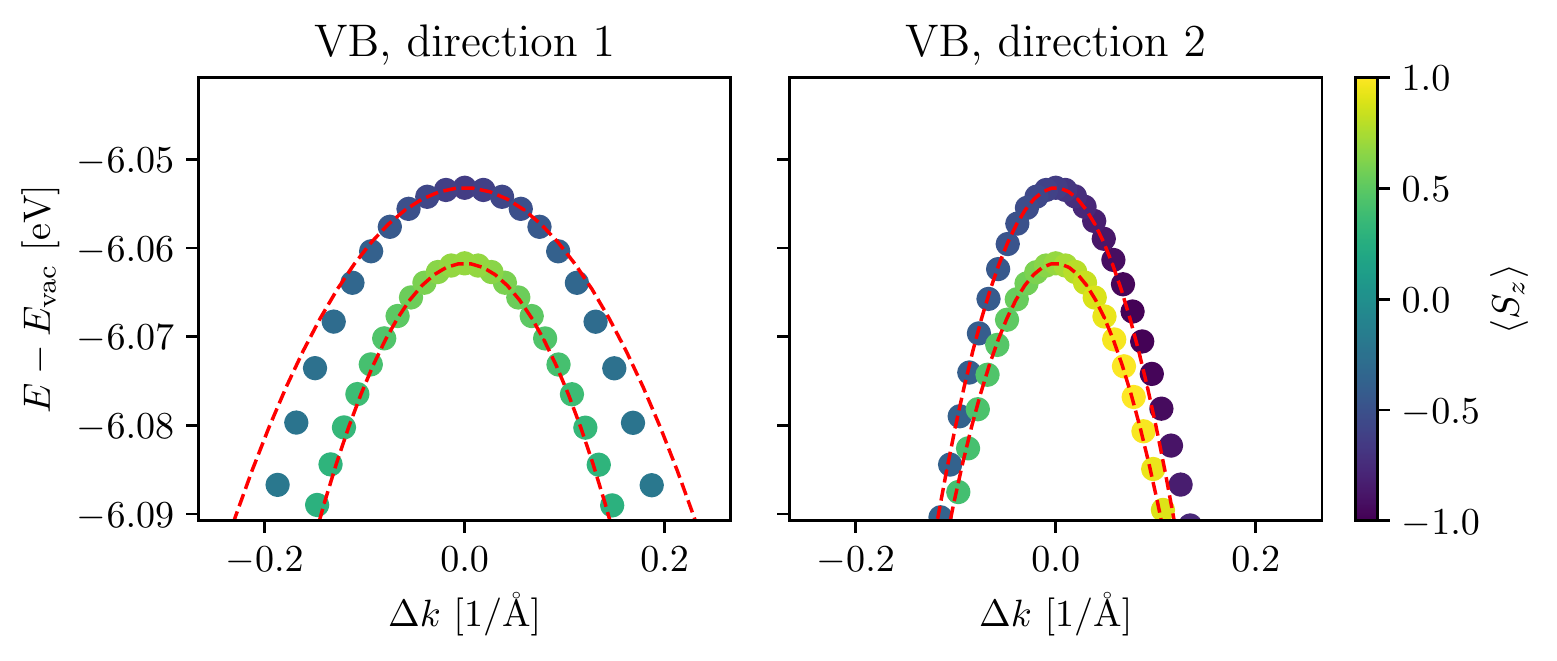}
    \end{subfigure}
  \caption{Zoom of the band structure of SnS in the GeSe crystal structure around the conduction and valence band extrema (upper and lower panels). Second order fits used to determine the effective masses are shown by red dashed lines. The peculiar band splitting in the conduction band minimum (upper left panel) is caused by a Rashba effect arising from the combination of spin-orbit coupling and the finite perpendicular electric field created by the asymmetric SnS structure.}
  \label{fig:masses_zoom}
\end{figure}

As an example, Figure~\ref{fig:masses_zoom} shows a zoom of the band structure of SnS-GeSe around the VBM and CBM (upper and lower panels). The second order fits to the band energies  (extracted from the fitted 3rd order polynomial) are shown by red dashed lines. It can be seen that both the conduction and valence bands are anisotropic leading to a heavy and light mass direction (left and right panels, respectively). The valence band is split by the SOC resulting in two bands separated by $\sim 10$ meV and with slightly different curvatures. The conduction band exhibits a non-trivial band splitting in one of the two directions. The peculiar band splitting is due to a Rashba effect arising from the combination of spin-orbit coupling and the finite perpendicular electric field created by the permanent dipole of the SnS structure where Sn and S atoms are displaced in the out of plane direction leading to a sizable vacuum level difference of 1.13 eV, see Figure \ref{fig:pot}.

Table~\ref{tab:masses} shows a comparison between selected effective masses from the C2DB and previously published data also obtained with the PBE xc-correlation functional and including SOC. Overall, the agreement is very satisfactory.

\begin{table}[htb]
  \begin{center}
\caption{Calculated PBE effective masses (in units of $m_0$), for the highest valence band and lowest conduction band, for different 2D materials. All C2DB values are calculated including spin-orbit coupling.\label{tab:masses}}
\begin{tabular}[h]{lrrrrr}
  \toprule
  & & \multicolumn{2}{c}{Electron mass ($m_0$)} & \multicolumn{2}{c}{Hole mass ($m_0$)} \\
  \cmidrule(l){3-4} \cmidrule(l){5-6}
  Material & $k$-point & C2DB & Literature & C2DB & Literature \\
  \midrule
  MoS$_2$                & K        & 0.42 & 0.44\cite{kormanyos2015a} & 0.53 & 0.54\cite{kormanyos2015a} \\
  WSe$_2$                & K        & 0.46 & 0.40\cite{kormanyos2015a} & 0.35 & 0.36\cite{kormanyos2015a} \\
  Phosphorene (zig-zag)  & $\Gamma$ & 1.24 & 1.24\cite{wang2015a}      & 6.56 & 6.48\cite{wang2015a}      \\
  Phosphorene (armchair) & $\Gamma$ & 0.14 & 0.13\cite{wang2015a}      & 0.13 & 0.12\cite{wang2015a}      \\
  \midrule
  MAD                    &          & 0.02 & -\phantom{0.0}            & 0.03 & -\phantom{0.0}            \\
  \bottomrule
\end{tabular}
\end{center}
\end{table}

\subsection{Work function}
For metallic compounds, the work function is obtained as the difference between the Fermi energy and the asymptotic value of the electrostatic potential in the vacuum region, see Figure~\ref{fig:pot}. The work function is determined for both PBE and HSE band structures (both including SOC) on a uniform $k$-point grid of density 12.0/\AA$^{-1}$. Since the SOC is evaluated non-selfconsistently, the Fermi energy is adjusted afterwards based on a charge neutrality condition.
 
\subsection{Deformation potentials}
For semiconductors, the deformation potentials quantify the shift in band edge energies (VBM or CBM) upon a linear deformation of the lattice. The uniaxial absolute deformation potential along axis $i$ ($i=x,y$) is defined as \cite{van1989a, resta1990a}
\begin{equation}\label{eq:deformation}
D_{ii}^\alpha=\frac{\Delta E_\alpha}{\varepsilon_{ii}},\quad \alpha=\mathrm{VBM,CBM} 
\end{equation}
where $\Delta E_\alpha$ is the energy shift upon strain and $\varepsilon_{ii}$ are the strains in the $i$-directions.

The deformation potentials are important physical quantities as they provide an estimate of the strength of the (acoustic) electron-phonon interaction, see Section \ref{sec:mobility}. Moreover, they are obviously of interest in the context of strain-engineering of band gaps and they can be used to can be used to infer an error bar on the band gap or band edge positions due to a known or estimated error bar on the lattice constant.

The calculation of $D^\alpha_{ii}$ is based on a central difference approximation to the derivative. A strain of $\pm 1\%$ is applied separately in the $x$ and $y$ directions and the ions are allowed to relax while keeping the unit cell fixed. Calculations are performed with the PBE xc-functional, a plane wave cutoff of 800 eV, and a $k$-point density of 12/Å$^{-1}$. 

\begin{figure}
  \centering
  \includegraphics{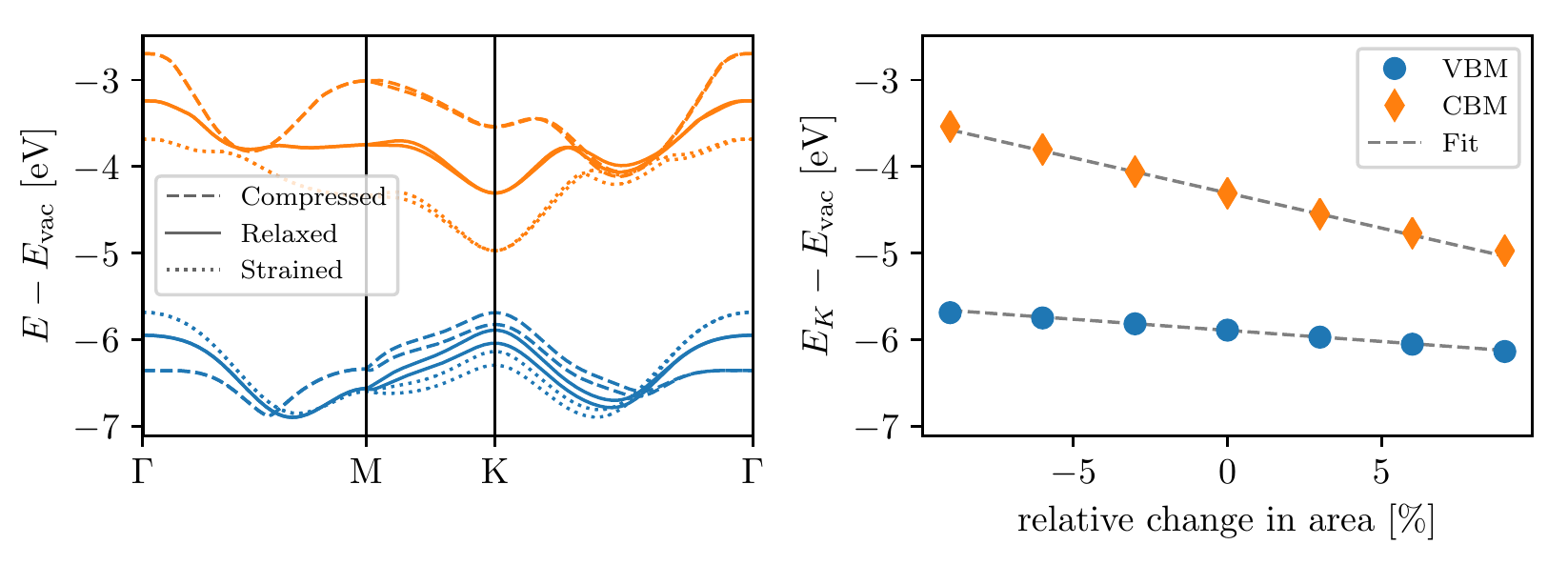}
\caption{Left: Valence- and conduction bands of MoS$_2$ for $\pm 4.5\%$ biaxial strain. Right: Energies of the VBM and CBM at the K point as function of strain. The symbols are the results of full DFT calculations, while the dashed lines are obtained from the deformation potentials evaluated at $\pm 1\%$ strain.\label{fig:strained-bands}}
\end{figure}

The change in band energy, $\Delta E_\alpha$, is measured relative to the vacuum level. In cases with nearly degenerate bands, care must be taken to track the correct bands as different bands might cross under strain. In this case, we use the expectation value $\langle \hat S_z \rangle$ to follow the correct band under strain. Figure~\ref{fig:strained-bands} shows how the band structure of MoS$_2$ changes as a function of strain. Both the VBM and the CBM shift down (relative to the vacuum level) when tensile strain is applied in the $x$ direction, but the conduction band shows a much larger shift, leading to an effective band gap closing under tensile strain.

Table~\ref{tab:defo_pot} shows a comparison between the deformation potentials in the C2DB, and literature values obtained using similar methods. There is generally good agreement, and part of the discrepancy can be ascribed to the fact that, in contrast to Ref. \citenum{wiktor2016}, our numbers include spin-orbit coupling.

\begin{table}[htb]
\begin{center}
  \caption{Absolute deformation potentials (in eV) of the VBM and CBM for different materials. All results are based on the PBE xc-functional.}
  \label{tab:defo_pot}
\begin{tabular}{lrrrrrr}
\toprule
              & & \multicolumn{2}{c}{Valence band} & \multicolumn{2}{c}{Conduction band}                            \\
\cmidrule(l){3-4} \cmidrule(l){5-6}
  Material     & $k$-point & C2DB & Ref.\citenum{wiktor2016} & C2DB & Ref. \citenum{wiktor2016} \\
\midrule

MoSe$_2$ & K & $-1.43$ & $-1.86$ & $-5.57$ & $-5.62$ \\ 
WS$_2$   & K & $-1.25$ & $-1.59$ & $-6.66$ & $-6.76$ \\
WSe$_2$  & K & $-1.21$ & $-1.43$ & $-6.21$ & $-6.35$ \\
\(h\)BN  & K & $-1.57$ & $-1.63$ & $-4.55$ & $-4.62$ \\
  \midrule
  MAD    &   & 0.26    & -       & 0.14    & -       \\
\bottomrule
\end{tabular}
\end{center}
\end{table}

\subsection{Plasma frequencies}
\label{sec:plasma}
The dielectric response of a 2D material is described by its 2D polarisability, $\alpha^\mathrm{2D}$ (see Section~\ref{sec:pol} for a general introduction of this quantity). For metals, it can be separated into contributions from intraband and interband transitions, i.e. $\alpha^\mathrm{2D} = \alpha^\mathrm{2D,intra} + \alpha^\mathrm{2D,inter}$. We have found that local field effects (LFEs) are negligible for the intraband component, which consequently can be treated separately and evaluated as an integral over the Fermi surface. Specifically, this leads to the Drude expression for the polarisability in the long wave length limit $\alpha^\mathrm{2D,intra}(\omega) = -\omega_\mathrm{P,2D}^2 / (2\pi\omega^2)$ where $\omega_\mathrm{P,2D}$ is the 2D plasma frequency, which in atomic units is given by
\begin{align}
  \omega^2_\mathrm{P,2D} = \frac{4\pi}{A} \sum_{sn\mathbf{k}} | \hat{\mathbf{q}} \cdot \mathbf{v}_{sn\mathbf{k}}|^2 \delta(\varepsilon_{sn\mathbf{k}} - \varepsilon_\mathrm{F}),
\end{align}
where $\mathbf{v}_{sn\mathbf{k}}= \langle sn\mathbf{k} | \mathbf{\hat{p}} / m_0 | sn\mathbf{k} \rangle$ is a velocity matrix element (with $m_0$ the electron mass), $\hat{\mathbf{q}} = \mathbf{q} / q$ is the polarisation direction, $s,n,\mathbf{k}$ denote spin, band and momentum indices, and $A$ is the supercell area. The 2D plasma frequency is related to the conventional 3D plasma frequency by $\omega_\mathrm{P,2D}^2(\omega) = \omega_\mathrm{P,3D}^2(\omega) L / 2$ where $L$ is the supercell height.

The plasma frequency defined above determines the intraband response of the 2D metal to external fields. In particular, it determines the dispersion relation of plasmon excitations in the metal. The latter are defined by the condition $\varepsilon^\mathrm{2D}(\omega_\text{P}) = 1 + 2 \pi q \alpha^{2\text{D}}(\omega_\text{P}) = 0$, where $q$ is the plasmon wave vector. Neglecting interband transitions (the effect of which is considered in Section~\ref{sec:stattrendsplasma}), the 2D plasmon dispersion relation becomes
\begin{align}
  \omega_\mathrm{P}(q) = \omega_{\mathrm{P,2D}} \sqrt{q} \label{eq:plasmonfrequency}
\end{align}
The plasma frequencies, $\omega_{\mathrm{P,2D}}$, for polarisation in the $x$ and $y$ directions, respectively, are calculated for all metals in the C2DB using the tetrahedral method\cite{MacDonald1979} to interpolate matrix elements and eigenvalues based on a PBE band structure calculation including SOC and with a $k$-point density of $20/\mathrm{\AA}^{-1}$.

\subsection{Electronic polarisability}
\label{sec:pol}
The polarisability tensor $\alpha_{ij}$ is defined by
\begin{align}
P_i(\mathbf{q},\omega)=\sum_j\alpha_{ij}(\mathbf{q},\omega)E_j(\mathbf{q},\omega),
\end{align}
where $P_i$ is the $i$'th component of the induced polarisation averaged over a unit cell and E\(_j\) is the $j$'th component of the macroscopic electric field. Using that $P_i=(D_i-E_i)/(4\pi)=\sum_j(\epsilon_{ij}-\delta_{ij})E_j/(4\pi)$ one observes that $\alpha_{ij}=(\epsilon_{ij}-\delta_{ij})/(4\pi)$, where $\epsilon_{ij}$ is the dielectric function. In contrast to the dielectric function, whose definition for a 2D material is not straightforward\cite{huser2013dielectric}, the polarisability allows for a natural generalisation to 2D by considering the induced dipole moment per unit area,
\begin{align}
P_i^{\mathrm{2D}}(\mathbf{q},\omega)=\sum_j\alpha^{\mathrm{2D}}_{ij}(\mathbf{q},\omega)E_j(\mathbf{q},\omega).
\end{align}
Since the $P_i$ is a full unit cell average and $P_i^{\mathrm{2D}}$ is integrated in the direction orthogonal to the slab, we have $P_i^{\mathrm{2D}}=LP_i$ and $\alpha^{\mathrm{2D}}_{ij}=L\alpha_{ij}$, where $L$ is the length of the unit cell in the direction orthogonal to the slab.

In the following, we focus on the longitudinal components of the polarisability and dielectric tensors, which are simply denoted by $\alpha$ and $\epsilon$. These are related to the density-density response function, $\chi$, via the relations 
\begin{align}
\alpha^{\mathrm{2D}}(\mathbf{q},\omega)&=\frac{L}{4\pi}(\epsilon(\mathbf{q},\omega)-1),\label{eq:2Dpol}\\
\epsilon^{-1}(\mathbf{q},\omega)&=1+\langle v_c(\mathbf{q})\chi(\omega)\rangle_\mathbf{q},
\end{align}
where $v_c$ is the Coulomb interaction and 
\begin{align}\label{eq:average}
\langle v_c\chi(\omega)\rangle_\mathbf{q}=\frac{1}{V}\int_{\mathrm{Cell}} d\mathbf{r}d\mathbf{r'}d\mathbf{r''}v_c(\mathbf{r},\mathbf{r'})\chi(\mathbf{r'},\mathbf{r''},\omega)e^{-i\mathbf{q}(\mathbf{r}-\mathbf{r''})},
\end{align}
where $\mathrm{Cell}$ is the supercell with volume $V$. The response function, $\chi$, satisfies the Dyson equation\cite{marques}
$\chi=\chi^{\mathrm{irr}} + \chi^{\mathrm{irr}}v_c\chi,$
where $\chi^{\mathrm{irr}}$ is the irreducible density-density response function. In the random phase approximation (RPA) $\chi^{\mathrm{irr}}$ is replaced by the non-interacting response function, $\chi^0$, whose plane wave representation is given by\cite{Hybertsen1987, Yan2011} 
\begin{align}\label{eq:chi0}
\chi^0_{\mathbf{G}\mathbf{G}'}(\mathbf{q}, \omega)=\frac{1}{\Omega}\sum_{\mathbf{k}\in\text{BZ}}\sum_{mn}(f_{n\mathbf{k}}-f_{m\mathbf{k+q}})\frac{\langle\psi_{n\mathbf{k}}|e^{-i (\mathbf{q} + \mathbf{G})\cdot{\mathbf{r}}}|\psi_{m\mathbf{k+q}}\rangle\langle\psi_{m\mathbf{k+q}}|e^{i (\mathbf{q} + \mathbf{G}')\cdot{\mathbf{r}}}|\psi_{n\mathbf{k}}\rangle}{\hbar\omega+\varepsilon_{n\mathbf{k}}-\varepsilon_{m\mathbf{k+q}}+i\eta},
\end{align}
where $\mathbf{G},\mathbf{G}'$ are reciprocal lattice vectors and $\Omega$ is the crystal volume.

For all materials in the database, we calculate the polarisability within the RPA for both in-plane and out-of-plane polarisation in the optical limit $q \rightarrow 0$. For metals, the interband contribution to the polarisability is obtained from Eq.~\eqref{eq:chi0} while the intraband contribution is treated separately as described in Section~\ref{sec:plasma}. The single-particle eigenvalues and eigenstates used in Eq. \ref{eq:chi0} are calculated with PBE, a $k$-point density of $20/\mathrm{\AA}^{-1}$ (corresponding to a $k$-point grid of $48\times48$ for MoS$_2$ and $60\times60$ for graphene), and 800 eV plane wave cutoff. The Dyson equation is solved using a truncated Coulomb potential\cite{Rozzi2006, Huser2013a} to avoid spurious interactions from neighboring images. We use the tetrahedron method to interpolate the eigenvalues and eigenstates and a peak broadening of $\eta=50$ meV. Local field effects are accounted for by including $\mathbf G$-vectors up to 50 eV. For the band summation we include 5 times as many unoccupied bands as occupied bands, which roughly corresponds to an energy cutoff of 50 eV. The calculations are performed without spin-orbit coupling.

\begin{figure}[h!]
\includegraphics[width=8cm]{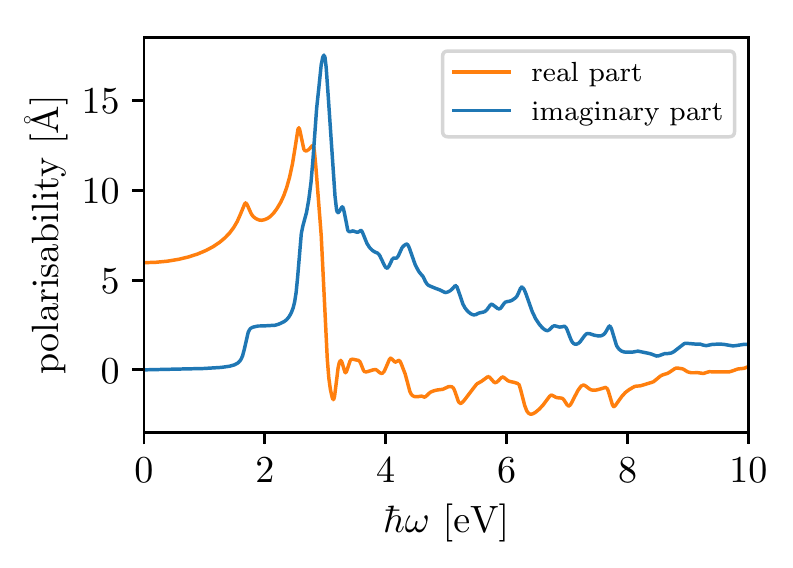}
\includegraphics[width=8cm]{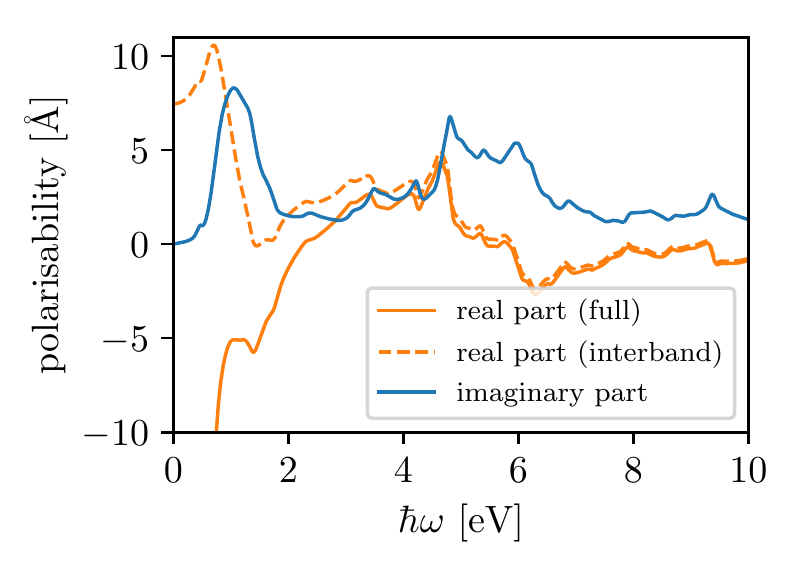}
\caption{Real and imaginary part of the RPA in-plane polarisability of monolayer MoS$_2$ in the H phase (left) and the metallic monolayer NbS$_2$ in the T phase (right). For metals, the real part is shown both with and without the intraband contributions.}
\label{fig:pol}
\end{figure}

In Figure~\ref{fig:pol} we show the real and imaginary part of $\alpha^{\mathrm{2D}}$ for the semiconductor MoS$_2$. The PBE band gap of this material is 1.6 eV and we see the onset of dissipation at that energy. We also see that the initial structure of $\mathrm{Im}\,{\alpha}$ is a constant, which is exactly what would be expected from the density of states in a 2D material with parabolic dispersion. Finally, we note that the static polarisability $\left.\mathrm{Re}\,\alpha\right|_{\omega=0}\approx6\mathrm{\AA}$, which can easily be read off the figure. The polarisability is also shown for the metallic 1T-NbS$_2$ where we display the real part with and without the intraband Drude contribution $\omega^2_\mathrm{P,2D}/(\hbar\omega+i\eta)^2$.

\subsection{Optical absorbance}

The power absorbed by a 2D material under illumination of a monochromatic light field with polarisation $\hat{ \mathbf{e}}$ is quantified by the dimensionless absorbance:
\begin{align}
  \label{eq:abs}
 \textrm{Abs}(\omega)=4\pi\omega\alpha^{\mathrm{2D}}(q\hat{\mathbf{e}}\rightarrow0,\omega) / c,
\end{align}
where $c$ is the speed of light. In Section \eqref{sec:pol} we gave a prescription for evaluating $\alpha^{\mathrm{2D}}$ in the RPA. However, absorption spectra of 2D semiconductors often display pronounced excitonic effects, which are not captured by the RPA. The Bethe-Salpeter equation (BSE) is a well-known method capable of describing excitonic effects and has been shown to provide good agreement with experimental absorption spectra for a wide range of materials.\cite{Onida2002}

For materials with finite band gap and up to four atoms per unit cell, we have calculated the RPA and the BSE absorption spectra for electric fields polarised parallel and perpendicular to the layers. The calculations are performed on top of PBE eigenstates and eigenvalues with spin-orbit coupling included and all unoccupied band energies shifted by a constant in order to reproduce the \GW  quasiparticle gap (the scissors operator method). If the \GW  gap is not available we use the GLLBSC gap for non-magnetic materials and the HSE gap for magnetic materials (since GLLBSC is not implemented in GPAW for spin-polarised systems). The screened interaction entering the BSE Hamiltonian is calculated within the RPA using a non-interacting response function evaluated from Eq. (\ref{eq:chi0}) with local field effects (i.e. $\mathbf G$-vectors) included up to 50 eV and 5 times as many unoccupied bands as occupied bands for the sum over states. We apply a peak broadening of $\eta=50$ meV and use a truncated Coulomb interaction. The BSE Hamiltonian is constructed from the four highest occupied and four lowest unoccupied bands on a $k$-point grid of density of $20/\mathrm{\AA}^{-1}$, and is diagonalised within the Tamm-Dancoff approximation. We note that the Tamm-Dancoff approximation has been found to be very accurate for bulk semiconductors\cite{sander2015beyond}. For monolayer MoS$_2$ we have checked that it reproduces the full solution of the BSE, but its general validity for 2D materials, in particular those with small band gaps, should be more thoroughly tested.

\begin{figure}
\includegraphics{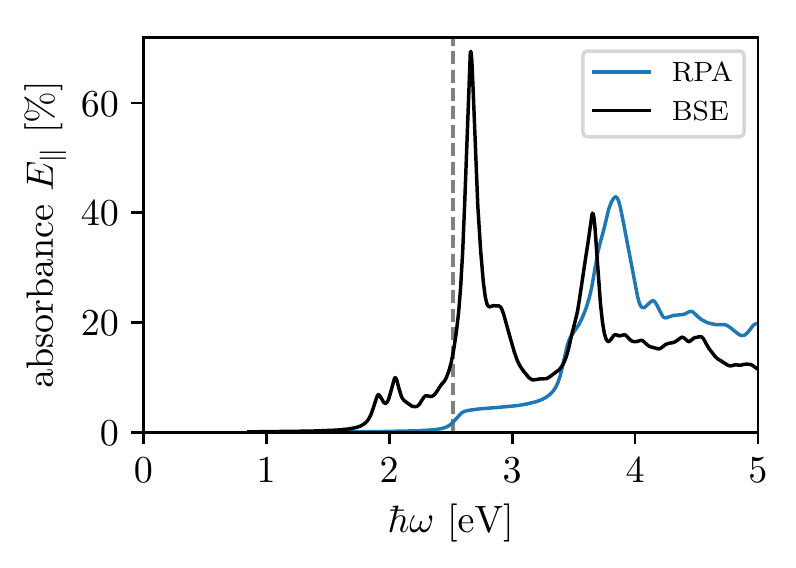}
\includegraphics{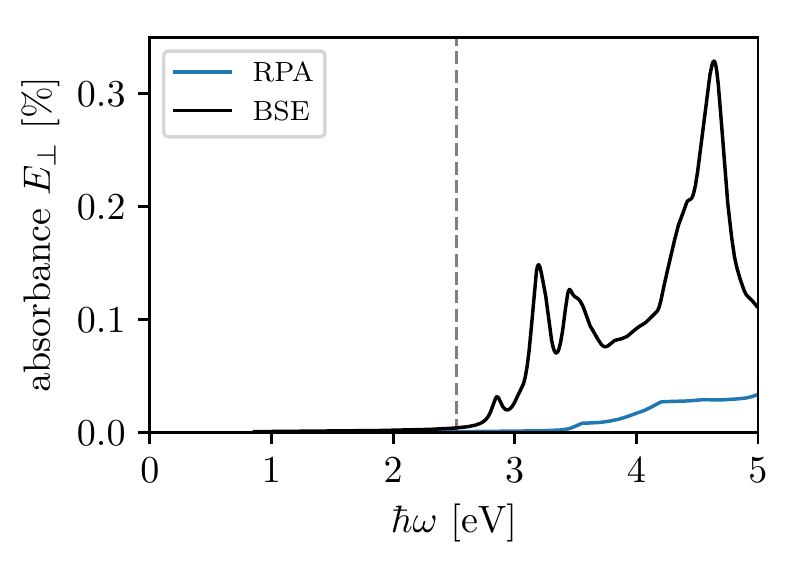}
\caption{Optical absorbance of MoS$_2$ obtained with the RPA and BSE methods. Left: Electric field polarisation parallel to the layer. The lowest energy part of the BSE spectrum is composed of two excitonic peaks separated by 0.15 eV, which originates from the spin-orbit splitting in the valence band at $\mathrm{K}$. Right: Electric field polarised perpendicular to the layer.}
\label{fig:abs}
\end{figure}

In Figure~\ref{fig:abs} we show the optical absorption spectrum of MoS$_2$ obtained with the electric field polarised parallel and perpendicular to the layer, respectively. Both RPA and BSE spectra are shown (the in-plane RPA absorbance equals the imaginary part of the RPA polarisability, see Figure \ref{fig:pol}(left), apart from the factor $4 \pi \omega$ and the scissors operator shift). The low energy part of the in-plane BSE spectrum is dominated by a double exciton peak (the so-called A and B excitons) and is in excellent agreement with experiments.\cite{Mak2010}

\begin{figure}
\includegraphics{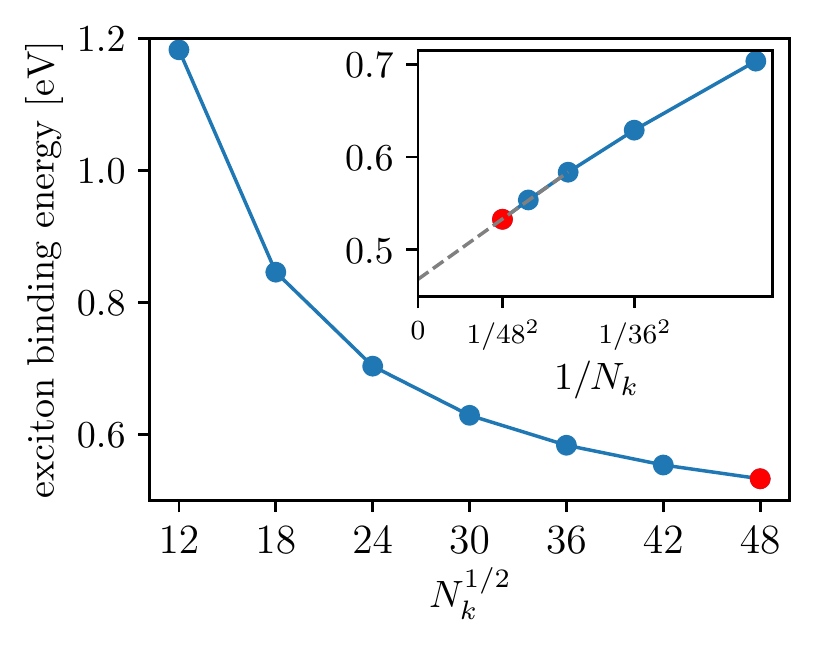}
\caption{Convergence of the binding energy of the lowest exciton in monolayer MoS$_2$ obtained from a BSE calculation as a function of $k$-point mesh. The quasiparticle energies entering the BSE Hamiltonian are obtained from a fully converged PBE calculations with a scissors operator applied to match the G$_0$W$_0$ band gap. The red point represents the $k$-point sampling applied in the database, which is seen to overestimate the extrapolated exciton binding energy by $\sim$0.06 eV (inset).}
\label{fig:bse_con}
\end{figure}

In general, calculations of electronic excitations of 2D materials converge rather slowly with respect to $k$-points due to the non-analytic behavior of the dielectric function in the vicinity of $q=0$.\cite{Huser2013b, Latini2015, DaJornada2017} In Figure~\ref{fig:bse_con} we show the $k$-point dependence of the binding energy of the A exciton in MoS$_2$ obtained as the difference between the direct band gap and the position of the first peak in Figure~\ref{fig:abs}. We observe a strong overestimation of the exciton binding energy at small $k$-point samplings, which converges slowly to a value of $\sim0.5$ eV at large $k$-point samplings. For $48\times 48$ $k$-points, corresponding to the $k$-point sampling used for the BSE calculations in the database, the exciton binding energy is 0.53 eV, whereas a $1 / N_k^2$ extrapolation to infinite $k$-point sampling gives 0.47 eV (see inset in Figure~\ref{fig:bse_con}). In general, the exciton binding energy decreases with increasing $k$-point sampling, and thus the exciton binding energies reported in the C2DB might be slightly overestimated. However, since \GW  band gaps also decrease when the $k$-point sampling is increased (see Figure~\ref{fig:gw-conv}) the two errors tend to cancel such that the absolute position of the absorption peak from BSE-\GW  converges faster than the band gap or exciton binding energy alone.

The BSE-\GW  method has previously been shown to provide good agreement with experimental photoluminescence and absorption measurements on 2D semiconductors. In Table~\ref{tab:bse} we show that our calculated position of the first excitonic peak agree well with experimental observations for four different TMDCs and phosphorene. Experimentally, the monolayers are typically supported by a substrate, which may alter the screening of excitons. However the resulting decrease in exciton binding energies is largely cancelled by a reduced quasiparticle gap such that the positions of the excitons are only slightly red-shifted as compared with the case of pristine monolayers\cite{drueppel2017a}.
\begin{center}
  \begin{table}
    \begin{tabular}{lcc}
      \toprule
      & \multicolumn{2}{c}{Energy of the first bright exciton (eV)}\\
      \cmidrule(l){2-3}
      Material & BSE@PBE-\GW  & Experiment \\
      \midrule
      MoS$_2$  & 2.00\hphantom{0} & 1.83\cite{Mouri2017}, 1.86\cite{Chen2016}, 1.87\cite{Fang2014a} \\
      MoSe$_2$ & 1.62\hphantom{0} & 1.57\cite{Mouri2017}, 1.57\cite{Wilson2017}, 1.58\cite{Ceballos2015} \\
      WS$_2$   & 2.07\hphantom{0} & 1.96\cite{Chen2016}, 2.02\cite{Ceballos2015} \\
      WSe$_2$  & 1.71\hphantom{0} & 1.64\cite{Fang2014a}, 1.66\cite{Wilson2017} \\
      P (phosphorene) & 1.45\hphantom{0} & 1.45\cite{Liu2014a}, 1.75\cite{Yang2015a} \\
      \midrule
      MAD w.r.t. experiment & 0.066& -\\
      \bottomrule
    \end{tabular}
    \caption{Comparison between calculated and experimental positions of the first excitonic peak for four different transition metal dichalcogenide monolayers and phosphorene.}
    \label{tab:bse}
  \end{table}
\end{center}

\section{Database overview}

\label{sec:overview}
Having described the computational workflow, we now turn to the content of the database itself. We first present a statistical overview of all the materials in the C2DB (i.e. without applying any stability filtering) by displaying their distribution over crystal structure prototypes and their basic properties. We also provide a short list with some of the most stable materials, which to our knowledge have not been previously studied. Next, the predicted stability of the total set of materials is discussed and visualised in terms of the descriptors for thermodynamic and dynamic stability introduced in Section \ref{sec:stabilityscale}. In Section~\ref{sec:data} we analyse selected properties in greater detail focusing on band gaps and band alignment, effective masses and mobility, magnetic properties, plasmons, and excitons. Throughout the sections we explore general trends and correlations in the data and identify several promising materials with interesting physical properties.  

 \subsection{Materials}
\label{sec:materials}
In Table~\ref{tab:classes} we list the major classes of materials currently included in the database. The materials are grouped according to their prototype, see Section~\ref{sec:prototypes}. For each prototype we list the corresponding space group, the total number of materials, and the number of materials satisfying a range of different conditions. The atomic structure of some of the different prototypes were shown in Figure~\ref{fig:prototype_overview}. The vast majority of the 2D materials that have been experimentally synthesised in monolayer form are contained in the C2DB (the 55 materials in Figure \ref{fig:exp} in addition to seven metal-organic perovskites). These materials are marked in the database and a literature reference is provided. Additionally, 80 of the monolayers in the C2DB could potentially be exfoliated from experimentally known layered bulk structures\cite{cheon2017data,ashton2017topology,mounet2016novel,choudhary2017a}. These materials are also marked and the ID of the bulk compound in the relevant experimental crystal structure database is provided.  

\begin{table}
  \caption{\label{tab:classes}Overview of the materials currently in the C2DB. The table shows the number of compounds listed by their crystal structure prototype and selected properties. $E_{\mathrm{gap}} > 0$ and ``direct gap'' refer to the PBE values, ``high stability'' refers to the stability scale defined in Section~\ref{sec:stabilityscale}, and the last three columns refer to the magnetic state, see Section~\ref{sec:relaxation}.  In this overview, separate magnetic phases of the same structure are considered different materials.}
\begin{center}
  \begin{tabular}{llrrrrrrr}
    \toprule
                      &          & \multicolumn{7}{c}{Number of materials}                                            \\
    \cmidrule(l){3-9}
Prototype & Symmetry & Total & $E_{\mathrm{gap}} > 0$ & Direct gap & High stability & NM & FM & AFM \\
\midrule

C        & P6/mmm            & 4 & 4 & 3 & 1 & 4 & 0 & 0 \\
CH       & P$\overline{3}$m1 & 8 & 7 & 6 & 1 & 8 & 0 & 0 \\
CH$_2$Si & P3m1              & 2 & 2 & 2 & 1 & 2 & 0 & 0 \\

BN   & P$\overline{3}$m2 & 10  & 9  & 5  & 1  & 10  & 0  & 0  \\
GaS  & P$\overline{3}$m2 & 125 & 34 & 95 & 8  & 100 & 18 & 7  \\
FeSe & P4/nmm            & 103 & 13 & 90 & 26 & 74  & 18 & 11 \\
GeSe & P3m1              & 20  & 19 & 5  & 6  & 20  & 0  & 0  \\
PbSe & P4/mmm            & 44  & 6  & 38 & 1  & 33  & 8  & 3  \\
P    & Pmna              & 9   & 9  & 0  & 1  & 9   & 0  & 0  \\

MoS$_2$ & P$\overline{3}$m2 & 241 & 85 & 176 & 53 & 156 & 85 & 0 \\
CdI$_2$  & P$\overline{3}$m1 & 315 & 95 & 231 & 90 & 218 & 80 & 17 \\
WTe$_2$  & P2$_1$/m          & 75  & 29 & 48  & 34 & 57  & 13 & 5  \\

FeOCl & Pmmn & 443  & 92 & 385 & 65 & 328 & 63 & 52    \\
MoSSe        & P3m1 & 9  & 6   & 6  & 5   & 8  & 1 & 0 \\

C$_3$N & P6/mmm & 25                & 1   & 24 & 0  & 25 & 0  & 0       \\
YBr$_3$         & P6/mmm            & 57  & 11 & 51 & 0  & 21 & 24 & 12 \\
TiCl$_3$        & P$\overline{3}$2m & 69  & 35 & 51 & 2  & 32 & 23 & 14 \\
BiI$_3$         & P$\overline{3}$m1 & 123 & 69 & 66 & 15 & 48 & 54 & 21 \\
TiS$_3$         & Pmmn              & 34  & 8  & 28 & 5  & 31 & 2  & 1  \\
MnTe$_3$        & P2$_1$/m          & 29  & 3  & 27 & 1  & 22 & 4  & 3  \\

Cr$_3$WS$_8$ & Pmm2   & 35                & 34 & 18 & 8  & 35 & 0  & 0     \\
CrWS$_4$              & Pmm2              & 18 & 17 & 7  & 8  & 18 & 0 & 0 \\
Ti$_2$CO$_2$          & P$\overline{3}$m1 & 28 & 8  & 20 & 12 & 19 & 6 & 3 \\
Ti$_2$CH$_2$O$_2$     & P$\overline{3}$m1 & 13 & 3  & 12 & 3  & 10 & 2 & 1 \\
Ti$_3$C$_2$           & P$\overline{3}$m2 & 12 & 0  & 12 & 0  & 7  & 5 & 0 \\
Ti$_3$C$_2$O$_2$      & P$\overline{3}$m2 & 26 & 0  & 26 & 0  & 20 & 6 & 0 \\
Ti$_3$C$_2$H$_2$O$_2$ & P$\overline{3}$m2 & 14 & 0  & 14 & 0  & 10 & 4 & 0 \\
PbA$_2$I$_4$          & P1               & 27 & 27 & 27 & 0  & 27 & 0 & 0 \\

\midrule
Sum &  & 1918 & 626 & 151 & 347 & 1352 & 416 & 150 \\
\bottomrule
\end{tabular}
\end{center}
\end{table}

\begin{figure}[htb]
  \centering
  \includegraphics{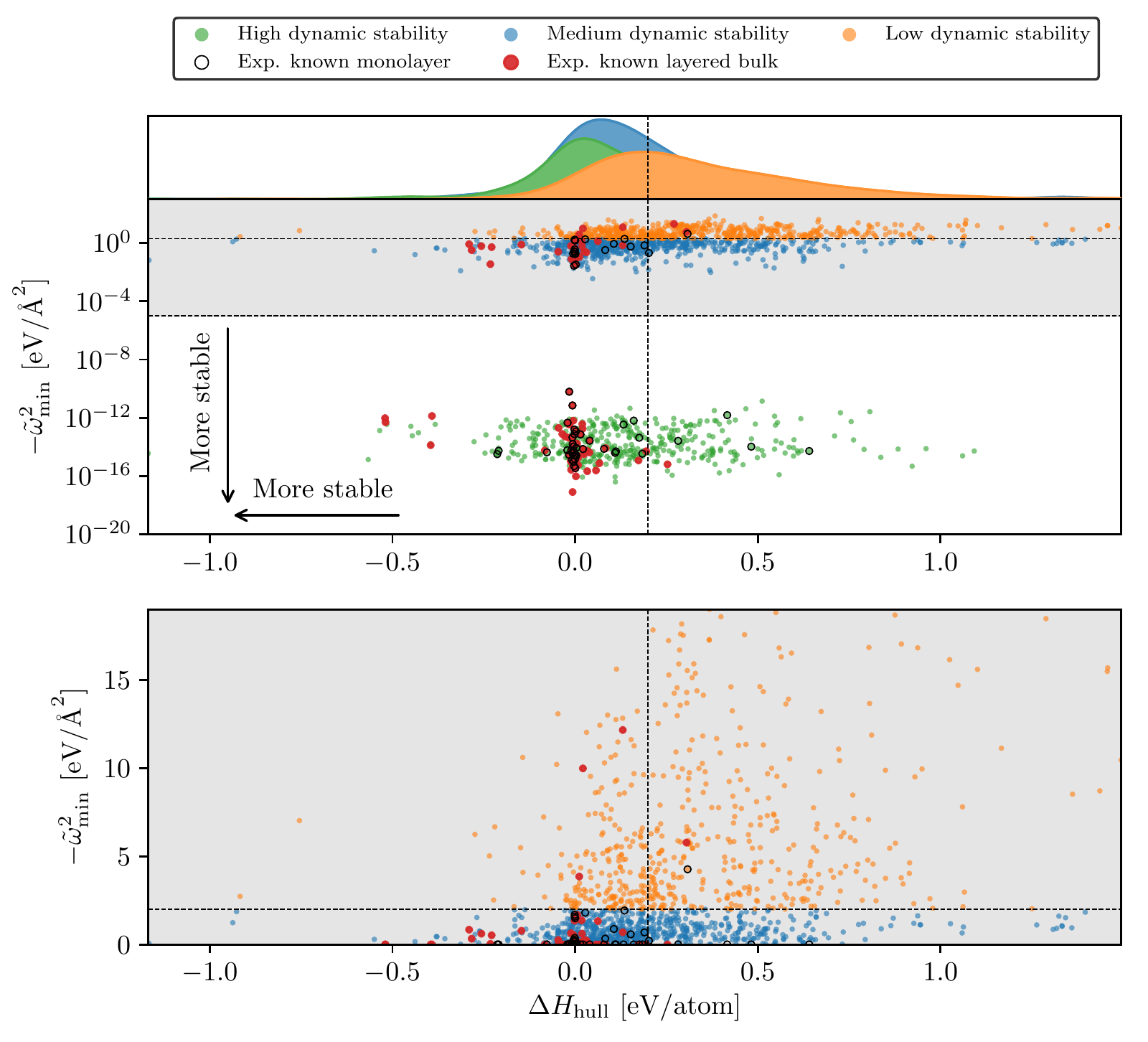}
\caption{The dynamic stability of the candidate materials as a function of the energy above the convex hull on a log scale (top), and a linear scale (bottom).  Experimentally synthesised monolayers are circled in black, while the known layered 3D structures are marked in red. The three different dynamic stability levels are indicated both by the horizontal dashed lines and the color of the points. The upper panel shows the marginal distribution of the energy over the convex hull for the points in each of the three stability levels.\label{fig:dynamic-thermodynamic}.}
\end{figure}

To illustrate how all the materials are distributed in terms of stability, we show the energy above the convex hull plotted against \(\tilde{\omega}^2_{\mathrm{min}}\) in Figure~\ref{fig:dynamic-thermodynamic}. It can be seen that the structures naturally sort themselves into two clusters according to the dynamic stability. The points have been colored according to the three levels for dynamic stability introduced in section 2.4. The lower panel shows the distribution of the materials in the grey region on a linear scale. While most of the experimentally known materials (red and black dots) have high dynamic stability, a significant part of them fall into the medium stability category. The marginal distributions on the plot show that the more dynamically stable materials are also more thermodynamically stable. The mean energy above the convex hull is 0.12 eV for the materials with high dynamical stability, while it is 0.25 eV for the others.

In Table~\ref{tab:stable} we show the key properties of a selected set of stable materials, distributed across a variety of different crystal structure prototypes. To our knowledge, these materials are not experimentally known, and they are therefore promising candidates further study and experimental synthesis.

\begin{table}
  \centering
  \caption{\label{tab:stable}Key properties of selected stable materials in the C2DB, which have not been previously synthesised. The calculated properties are the magnetic state, formation energy, energy above the convex hull, work function, PBE gap and the nature of the gap (direct or indirect).}
\begin{tabular}{llC{2cm}R{1.2cm}R{1.2cm}R{1.2cm}C{1.7cm}L{1.5cm}}
\toprule
  Prototype & Formula & Magnetic state &  $\Delta H $ (eV) &  $\Delta H_{\mathrm{hull}}$ (eV) &  \(\Phi\) (eV) &  PBE gap (eV) & Direct gap\\

  \midrule
BiI$_3$      & VI$_3$       & FM  & $-0.51$ & $-0.15$ & 5.3  &      &     \\
             & CoCl$_3$     & NM  & $-0.65$ & $-0.21$ &      & 1.13 & No  \\
             & CoBr$_3$     & NM  & $-0.41$ & $-0.16$ &      & 0.96 & No  \\
             & CoI$_3$      & NM  & $-0.14$ & $-0.14$ &      & 0.53 & No  \\
   \midrule
CdI$_2$      & FeO$_2$      & FM  & $-1.14$ & $-0.36$ & 7.31 &      &     \\
             & MnSe$_2$     & FM  & $-0.47$ & $-0.18$ & 5.09 &      &     \\
             & MnS$_2$      & FM  & $-0.57$ & $-0.12$ & 5.74 &      &     \\
             & PdO$_2$      & NM  & $-0.40$ & $-0.08$ &      & 1.38 & No  \\
             & CaBr$_2$     & NM  & $-2.09$ & $-0.02$ &      & 4.86 & No  \\
      \midrule
FeOCl        & RhClO        & NM  & $-0.65$ & $-0.18$ & 5.49 &      &     \\
             & NiClO        & AFM & $-0.64$ & $-0.17$ & 6.32 &      &     \\
             & NiBrO        & AFM & $-0.52$ & $-0.16$ & 5.78 &      &     \\
             & ScIS         & NM  & $-1.68$ & $-0.14$ &      & 1.66 & Yes \\
     \midrule
FeSe         & CoSe         & FM  & $-0.27$ & 0.02    & 4.22 &      &     \\
             & RuS          & NM  & $-0.38$ & 0.05    & 4.72 &      &     \\
             & MnSe         & AFM & $-0.50$ & $-0.20$ &      & 0.90 & No  \\
             & MnS          & AFM & $-0.64$ & $-0.19$ &      & 0.78 & No  \\
    \midrule
GaS          & AlSe         & NM  & $-0.72$ & $-0.02$ &      & 2.00 & No  \\
             & AlS          & NM  & $-0.89$ & 0.00    &      & 2.09 & No  \\
     \midrule
GeSe         & GeSe         & NM  & $-0.19$ & 0.04    &      & 2.22 & No  \\
             & GeS          & NM  & $-0.22$ & 0.05    &      & 2.45 & No  \\
             & GeTe         & NM  & $-0.01$ & 0.09    &      & 1.47 & No  \\
             & SnSe         & NM  & $-0.33$ & 0.10    &      & 2.15 & No  \\
     \midrule
MoS$_2$      & VS$_2$       & FM  & $-0.88$ & $-0.02$ & 5.95 &      &     \\
             & ScBr$_2$     & FM  & $-1.59$ & $-0.40$ &      & 0.16 & No  \\
             & YBr$_2$      & FM  & $-1.73$ & $-0.23$ &      & 0.34 & No  \\
             & FeCl$_2$     & FM  & $-0.67$ & $-0.16$ &      & 0.35 & Yes \\
             & TiBr$_2$     & NM  & $-1.14$ & $-0.04$ &      & 0.76 & No  \\
             & ZrBr$_2$     & NM  & $-1.34$ & $-0.04$ &      & 0.83 & No  \\
       \midrule
Ti$_2$CO$_2$ & Zr$_2$CF$_2$ & NM  & $-2.36$ & $-0.08$ & 3.92 &      &     \\
             & Hf$_2$CF$_2$ & NM  & $-2.26$ & 0.03    & 3.62 &      &     \\
             & Y$_2$CF$_2$  & NM  & $-2.50$ & $-0.17$ &      & 1.12 & No  \\
     \midrule
WTe$_2$      & NbI$_2$      & NM  & $-0.37$ & 0.04    & 3.01 &      &     \\
             & HfBr$_2$     & NM  & $-1.16$ & $-0.18$ &      & 0.85 & No  \\
             & OsSe$_2$     & NM  & $-0.17$ & 0.00    &      & 0.57 & No  \\
     \bottomrule
\end{tabular}
\end{table}
\clearpage
 
\subsection{Properties: Example applications}
\label{sec:data}
In the following sections we present a series of case studies focusing on different properties of 2D materials including band gaps and band alignment, effective masses and mobility, magnetic order, plasmons and excitons. The purpose is not to provide an in-depth nor material specific analysis, but rather to explore trends and correlations in the data and showcase some potential applications of the C2DB. Along the way, we report some of the novel candidate materials revealed by this analysis, which could be interesting to explore closer in the future.

\subsubsection{Band gaps and band alignment}
\label{sec:trend-band-gap}
The band gaps and band edge positions of all semiconductors and insulators in the C2DB have been calculated with the PBE, HSE06, and GLLBSC xc-functionals while \GW calculations have been performed for the $\sim$250 simplest materials. The relatively large size of these datasets and the high degree of consistency in the way they are generated (all calculations performed with the same code using same PAW potentials and basis set etc.) provide a unique opportunity to benchmark the performance of the different xc-functionals against the more accurate \GW method.

\begin{figure*}
  \begin{subfigure}[b]{0.5\columnwidth}
    \includegraphics{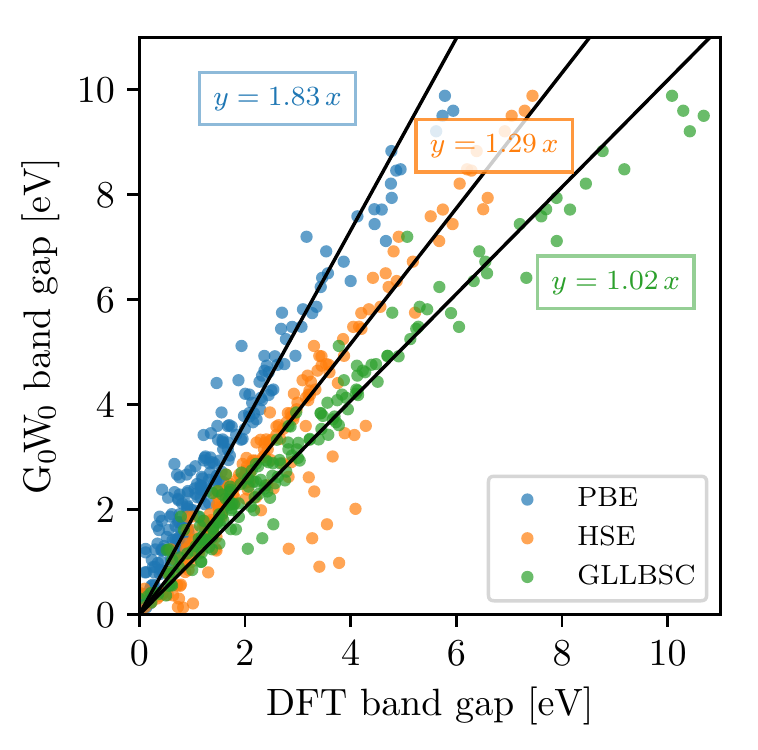}
  \end{subfigure}%
  \begin{subfigure}[b]{0.5\columnwidth}
    \includegraphics{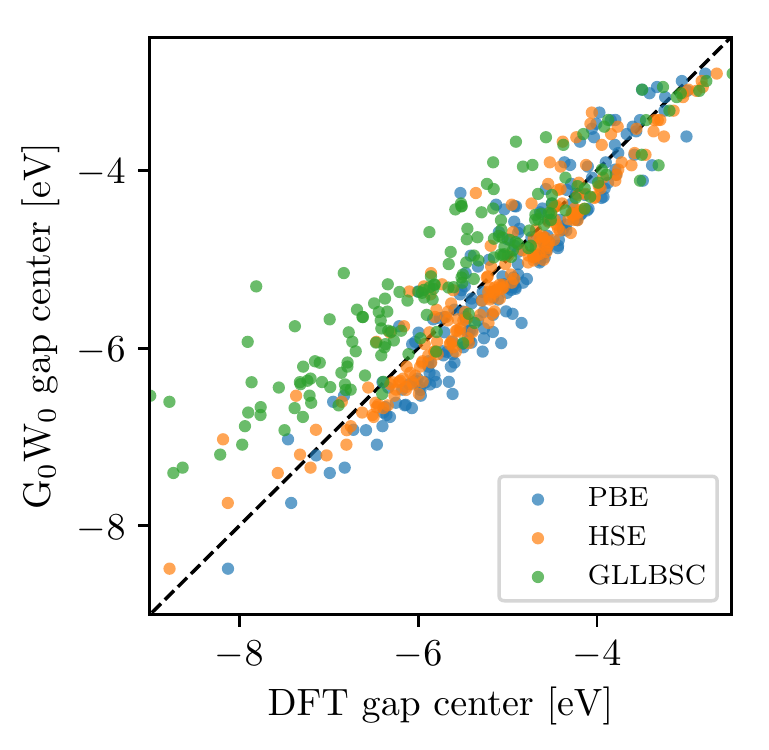}
  \end{subfigure}
\caption{Band gaps (left) and gap centers (right) obtained with different DFT functionals plotted against the \GW  results.
The band gaps of PBE (blue), HSE06 (orange), and GLLBSC (green) are fitted linearly to \GW\!\!.
For the gap centers the bisector is shown.}
\label{fig:bs-trends}
\end{figure*}

\begin{table}[h]\centering
  \begin{tabular}{lccc}
  \toprule
     & PBE & HSE06 & GLLBSC \\\hline
    MAD w.r.t. \GW  (band gap)   & 1.49 & 0.82 & 0.38 \\
    MAD w.r.t. \GW  (gap center) & 0.37 & 0.32 & 0.76 \\
  \bottomrule
  \end{tabular}
\caption{The mean absolute deviation (in eV) of the band gap and band gap center calculated with three different xc-functionals with respect to \GW\!. 
}\label{tab:MAE}
\end{table}

Figure~\ref{fig:bs-trends} compares the size and center of the band gaps obtained with the density functionals
to the \GW  results. Relative to \GW  the PBE functional underestimates the gaps by 45\%, i.e. on average the PBE values must be scaled by $1.83$ to reproduce the \GW  results. The HSE06 band gaps are closer to \GW  but are nevertheless systematically underestimated by more than 20\%. On the other hand, GLLBSC shows very good performance with band gaps only 2\% smaller than \GW  on average. Table \ref{tab:MAE} shows the mean absolute deviations of the DFT methods relative to \GW. We note that although GLLBSC provides an excellent description of the \GW  band gaps \emph{on average} the spread is sizable with a mean absolute deviation of 0.4 eV.

We note a handful of outliers in Figure~\ref{fig:bs-trends} with large HSE band gaps compared to PBE and \GW. For one of these, namely the ferromagnetic CoBr$_2$-CdI$_2$, we obtain the band gaps: 0.30 eV (PBE), 3.41 eV (HSE), and 0.91 eV (G$_0$W$_0$). For validation, we have performed GPAW and QuantumEspresso calculations with the norm-conserving HGH pseudopotentials and plane wave cutoff up to 1600 eV. The converged band gaps are 0.49 eV (GPAW-HGH-PBE), 0.51 eV (QE-HGH-PBE) and 3.69 eV (GPAW-HGH-HSE), 3.52 eV (QE-HGH-HSE), which are all in reasonable agreement with the C2DB results. It should be interesting to explore the reason for the anomalous behavior of the HSE band gap in these materials. 

Compared to the band gaps, the gap centers predicted by PBE and HSE06 are in overall better agreement with the \GW  results. This implies that, on average, the \GW  correction of the DFT band energies is symmetric on the valence and conduction band. In contrast, the GLLBSC predicts less accurate results for the gap center. This suggests that an accurate and efficient prediction of absolute band energies is obtained by combining the GLLBSC band gap with the PBE band gap center.

Next, we consider the band alignment at the interface between different 2D materials. Assuming that the bands line up with a common vacuum level and neglecting hybridisation/charge transfer at the interface, the band alignment is directly given by the VBM and CBM positions relative to vacuum.

\begin{figure*}
  \begin{subfigure}[b]{0.15\columnwidth}
    \includegraphics[height=5.4cm]{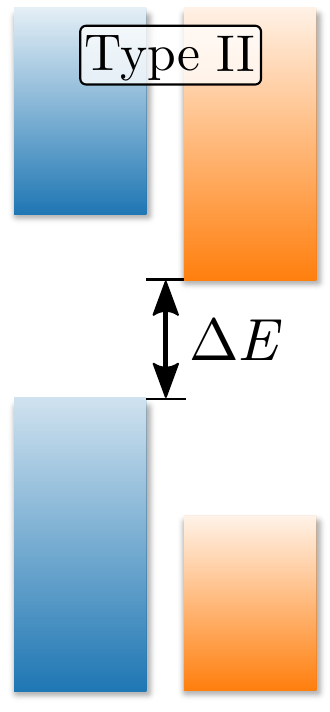}
    \vspace*{.7cm}~
  \end{subfigure}%
  \begin{subfigure}[b]{0.43\columnwidth}
    \includegraphics[width=\columnwidth]{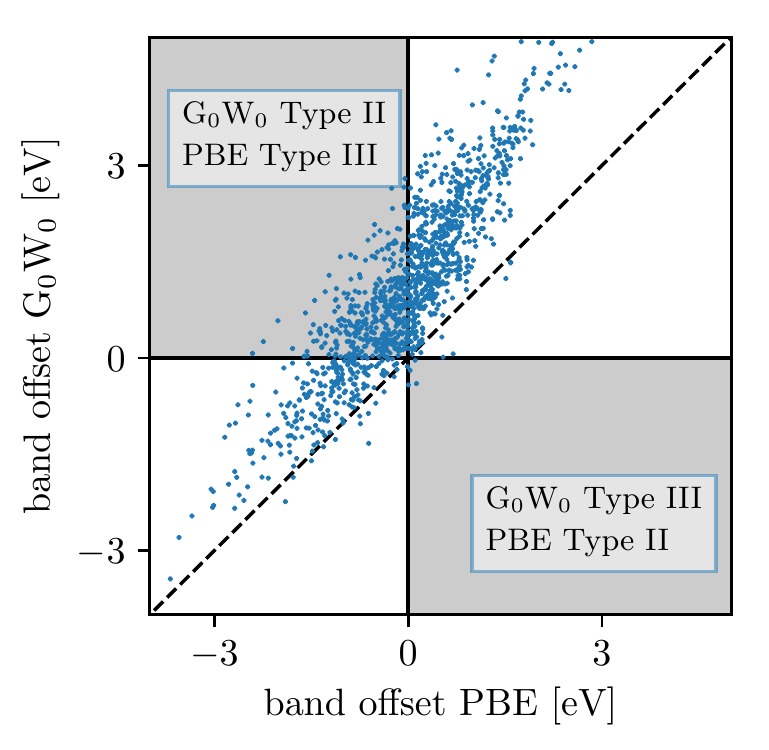}
  \end{subfigure}%
  \begin{subfigure}[b]{0.43\columnwidth}
    \includegraphics[width=\columnwidth]{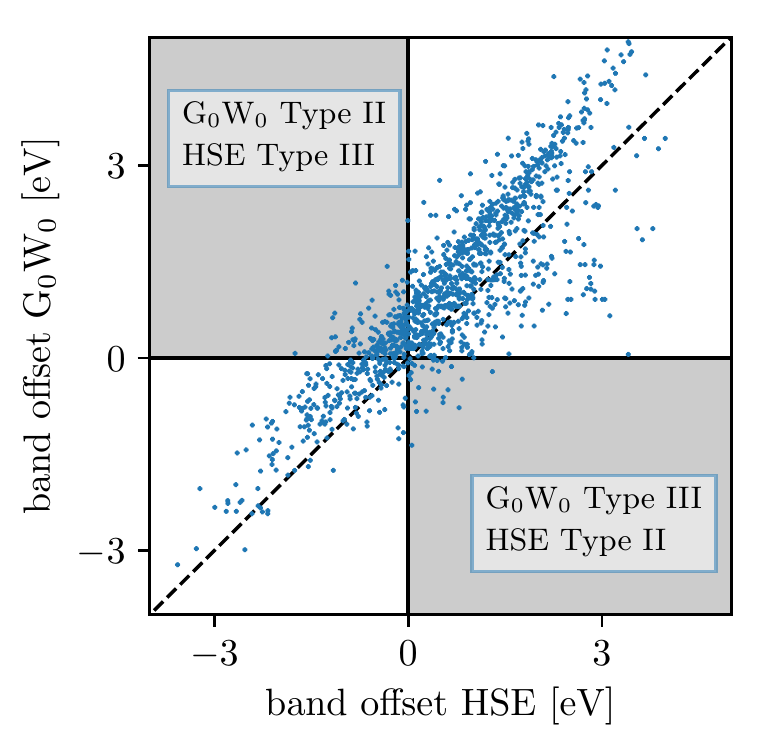}
  \end{subfigure}
\caption{Band alignment at heterobilayers, assuming that the bands line up with a common vacuum level.
The left panel shows the definition of the band offset $\Delta E$.
The middle and right panels show randomly selected bilayer combinations
with Type II ($\Delta E > 0$) or Type III ($\Delta E < 0$) band alignment.
The black line indicates the bisector,
the grey areas indicate qualitatively wrong DFT band alignments compared to \GW\!.}
\label{fig:ba-trends}
\end{figure*}

We focus on pairs of 2D semiconductors for which the \GW  band alignment is either Type II ($\Delta E > 0$) or Type III ($\Delta E < 0$), see Figure~\ref{fig:ba-trends}(left). Out of approximately 10000 bilayers predicted to have Type II band alignment by \GW\!, the PBE and HSE06 functionals predict qualitatively wrong band alignment (i.e. Type III) in 44\% and 21\% cases, respectively (grey shaded areas). In particular, PBE shows a sizable and systematic underestimation of $\Delta E$ as a direct consequence of the underestimation of the band gaps in both monolayers.

\subsubsection{Effective masses and mobilities}\label{sec:mobility}
The carrier mobility relates the drift velocity of electrons or holes to the strength of an applied electric field and is among the most important parameters for a semiconductor material. In general, the mobility is a sample specific property which is highly dependent on the sample purity and geometry, and (for 2D materials) interactions with substrate or embedding layers. Here we consider the phonon-limited mobility, which can be considered as the intrinsic mobility of the material, i.e. the mobility that would be measured in the absence of any sample specific- or external scattering sources.       

\begin{figure}
  \begin{subfigure}[b]{0.5\columnwidth}
    \includegraphics{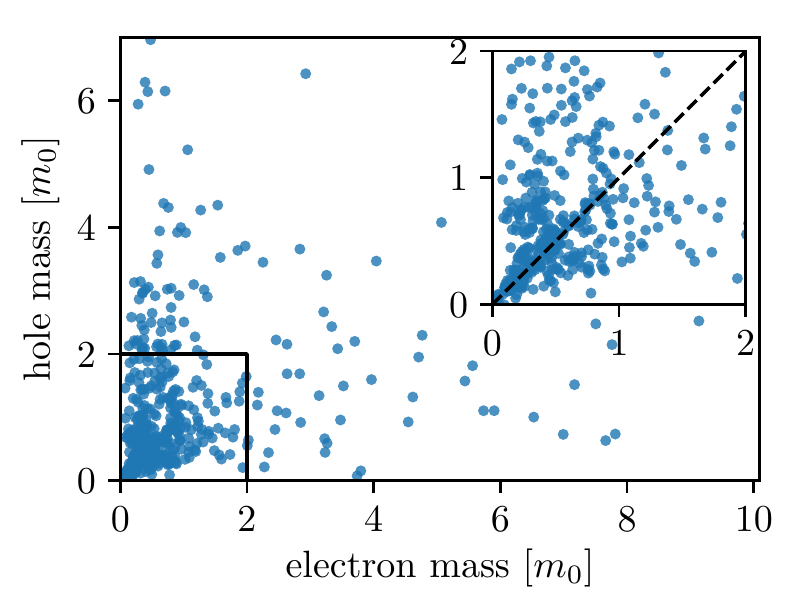}
  \end{subfigure}%
  \begin{subfigure}[b]{0.5\columnwidth}
    \includegraphics{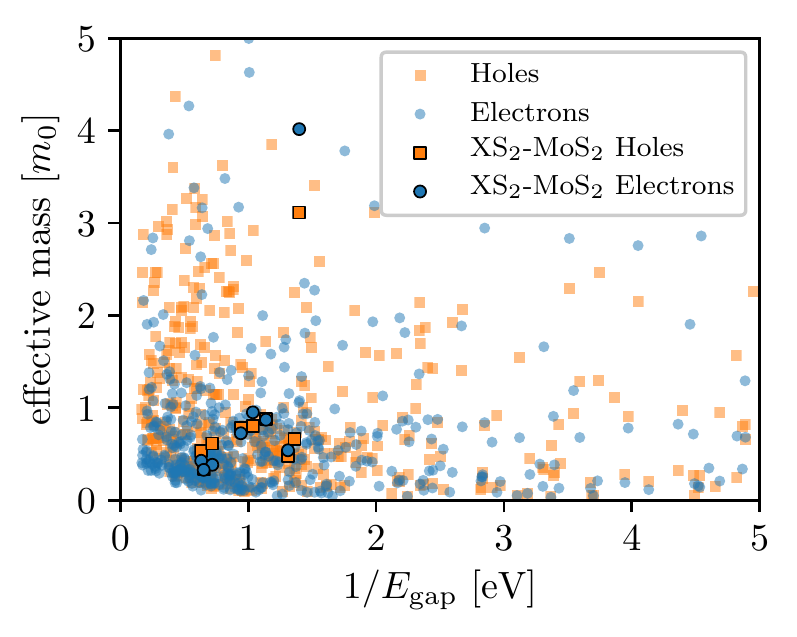}
  \end{subfigure}\caption{(left) Effective electron mass versus hole mass. On average holes are slightly heavier than electrons. (right) Effective carrier mass versus the inverse band gap. The linear correlation between the two quantities, expected from $\mathbf{k}\cdot\mathbf{p}$ perturbation theory, is seen not to hold in general. Materials with the formula XS\(_2\) in the MoS\(_2\) crystal structure are highlighted.}
\label{fig:ehmasses}
\end{figure}

The effective masses of the charge carriers have been calculated both with and without SOC for $\sim 600$ semiconductors. Figure~\ref{fig:ehmasses} shows the electron mass plotted against the hole mass. The data points are scattered, with no clear correlation between the electron and hole masses. Overall, the electron masses are generally slightly smaller than the hole masses. The mean electron mass is 0.9 $m_0$, while the mean hole mass is 1.1 $m_0$, and 80\% of the electron masses are below $m_0$ while the fraction is only 65\% for the holes. This is not too surprising, since, on average, the energetically lower valence band states are expected to be more localised and thus less dispersive than the conduction band states.

The right panel of \figref{fig:ehmasses} shows the effective mass for electrons and holes plotted as a function of the inverse band gap. It can be seen that there is no clear correlation between the two quantities, which is confirmed by calculating the cross-correlation coefficient: for both electrons and holes it is less than 0.02. This provides empirical evidence against the linear relation between effective masses and inverse band gaps derived from $\mathbf{k} \cdot \mathbf{p}$ perturbation theory. The relation is based on the assumption that the perturbative expansion is dominated by the conduction and valence band and that the momentum matrix element between these states, \(\left\langle u_{\mathrm{c}} \middle| \mathbf{\hat{p}}\middle| u_{\mathrm{v}}\right\rangle \), does not vary too much as function of the considered parameter (here the type of material). These assumptions clearly do not hold across a large set of different semiconductors. If we focus on a specific class of materials, e.g. sulfides in the MoS\(_2\) structure indicated by the highlighted symbols, we see a slightly improved trend but still with significant fluctuations.

If one assumes energetically isolated and parabolic bands, the intrinsic mobility limited only by scattering on acoustic phonons can be estimated from the Takagi formula~\cite{Takagi:1994cd}, 
\begin{equation}
\mu_i = \frac{e \hbar^3 \rho v_{i}^2}{k_{\mathrm{B}} T m_i^{*}m^{*}_{\mathrm{d}} D_{i}^2}\label{eq:takagi}.
\end{equation}

Here \(i\) refers to the transport direction, $\rho$ is the mass density, $v_{i}$ is the speed of sound in the material, $m_{i}^*$ is the carrier mass, $m_{\mathrm{d}}^*$ is the equivalent isotropic density-of-state mass defined as $m^*_{\mathrm{d}}=\sqrt{m_x^*m_y^*}$, and $D_{i}$ is the deformation potential. We stress that the simple Takagi formula is only valid for temperatures high enough that the acoustic phonon population can be approximated by the Rayleigh-Jeans law, $n\approx \hbar \omega_{\mathrm{ac}} / k_{\mathrm{B}} T$, but low enough that scattering on optical phonons can be neglected.

\begin{figure}
  \begin{subfigure}[b]{0.5\columnwidth}
    \includegraphics{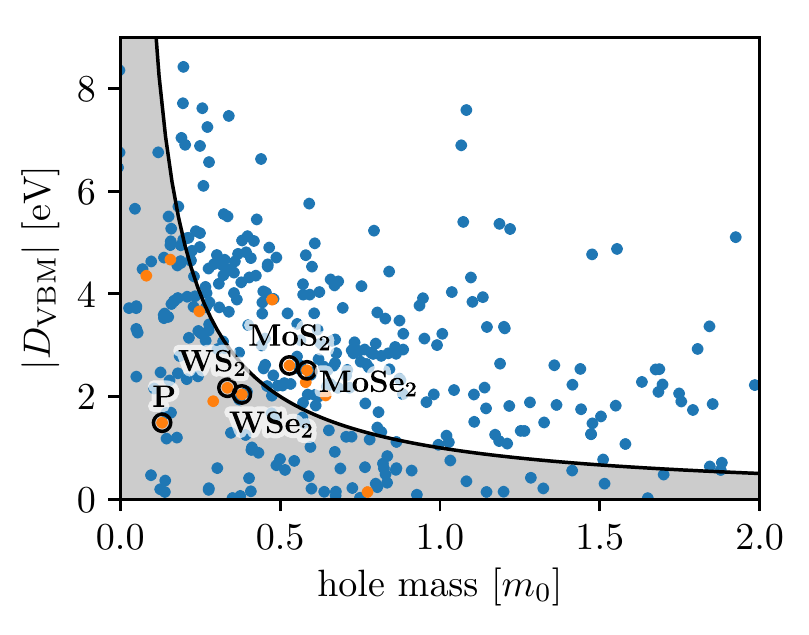}
  \end{subfigure}%
  \begin{subfigure}[b]{0.5\columnwidth}
    \includegraphics{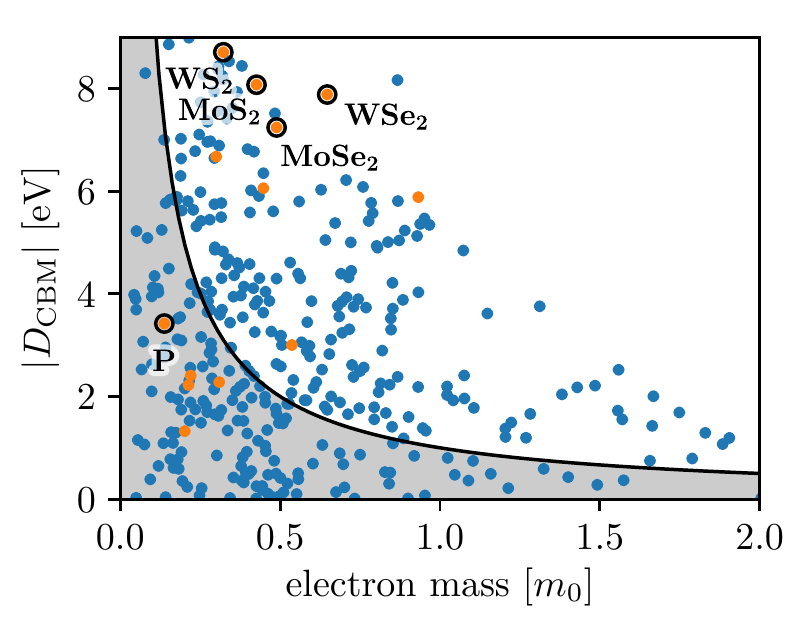}
  \end{subfigure}
\caption{The deformation potential of the materials plotted against the carrier effective masses for the valence band (left) and the conduction band (right). The shaded region in each panel indicates a region of interest in the search for high-mobility semiconductors.\label{fig:D_masses} The 2D materials which have previously been synthesised in monolayer form are highlighted in orange and selected structures for which mobilities have been measured are labelled.}
\end{figure}

For the semiconductors in the C2DB we have found that the denominator of Equation~\eqref{eq:takagi} varies more than the numerator. Consequently, a small product of deformation potential and effective mass is expected to correlate with high mobility. Figure~\ref{fig:D_masses} shows the deformation potential plotted against the carrier mass for the valence and conduction bands, respectively. The shaded area corresponds, somewhat arbitrarily, to the region for which $m_i^* D_i < m_0(1 \,\mathrm{eV})$. The 2D semiconductors which have been synthesised in monolayer form are indicated with orange symbols while those which have been used in field effect transistors are labeled. Consistent with experimental findings, phosphorene (P) is predicted to be among the materials with the highest mobility for both electrons and holes.

Interestingly, a number of previously unknown 2D materials lie in this shaded region and could be candidates for high mobility 2D semiconductors. Table~\ref{tab:mobilities} lists a few selected materials with high intrinsic mobility according to Equation~\eqref{eq:takagi}, which all have ``high'' overall stability (see Section \ref{sec:stabilityscale}). In the future, it will be interesting to explore the transport properties of these candidate materials in greater detail.

To give a sense of scale for the numbers in Table~\ref{tab:mobilities} to scale, we consider the well studied example of MoS$_2$. For this material we obtain an electron mobility of 240 cm\(^2\)V\(^{-1}\)s\(^{-1}\) while a full \emph{ab-initio} calculation found a phonon-limited mobility of 400 cm\(^2\)V\(^{-1}\)s\(^{-1}\) (in good agreement with experiments on hBN encapsulated MoS$_2$\cite{cui2015multi}), with the acoustic phonon contribution corresponding to a mobility of 1000 cm\(^2\)V\(^{-1}\)s\(^{-1}\). Similarly, for the series MX\(_2\) (M = W, Mo, X = S, Se), we calculate room-temperature electron mobilities between 200 cm\(^2\)V\(^{-1}\)s\(^{-1}\) and 400 cm\(^2\)V\(^{-1}\)s\(^{-1}\), which are all within 50\% of the \emph{ab-initio} results\cite{jin2014a}. Presumably, as in the case for MoS$_2$, the good quantitative agreement is partly a result of error cancellation between an overestimated acoustic phonon scattering and the neglect of optical phonon scattering. Importantly, however, the relative ordering of the mobilities of the four MX\(_2\) monolayers is correctly predicted by Equation~\eqref{eq:takagi} for all but one pair (MoS$_2$ and WSe$_2$) out of the six pairs. These results illustrate that Equation~\eqref{eq:takagi} should only be used for "order of magnitude" estimates of the mobility whereas relative comparisons of mobilities in different materials are probably reliable. 

\begin{table}[htb]
  \centering
  \begin{tabular}{lllC{2cm}R{2cm}C{1.0cm}cC{1.5cm}}
    \toprule
    Carrier & Formula & Prototype & PBE gap (eV) & \multicolumn{1}{C{2.2cm}}{\(\mu_{\mathrm{high}}\) (cm\(^2\)V\(^{-1}\)s\(^{-1}\))} & {\(m^*\) (\(m_0\))} & \(\dfrac{\mu_{\mathrm{high}}}{\mu_{\mathrm{low}}}\) \\
    \midrule

Holes     & PbS$_2$  & MoS$_2$ & 1.39 & 30000 & 0.62 & 1.4 \\
          & OsO$_2$  & WTe$_2$ & 0.17 & 23000 & 0.23 & 3.0 \\
          & ZrCl$_2$ & MoS$_2$ & 0.98 & 12000 & 0.43 & 1.1 \\
          & WSSe     & MoSSe   & 1.40 & 5500  & 0.37 & 1.0 \\
    \midrule
Electrons & PtTe$_2$ & CdI$_2$ & 0.30 & 9600  & 0.17 & 1.3 \\
          & GaO      & GaS     & 1.56 & 7200  & 0.32 & 1.0 \\
          & NiS$_2$  & CdI$_2$ & 0.58 & 6000  & 0.29 & 1.5 \\
            & RuTe$_2$ & WTe$_2$ & 0.64 & 4600  & 1.55 & 8.5 \\
    \bottomrule
  \end{tabular}
  \caption{Key transport properties of selected materials with high intrinsic room-temperature mobility according to Equation~(\ref{eq:takagi}). All the materials shown have "high" overall stability as defined in Section~\ref{sec:stabilityscale}. \(\mu_{\mathrm{high}}\) is the larger value of the mobilities in the \(x\) or \(y\) directions, \(m^{*}\) is the corresponding effective mass, and \(\mu_{\mathrm{high}} / \mu_{\mathrm{low}}\) is the ratio of the mobilities in the two directions.}
  \label{tab:mobilities}
\end{table}

\subsubsection{Magnetic properties}
Recently, a single layer of CrI$_3$ was reported to exhibit ferromagnetic order with a Curie temperature of 45 K.\cite{huang2017layer} This comprises the first example of a pure 2D material exhibiting magnetic order and there is currently an intense search for new 2D materials with magnetic order persisting above room temperature.\cite{Gong2017b, Bonilla2018, Huang2018}

For 2D materials, magnetic order will only persist at finite temperatures in the presence of magnetic anisotropy (MA). Indeed, by virtue of the Mermin-Wagner theorem, magnetic order is impossible in 2D unless the rotational symmetry of the spins is broken\cite{mermin-wagner}. A finite MA with an out of plane easy axis breaks the assumption behind the Mermin-Wagner theorem and makes magnetic order possible at finite temperature. The critical temperature for magnetic order in 2D materials will thus have a strong dependence on the anisotropy.

The MA originates from spin-orbit coupling and is here defined as the energy difference between in-plane and out-of-plane orientation of the magnetic moments, see Eq.~\eqref{eq:anisotropy}. With our definition, a negative MA corresponds to an out-of-plane easy axis. We note that most of the materials in the C2DB are nearly isotropic in-plane. Consequently, if the easy axis lies in the plane, the spins will exhibit an approximate in-plane rotational symmetry, which prohibits magnetic order at finite temperatures. Since spin-orbit coupling becomes large for heavy elements, we generally expect to find larger MA for materials containing heavier elements. In general the magnitude of the MA is small. For example, for monolayer CrI$_3$ with a Curie temperature of 45 K\cite{huang2017layer} we find a MA of \(-0.85\) meV per Cr atom in agreement with previous calculations\cite{Lado2017}. Although small, the MA is, however, crucial for magnetic order to emerge at finite temperature.

\begin{figure}
\centering
\includegraphics{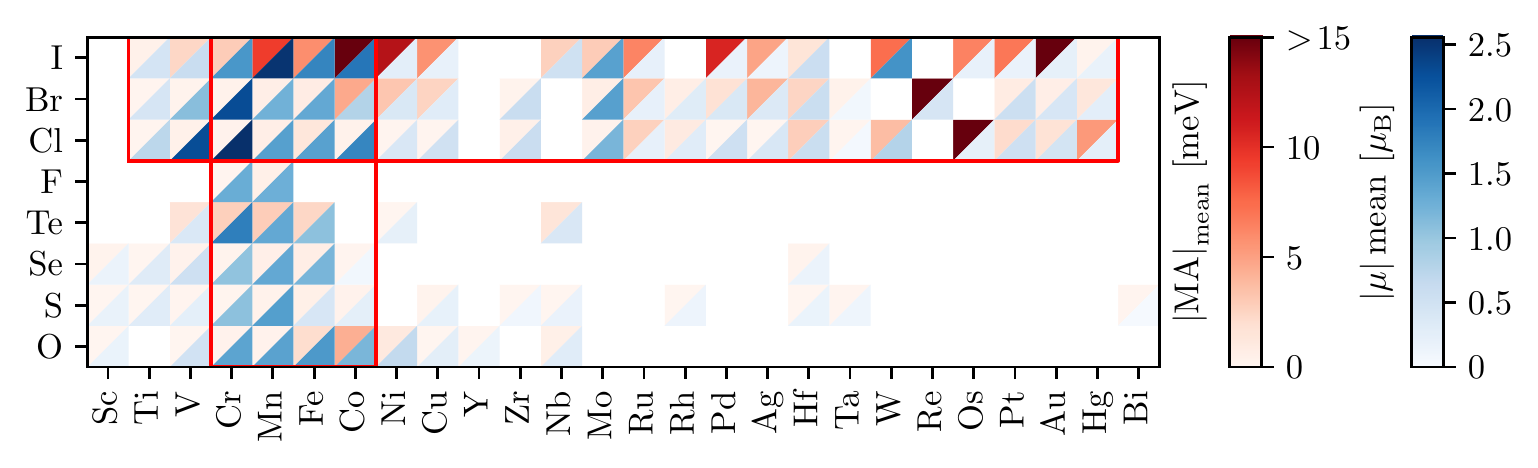}
\caption{Absolute magnetic anisotropy (red) and magnetic moment (blue) averaged over all materials in the C2DB with a given composition. The two red boxes highlight the halides and \(3d\) metals.}
\label{fig:maganis_magmom}
\end{figure}

In Figure~\ref{fig:maganis_magmom} we show the magnitude of the magnetic anisotropy (red triangles) and the magnetic moment per metal atom (blue triangles) averaged over all materials with a given chemical composition. The plot is based on data for around 1200 materials in the medium to high stability categories (see Table \ref{tab:stabilityscale}) out of which around 350 are magnetic. It is interesting to note that while the magnetic moment is mainly determined by the metal atom, the MA depends strongly on the non-metal atom. For example, the halides (Cl, Br, I) generally exhibit much larger MAs than the chalcogenides (S, Se, Te). Overall, iodine (I) stands out as the most significant element for a large MA while the $3d$ metals Cr, Mn, Fe and Co are the most important elements for a large magnetic moment. Since the MA is driven by spin-orbit coupling (SOC) and the spin is mainly located on the metal atom, one would expect a large MA to correlate with a heavy metal atom. However, it is clear from the figure that it is not essential that the spin-carrying metal atom should also host the large SOC. For example, we find large MA for several $3d$ metal-iodides despite of the relatively weak SOC on the $3d$ metals. This shows that the MA is governed by a rather complex interplay between the spins, orbital hybridisation and crystal field.  

A selection of materials predicted to have high overall stability (see Section~\ref{sec:stabilityscale}) and high out-of-plane magnetic anisotropy ($\textrm{MA}< -0.3$ meV/magnetic atom) is listed in Tab.~\ref{tab:maganis}. We find several semiconductors with anisotropies comparable to CrI\(_3\) and several metals with even higher values. Looking also at materials with medium stability, we find semiconductors with anisotropies up to 2 meV. It is likely that some of these materials will have Curie temperatures similar to, or even higher than, CrI$_3$.

\begin{table}[htb]
  \centering
  \definecolor{Grey}{gray}{0.9}
\begin{tabular}[h]{llC{2.2cm}C{2cm}C{2.5cm}C{2cm}r}
  	\toprule
	Formula & Prototype & Magnetic moment (\(\mu_{\mathrm{B}}\)) & PBE gap (eV) & MA (meV/atom)& Magnetic state & \multicolumn{1}{C{2cm}}{\(\Delta H_{\mathrm{hull}}\)  (eV/atom)}\\
	\midrule

  OsI$_3$  & BiI$_3$ & 0.9 & 0.0\phantom{0} & $-3.17$ &	FM	& 0.18    \\
  CrTe	   & FeSe    & 2.6 & 0.0\phantom{0} & $-1.85$ &	AFM	& 0.15    \\
  FeCl$_3$ & BiI$_3$ & 1.0 & 0.01           & $-1.84$ &	FM	& $-0.08$ \\
  FeTe	   & FeSe    & 1.9 & 0.0\phantom{0} & $-1.06$ &	FM	& 0.08    \\
  MnTe$_2$ & CdI$_2$ & 2.7 & 0.0\phantom{0} & $-0.94$ &	FM	& $-0.10$ \\
  FeBr$_3$ & BiI$_3$ & 1.0 & 0.04           & $-0.88$ &	FM	& $-0.04$ \\
\rowcolor{Grey}  \bf{CrI$_{\mathbf{3}}$} & \bf{BiI$_{\mathbf{3}}$} & \bf{3.0} & \bf{0.90} & $-\bf{0.85}$ & \bf{FM} & $-\bf{0.21}$ \\
  MnTe	   & FeSe    & 3.8 & 0.69           & $-0.75$ &	AFM	& $-0.15$ \\
  NiO	   & PbSe    & 1.1 & 0.0\phantom{0} & $-0.53$ &	FM	& 0.05    \\
  FeBrO    & FeOCl   & 1.1 & 0.0\phantom{0} & $-0.47$ &	FM	& $-0.05$ \\
  CrISe	   & FeOCl   & 3.0 & 0.0\phantom{0} & $-0.45$ & FM 	& $-0.10$ \\
  MnSe$_2$ & CdI$_2$ & 2.8 & 0.0\phantom{0} & $-0.40$ &	FM	& $-0.18$ \\
  CrIS	   & FeOCl   & 3.0 & 0.35           & $-0.36$ & FM 	& $-0.10$ \\
  MnO$_2$  & CdI$_2$ & 3.0 & 1.13           & $-0.36$ &	FM	& 0.02    \\
  VCl$_3$  & BiI$_3$ & 2.0 & 0.0\phantom{0} & $-0.35$ &	FM	& $-0.01$ \\
  MnSe	   & FeSe    & 3.7 & 0.90           & $-0.31$ &	AFM	& $-0.20$ \\
  CrSe	   & FeSe    & 2.0 & 0.0\phantom{0} & $-0.31$ &	AFM	& 0.12    \\

  	\bottomrule
\end{tabular}
\caption{Selection of magnetic materials with a negative MA per magnetic atom. The prototype, the magnetic moment of the magnetic atom, the energy gap calculated with PBE xc-functional and the magnetic state are also shown. The experimentally synthesised ferromagnetic monolayer CrI$_3$ is highlighted.}\label{tab:maganis}
\end{table}

In addition to the MA, the critical temperature depends sensitively on the magnetic exchange couplings. We are presently developing a workflow for systematic calculation of exchange coupling constants, which will allow us to estimate the Curie temperature of all the magnetically ordered 2D materials. The database contains several 2D materials with anti-ferromagnetic order. As a note of caution, we mention that the magnetic interactions in AFM materials typically arise from the super-exchange mechanism, which is poorly described by PBE and requires a careful verification using a PBE+U scheme\cite{Olsen2017}.

\subsubsection{Plasmons}\label{sec:stattrendsplasma}
The unique optical properties of 2D materials make them highly interesting as building blocks for nanophotonic applications\cite{low2017polaritons,gjerding2017layered}. Many of these applications involve electron rich components which can capture, focus, and manipulate light via plasmons or plasmon-polaritons. Graphene sheets can host plasmons that are long lived, can be easily tuned via electrostatic or chemical doping, and can be used to confine light to extremely small volumes\cite{chen2012optical}. However, due to the limited charge carrier density achievable in graphene, its plasmons are limited to the mid-infrared regime. Here we show that some metallic monolayers support plasmons with significantly higher energies than graphene and could potentially push 2D plasmonics into the optical regime.

\begin{figure*}
  \begin{subfigure}[b]{0.51\columnwidth}
    \includegraphics{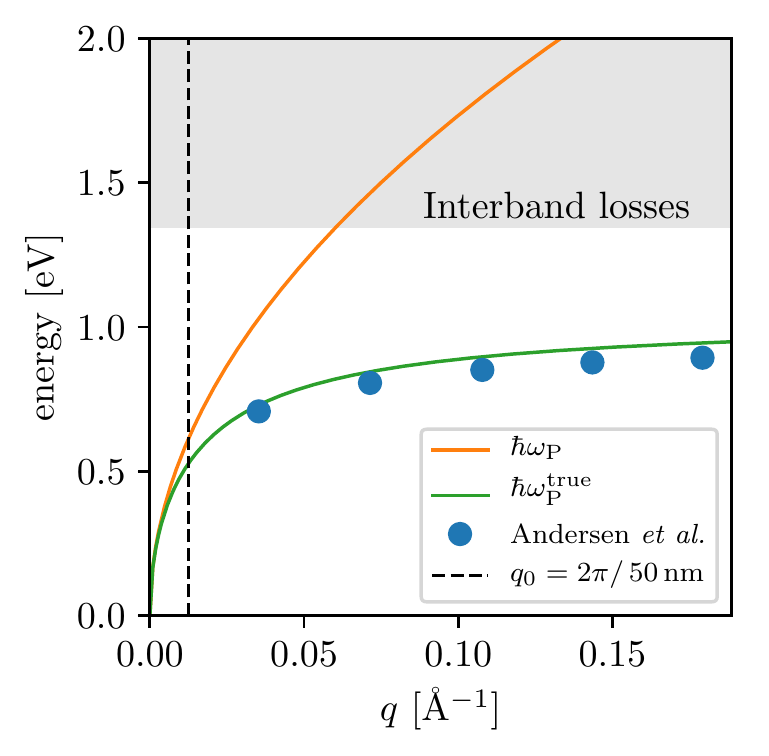}
  \end{subfigure}%
  \begin{subfigure}[b]{0.51\columnwidth}
    \includegraphics{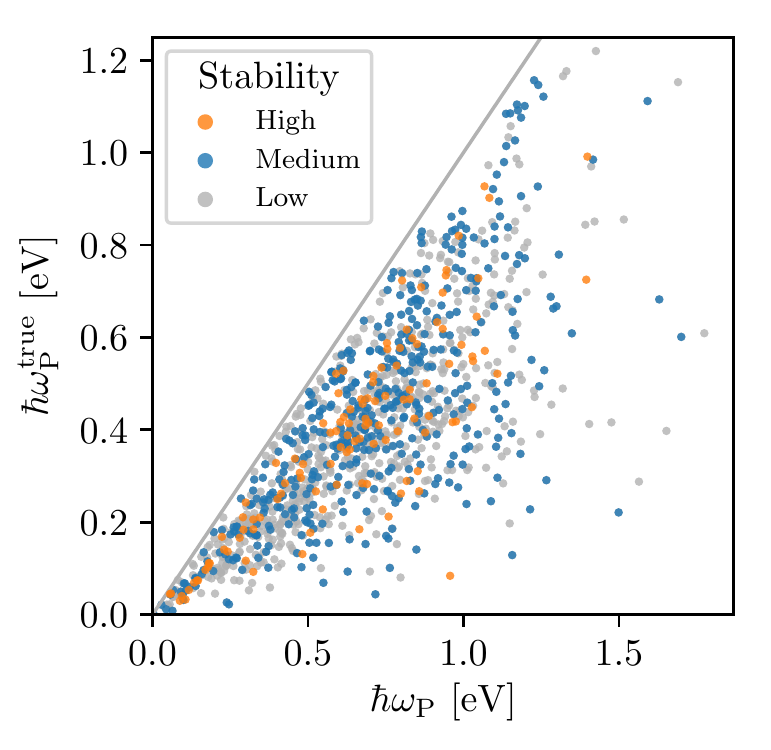}
  \end{subfigure}
\caption{(left) Plasmon dispersion relations for the unscreened (i.e. intraband) and true plasmons, $\hbar\omega_\mathrm{P}$ and $\hbar\omega_\mathrm{P}^\mathrm{true}$, respectively, for NbS$_2$ in the H phase (the MoS$_2$ crystal structure prototype). This is compared to the full first principles calculations of the plasmons in NbS$_2$ by Andersen \textit{et al.} (data points).\cite{andersen2013plasmons} (right) The in-plane averaged true plasmon frequency versus the unscreened plasmon frequency for all metals in C2DB at a plasmon wavelength of $\lambda_\mathrm{P}=50$ nm corresponding to $q_0$ in the left panel. The data points are colored by the overall stability level as defined in Section \ref{sec:stabilityscale}, and the straight line corresponds to $\hbar\omega_\mathrm{P} = \hbar\omega_\mathrm{P}^\mathrm{true}$.}
\label{fig:plasma}
\end{figure*}

Figure~\ref{fig:plasma} (left) shows the plasmon dispersion for monolayer NbS$_2$ in the MoS$_2$ crystal structure. The effect of interband transitions on the plasmon is significant as can be seen by comparison to the pure intraband plasmon ($\hbar\omega_\mathrm{P}$). The true plasmon energies are obtained from the poles of the (long wave length limit) dielectric function including the interband transitions, $\epsilon = 1 + 2 \pi q (\alpha^{2\text{D},\mathrm{intra}} + \alpha^{2\text{D},\mathrm{inter}})$, yielding $\omega^\mathrm{true}_\mathrm{P} = \omega_\mathrm{P,2D}[1 + 2\pi q \alpha^{2\text{D},\mathrm{inter}}(\omega_\mathrm{P}^\mathrm{true})]^{-1/2}$. For simplicity we ignore the frequency dependence of the interband polarisability, i.e. we set $\alpha^{2\text{D},\mathrm{inter}}(\omega_\mathrm{P}^\mathrm{true})\approx\alpha^{2\text{D},\mathrm{inter}}(\omega=0)$, which should be valid for small plasmon energies (far from the onset of interband transitions). The validity of this approximation is confirmed by comparing to the full \emph{ab initio} calculations of Andersen \textit{et al.} (blue dots) which include the full $q$- and $\omega$-dependence.\cite{andersen2013plasmons} The figure shows that interband screening generally reduces the plasmon energy and becomes increasingly important for larger $q$.

Figure~\ref{fig:plasma} (right) shows the in-plane averaged true plasmon energy of all metals in the C2DB plotted against the intraband plasmon energy at a fixed plasmon wavelength of $\lambda_\mathrm{P}=50$ nm (corresponding to $q_0$ at the dashed vertical line in the left panel). For comparison, the plasmon energy of freestanding graphene at this wavelength and for the highest achievable doping level ($E_F=\pm 0.5$ eV relative to the Dirac point) is around 0.4 eV. The data points are colored according to the overall stability level as defined in Section \ref{sec:stabilityscale}. Table \ref{tab:plasmafrequencies} shows a selection of the 2D metals with "high" overall stability (see Section \ref{sec:stabilityscale}) and large plasmon frequencies. We briefly note the interesting band structures of the metals in the FeOCl prototype (not shown) which exhibits band gaps above or below the partially occupied metallic band(s), which is likely to lead to reduced losses in plasmonic applications.\cite{gjerding2017band} A detailed study of the plasmonic properties of the lead candidate materials will be published elsewhere. However, from Figure~\ref{fig:plasma} (right) it is already clear that several of the 2D metals have plasmon energies around 1 eV at $\lambda_\mathrm{P}=50$ nm, which significantly exceeds the plasmon energies in highly doped graphene.

\begin{table}[htb]
\centering
\caption{Selection of 2D metals with high plasmon energies $\omega_\mathrm{P}^\mathrm{true}$ for a plasmon wavelength of $\lambda_\mathrm{P}=50$ nm. The interband screening $\alpha^{2\text{D},\mathrm{inter}}$ at $\omega=0$ and the in-plane averaged 2D plasma frequency $\omega_\mathrm{P,2D}$, which are required to reproduce $\omega_\mathrm{P}^\mathrm{true}$, are also reported.}
\label{tab:plasmafrequencies}
\begin{tabular}{llC{2cm}R{1.2cm}R{1.5cm}R{1.5cm}}
\toprule
  Prototype & Formula & Magnetic state & $\omega_\mathrm{P}^\mathrm{true}$ [eV] & \multicolumn{1}{C{2cm}}{$\alpha^{2\text{D},\mathrm{inter}}$ [\AA]}& $\omega_\mathrm{P,2D}$ [eV\AA]\\
  \midrule

TiS$_3$ & TaSe$_3$ & NM & 0.99 & 12.54           & 12.48 \\
FeOCl   & PdClS    & NM & 0.93 & 4.13            & 9.52  \\
FeOCl   & NiClS    & NM & 0.90 & 5.60            & 9.66  \\
CdI$_2$ & TaS$_2$  & NM & 0.82 & 5.59            & 8.79  \\
FeOCl   & ZrIS     & NM & 0.75 & 7.68            & 8.43  \\
CdI$_2$ & NbS$_2$  & NM & 0.73 & 8.2\phantom{0}  & 8.42  \\
FeOCl   & ZrClS    & NM & 0.73 & 13.6\phantom{0} & 9.35  \\
TiS$_3$ & TaS$_3$  & NM & 0.73 & 34.22           & 12.44 \\
PbSe    & NiO      & FM & 0.72 & 2.9\phantom{0}  & 7.16  \\ \bottomrule
\end{tabular}
\end{table}

\subsubsection{Excitons}
Two-dimensional materials generally exhibit pronounced many-body effects due to weak screening and strong quantum confinement. In particular, exciton binding energies in monolayers are typically an order of magnitude larger than in the corresponding layered bulk phase and it is absolutely crucial to include excitonic effects in order to reproduce experimental absorption spectra.

The electronic screening is characterised by the in-plane 2D polarisability, see Eq.~\eqref{eq:2Dpol}. For a strictly 2D insulator, the screened Coulomb potential can be written as $W^{\mathrm{2D}}(q)=v_c^{\mathrm{2D}}(q)/\epsilon^{\mathrm{2D}}(q)$ with $\epsilon^{\mathrm{2D}}(q)=1+2\pi\alpha^{\mathrm{2D}} q$ and $v_c^{\mathrm{2D}}(q)=2\pi/q$ is the 2D Fourier transform of the Coulomb interaction. The $q$-dependence of $\epsilon^{\mathrm{2D}}$ indicates that the screening is non-local, i.e. it cannot be represented by a $q$-independent dielectric constant, and that Coulomb interactions tend to be weakly screened at large distances (small $q$ vectors).\cite{Cudazzo2011, chernikov2014exciton, Huser2013b} This is in sharp contrast to the case of 3D semiconductors/insulators where screening is most effective at large distances where its effect can be accounted for by a $q$-independent dielectric constant. For a two-band model with isotropic parabolic bands, the excitons can be modeled by a 2D Hydrogen-like (Mott-Wannier) Hamiltonian where the Coulomb interaction is replaced by $W=1/\epsilon r$ and the electron mass is replaced by a reduced excitonic mass $\mu_{\mathrm{ex}}$ derived from the curvature of conduction and valence bands. It has previously been shown that the excitonic Rydberg series of a 2D semiconductor can be accurately reproduced by this model if the dielectric constant, $\epsilon$, is obtained by averaging $\epsilon^{\mathrm{2D}}(q)$ over the extent of the exciton in reciprocal space\cite{olsen2016simple}. For the lowest exciton ($n=1$), the binding energy can then be expressed as
\begin{align}\label{eq:E_B}
E_{\mathrm{B}}=\frac{8\mu_{\mathrm{ex}}}{(1+\sqrt{1+32\pi\alpha^{\mathrm{2D}}\mu_{\mathrm{ex}}/3})^2}.
\end{align}
It has furthermore been demonstrated that this expression gives excellent agreement with a numerical solution of the Mott-Wannier model employing the full $q$-dependent dielectric function, $\epsilon^{\mathrm{2D}}(q)=1+2\pi\alpha^{\mathrm{2D}} q$, for 51 transition metal dichalcogenides.\cite{olsen2016simple} We note that the previous calculations were based on LDA and we generally find that the PBE values for $\alpha^{\mathrm{2D}}$ obtained in the present work are 10-20 \% smaller compared with LDA.

\begin{figure}[h!]
  \begin{subfigure}[b]{0.5\textwidth}
\includegraphics{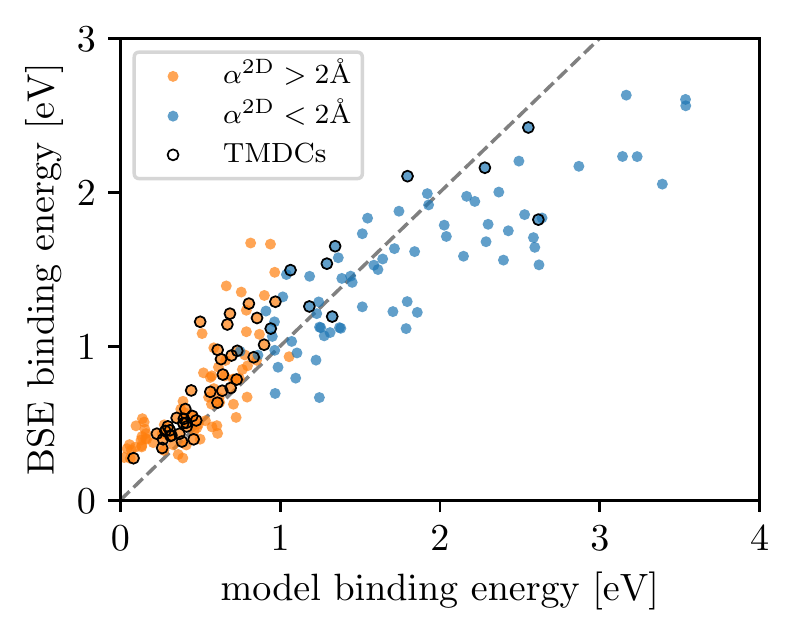}    
\end{subfigure}%
\begin{subfigure}[b]{0.5\textwidth}
\includegraphics{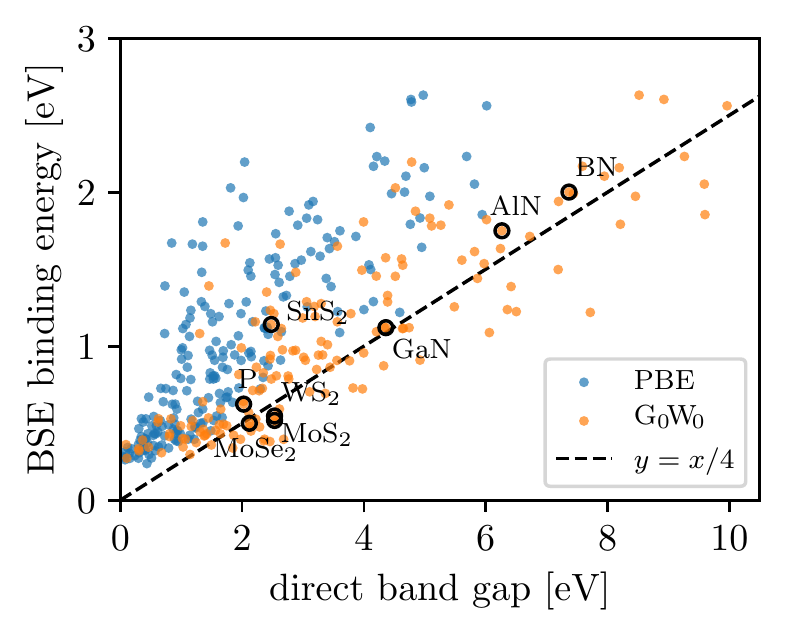}    
  \end{subfigure}

\caption{(left) Binding energy of the lowest exciton in 194 semiconducting monolayers calculated from the Bethe-Salpeter Equation (BSE) and the screened 2D Mott-Wannier model Eq. \eqref{eq:E_B}. The colors of the symbols indicate if screening in the material is weak \((\alpha^{\mathrm{2D}}<2 \mathrm{\AA})\). The transition metal dichalcogenides (TMDCs) are highlighted with black circles. (right) The exciton binding energy as a function of the direct band gap, with selected materials labelled.}
\label{fig:excitons}
\end{figure}

In Figure~\ref{fig:excitons}, we compare the binding energy of the lowest exciton obtained from BSE-PBE with G$_0$W$_0$ scissors operator and the Mott-Wannier model Eq. \eqref{eq:E_B}, respectively. Results are shown for the 194 non-magnetic semiconductors out of the total set of $\sim 250$ materials for which BSE calculations have been performed. We focus on the optically active zero-momentum excitons and compute the exciton masses by evaluating the curvature of the band energies at the \emph{direct gap}, see Section \ref{sec:mass}. For anisotropic materials we average the heavy and light exciton masses as well as the $x$ and $y$ components of the polarisability, $\alpha^{\mathrm{2D}}$, to generate input parameters for the isotropic model Eq. (\ref{eq:E_B}). 

Although a clear correlation with the BSE results is observed, it is also evident that the Mott-Wannier model can produce significant errors. The mean absolute deviation between BSE and the model is $0.28$ eV for all materials and 0.20 eV for the subset of transition metal dichalcogenides (TMDCs). Furthermore, the Mott-Wannier model seems to overestimate $E_{\mathrm{B}}$ for more strongly bound excitons while the opposite trend is seen for weakly bound excitons. As explained below these trends are a consequence of systematic errors in the Mott-Wannier model which can be traced to two distinct sources.

(i) \emph{Weak screening}: If $\alpha^{\mathrm{2D}}$ is small (on the order of 1 \AA), the exciton becomes strongly localised and the orbital character of the states comprising the exciton plays a significant role. In general, the Mott-Wannier model tends to overestimate the exciton binding energy in this case as can be seen from the relatively large deviation of points with model binding energies $>2.0$ eV in Figure~\ref{fig:excitons}. The overestimated binding energy results from the homogeneous electron and hole distributions implicitly assumed in the Mott-Wannier model. In reality, the short range variation of the electron and hole distributions is determined by the shape of the conduction and valence band states. In general these will differ leading to a reduced spatial overlap of the electron and hole and thus a lower Coulomb interaction. For example, SrCl$_2$ in the CdI$_2$ prototype ($\alpha^{\mathrm{2D}}=0.68$ \AA) has a BSE binding energy of 2.1 eV and a model binding energy of 3.4 eV. From the PDOS of this material (see the C2DB webpage) it is evident that the electron and hole are mainly residing on the Sr and Cl atoms, respectively. 

(ii) \emph{Breakdown of the parabolic band approximation}: Materials with small band gaps often exhibit hyperbolic rather than parabolic band structures in the vicinity of the band gap. 
This typically happens in materials with small band gaps such as BSb in the BN prototype.
In Figure~\ref{fig:excitons} these materials can be identified as the cluster of points with model binding energies $<0.25$ eV and BSE binding energies $>0.25$ eV. A similar situation occurs if the conduction and valence bands flatten out away from the band gap region. In both of these cases, the excitons tend to be delocalised over a larger area in the Brillouin zone than predicted by the parabolic band approximation of the Mott-Wannier model. Typically, such delocalisation will result in larger binding energies than predicted by the model. For example, FeI$_2$ in the CdI$_2$ prototype exhibits shallow band minima in a ring around the $\Gamma$-point and has a BSE binding energy of 1.1 eV and a model binding energy of 0.5 eV because the model assumes that the exciton will be located in the vicinity of the shallow minimum (and thus more delocalised in real space). A detailed inspection reveals that such break down of the parabolic band approximation is responsible for most of the cases where the model underestimates the binding energy. 

Other sources of errors come from contributions to the exciton from higher/lower lying bands, i.e. break down of the two-band approximation, and anisotropic exciton masses not explicitly accounted for by Eq. (\ref{eq:E_B}). 

Based on this comprehensive and unbiased assessment of the Mott-Wannier model, we conclude that while the model can be useful for understanding trends and qualitative properties of excitons, its quantitative accuracy is rather limited when applied to a broad set of materials without any parameter tuning. For quantitative estimates $\alpha^{\mathrm{2D}}$ should not be too small (certainly not less than 2 \AA) and the the validity of the effective mass approximation should be carefully checked by inspection of the band structure. 

It has been argued that there should exist a robust and universal scaling between the exciton binding energy and the quasiparticle band gap of 2D semiconductors, namely $E_\mathrm{B} \approx E_{\mathrm{gap}}/4$.\cite{Jiang2017a} This scaling relation was deduced empirically based on BSE-GW calculations for around 20 monolayers and explained from Equation~\eqref{eq:E_B} and the relation $\alpha^{\mathrm{2D}}\propto 1/E_{\mathrm{gap}}$ from $\mathbf k \cdot \mathbf p$ perturbation theory. Another work observed a similar trend\cite{Choi2015} but explained it from the $1/E_{\textrm{gap}}$ dependence of the exciton effective mass expected from $\mathbf k \cdot \mathbf p$ perturbation theory. Based on our results we can completely refute the latter explanation (see Figure \ref{fig:ehmasses}(right)). In Figure \ref{fig:excitons}(right) we show the exciton binding energy plotted versus the direct PBE and \GW band gaps, respectively. While there is a correlation, it is by no means as clear as found in Ref. \citenum{Jiang2017a}.

\section{Conclusions and Outlook}
\label{sec:conclusion}
The C2DB is an open database with calculated properties of two-dimensional materials. It currently contains more than 1500 materials distributed over 32 different crystal structures. A variety of structural, elastic, thermodynamic, electronic, magnetic and optical properties are computed following a high-throughput, semi-automated workflow employing state of the art DFT and many-body methods. The C2DB is growing continuously as new structures and properties are being added; thus the present paper provides a snapshot of the present state of the database. The C2DB can be browsed online using simple and advanced queries, and it can be downloaded freely at \url{https://c2db.fysik.dtu.dk/} under a Creative Commons license. 

The materials in the C2DB comprise both experimentally known and not previously synthesised structures. They have been generated in a systematic fashion by combinatorial decoration of different 2D crystal lattices. The full property workflow is performed only for structures that are found to be dynamically stable and have a negative heat of formation. We employ a liberal stability criterion in order not to exclude potentially interesting materials that could be stabilised by external means like substrate interactions or doping even if they are unstable in freestanding form. As an important and rather unique feature, the C2DB employs beyond-DFT methods, such as the many-body GW approximation, the random phase approximation (RPA) and the Bethe-Salpeter Equation (BSE). Such methods are essential for obtaining quantitatively accurate descriptions of key properties like band gaps and optical spectra. This is particularly important for 2D materials due the weak dielectric screening in reduced dimensions, which tends to enhance many-body effects. For maximal transparency and reproducibility of the data in the C2DB, all relevant parameters have been provided in this paper. Additionally, all scripts used to generate the data are freely available for download under a GPL license.

Beyond its obvious use as a look-up table, the C2DB offers access to numerous well documented, high-quality calculations, making it ideally suited for benchmarking and comparison of different codes and methodologies. The large set of different available properties makes the C2DB interesting as a playground for exploring structure-property relations and for applying and advancing machine learning approaches in materials science. Moreover, the C2DB should be useful as a stepping stone towards the development of theoretical models for more complex 2D structures such as van der Waals heterostructures (see below).

As reported in this work, based on the combinatorial screening approach, we have identified a number of new, potentially synthesisable 2D materials with interesting properties including ferromagnets with large magnetic anisotropy, semiconductors with high intrinsic carrier mobility, and metals with plasmons in the visible frequency range. These predictions are all based on the computed properties of the perfect crystalline materials. While the pristine crystal constitutes an important baseline reference it remains an idealised model of any real material. In the future, it would be interesting to extend the database to monolayers with adsorbed species and/or point defects. Not only would this allow for a more realistic assessment of the magnetic and (opto)electronic properties, it would also facilitate the design and discovery of 2D materials for e.g. battery electrodes and (electro)catalysis\cite{seh2016two,xie2013defect}.     

The C2DB should also be useful as a platform for establishing parametrisations of computationally less expensive methods such as tight-binding models\cite{ridolfi2015tight} and \(\mathbf k\cdot \mathbf p\) perturbation theory\cite{kormanyos2015a}. Such methods are required e.g. for device modeling, description of magnetic field effects, and van der Waals heterostructures. The database already provides band structures, spin orbit-induced band splittings, and effective masses, which can be directly used to determine model parameters. It would be straightforward to complement these with momentum matrix elements at band extrema for modeling of optical properties and construction of full \(\mathbf k\cdot \mathbf p\) Hamiltonians. Similarly, the spread functional required as input for the construction of Wannier functions e.g. by the ASE\cite{larsen2017atomic} or the Wannier90\cite{mostofi2008wannier90} packages, could be easily and systematically produced. This would enable direct construction of minimal basis set Hamiltonians and would allow for the calculation of Born charges and piezoelectric coefficients as well as certain topological invariants\cite{taherinejad2014wannier}. A workflow to calculate exchange couplings of magnetic 2D materials is currently being developed with the aim of predicting magnetic phase transitions and critical temperatures.            

Of specific interest is the modeling of the electronic and optical properties of vdW heterostructures. Due to lattice mismatch or rotational misalignment between stacked 2D layers, such structures are difficult or even impossible to treat by conventional \emph{ab-initio} techniques. Different simplified models have been proposed to describe the electronic bands, including tight-binding Hamiltonians derived from strained lattice configurations\cite{bokdam2014band} and perturbative treatments of the interlayer coupling\cite{tritsaris2016perturbation}. In both cases, the data in the C2DB represents a good starting point for constructing such models. The effect of dielectric screening in vdW heterostructures can be incorporated e.g. by the quantum electrostatic heterostructure (QEH) model\cite{andersen2015dielectric} which computes the dielectric function of the vdW heterostructure from the polarisabilities of the isolated monolayers. The latter are directly available in the C2DB, at least in the long wavelength limit. 

Finally, it would be relevant to supplement the current optical absorbance spectra by other types of spectra, such as Raman spectra, infrared absorption or XPS, in order to assist experimentalists in characterising their synthesised samples. The automatic first-principles calculation of such spectra is, however, not straightforward and will require significant computational investments.   
 
\section{Acknowledgements}
The Center for Nanostructured Graphene is sponsored by the Danish National Research Foundation, Project DNRF103. The project also received funding from the European Unions Horizon 2020 research and innovation programme under Grant Agreement No. 676580 with The Novel Materials Discovery (NOMAD) Laboratory, a European Center of Excellence. This work was also supported by a research grant (9455) from VILLUM FONDEN. This project has received funding from the European Research Council (ERC) under the European Union's Horizon 2020 research and innovation programme (grant agreement No 773122, LIMA).
 
\bibliography{references}
\clearpage

\end{document}